\documentclass[aps,pre,reprint,amsmath,amssymb,longbibliography,lengthcheck,showpacs,superscriptaddress,floatfix]{revtex4-2}
\usepackage[dvipdfmx]{graphicx}
\usepackage{enumerate}
\usepackage{color}
\usepackage{siunitx}
\usepackage[maxfloats=256]{morefloats}
\begin{document}
%
%%%%%%%%%%%%%%%%%%%%%%%%%%%%%%%%%%%%%%%%%%%%%%%%%%%%%%%%%%%%%%%%%%%%%%%%%%%%%%%%%%%%%%%%%%%%%%%%%
\title{Boson peak in covalent network glasses: Isostaticity and marginal stability}
%
%%%%%%%%%%%%%%%%%%%%%%%%%%%%%%%%%%%%%%%%%%%%%%%%%%%%%%%%%%%%%%%%%%%%%%%%%%%%%%%%%%%%%%%%%%%%%%%%%
\author{Hideyuki Mizuno}
\email{hideyuki.mizuno@phys.c.u-tokyo.ac.jp}
\affiliation{Graduate School of Arts and Sciences, The University of Tokyo, Tokyo 153-8902, Japan}
\author{Tatsuya Mori}
\affiliation{Department of Materials Science, University of Tsukuba, Ibaraki 305-8573, Japan}
\author{Giacomo Baldi}
\affiliation{Department of Physics, University of Trento, Via Sommarive 14, 38123 Povo (Tn), Italy}
\author{Emi Minamitani}
\affiliation{SANKEN, Osaka University, Osaka 567-0047, Japan}
%
%%%%%%%%%%%%%%%%%%%%%%%%%%%%%%%%%%%%%%%%%%%%%%%%%%%%%%%%%%%%%%%%%%%%%%%%%%%%%%%%%%%%%%%%%%%%%%%%%
\date{\today}
%
%%%%%%%%%%%%%%%%%%%%%%%%%%%%%%%%%%%%%%%%%%%%%%%%%%%%%%%%%%%%%%%%%%%%%%%%%%%%%%%%%%%%%%%%%%%%%%%%%
\begin{abstract}
The boson peak (BP) stands as a key feature in understanding glasses and amorphous materials.
It directly underlies their anomalous material properties, including thermal behaviors such as excess specific heat and low thermal conductivity, as well as mechanical characteristics such as nonaffine elasticity and fragile plasticity.
Despite its importance, understanding of the BP remains limited in covalent network glasses.
The most promising concepts are isostaticity and marginal stability, which have been established in theories of rigidity percolation and the jamming transition.
While these concepts, supported by comprehensive data, account for the BP in packing-based glasses, comparable explanations have not yet been demonstrated for covalent network glasses.
Here we study silica glass, a prototypical covalent network glass, using molecular dynamics simulations.
We show that the BP in silica glass is governed by near-isostatic constraints and marginal stability, supporting their universality across diverse glassy systems.
Furthermore, we reveal that these principles manifest as a wavenumber-independent band in the dynamical structure factor, and we demonstrate consistency with inelastic X-ray scattering data on silica glass.
Our results provide a unified, experimentally testable framework for deciphering the BP and for refining the interpretation of scattering data in amorphous materials.
\end{abstract}
%
%%%%%%%%%%%%%%%%%%%%%%%%%%%%%%%%%%%%%%%%%%%%%%%%%%%%%%%%%%%%%%%%%%%%%%%%%%%%%%%%%%%%%%%%%%%%%%%%%
\maketitle
%
%%%%%%%%%%%%%%%%%%%%%%%%%%%%%%%%%%%%%%%%%%%%%%%%%%%%%%%%%%%%%%%%%%%%%%%%%%%%%%%%%%%%%%%%%%%%%%%%%
\section{Introduction}
{T}he vibrational states of atoms and molecules in crystals are well characterized as phonons, providing a clear understanding of their material properties~\cite{Ashcroft_1976}.
In contrast, the vibrational states in glasses deviate markedly from phonons, producing anomalous properties relative to crystals~\cite{Ramos_2022}.
Whereas the vibrational density of states~(vDOS) in crystals is accurately captured by Debye theory based on phonons, glasses exhibit an excess over the Debye prediction, an anomaly known as the boson peak~(BP)~\cite{Ramos_2022,Nakayama_2002}.
The BP has been observed in the dynamical structure factor by light, inelastic X-ray, and neutron scattering, establishing it as an experimental hallmark of glasses~\cite{Buchenau_1986,Yamamuro_1996,Arai_1999,Harris_2000,Monaco_2006,Niss_2007,Ruffle_2008,Monaco_2009,Baldi_2008,Baldi_2010,Baldi_2011,Ruta_2012,Baldi_2016}.
The BP is directly linked to the excess specific heat of glasses and strongly influences transport properties including sound propagation and heat conduction~\cite{Zeller_1971,Anderson_1972,Cahill_1988,Monaco2_2009,Mizuno_2014,Tanguy_2023}.
It also affects nonaffine elastic responses and plastic deformation in glasses~\cite{Leonforte_2005,Leonforte_2006,Tanguy_2010,Manning_2011}.
Thus, the BP is central to understanding glasses and amorphous materials.

Understanding of the BP has advanced over decades.
Several theoretical frameworks have been proposed, including elastic heterogeneities~\cite{Schirmacher_2006,Marruzzo_2013,Mizuno2_2013,Schirmacher_2015,Zhang_2017}, soft-potential models~\cite{Karpov_1983,Buchenau_1991,Gurevich_2003,Bouchbinder_2021}, and random-matrix approaches~\cite{Beltukov_2013,Vogel_2023}.
Among these, early works~\cite{Phillips_1979,Dohler_1980,Phillips2_1981,Thorpe_1983} introduced the concept of isostaticity based on the topology of random networks.
They accounted for the rigidity of covalent network glasses by counting constraints~(bonds) against degrees of freedom.
When the number of constraints falls below the number of degrees of freedom, the system can deform continuously at no energy cost, and zero-energy vibrational modes, known as floppy modes, appear.
Conversely, when constraints exceed degrees of freedom, excess constraints render energy-free deformations costly, shifting floppy modes to finite frequencies.
The boundary where constraints equal degrees of freedom defines the isostatic state, the rigidity threshold known as Maxwell's criterion~\cite{Maxwell_1864}.
Thus, by tracking how constraints surpass degrees of freedom, one can assess rigidity and the character of low-frequency vibrational states.
This concept was developed within the framework of rigidity percolation~\cite{Feng_1985,He_1985,Cai_1989,Jacobs_1995} and has also been employed to explain low-energy dynamics and two-level tunneling states in glasses~\cite{Trachenko_1998,Trachenko_2000}.

Subsequently, isostaticity was incorporated into the physics of jammed materials, providing insights into the jamming and glass transitions~\cite{Durian_1995,Hern_2003,Hecke_2010}.
In this context, the topological picture above was unified with energetic considerations through the notion of frustration~\cite{Wyart_2005,Wyart2_2005,Wyart3_2005}.
Dense and disordered packings generate strong repulsive forces, giving rise to frustration within the system.
It has been proposed that glasses are solids that marginally counteract this frustration.
From the vibrational perspective, frustration shifts modes that would be floppy at isostaticity but are pushed to finite frequencies by excess constraints back toward low frequencies.
This produces a gapless vDOS that extends down to zero frequency and signals marginal stability.
These modes contribute to the excess low-frequency vDOS and form the BP.
Consequently, the BP can be viewed as a manifestation of isostaticity and marginal stability.
Building upon this scenario, a mean-field theory was constructed using random spring networks, which successfully explains the BP~\cite{Wyart_2010,Degiuli_2014}.

Numerical simulations have tested this scenario and its mean-field predictions.
In particular, the packing model of harmonic spheres~(HS) has been extensively studied in the context of the jamming transition~\cite{Durian_1995,Hern_2003,Hecke_2010}.
Jamming scaling laws for the vDOS and the BP have been established~\cite{Silbert_2005,Xu_2010,Charbonneau_2016,Mizuno_2017,Mizuno_2018}, consistent with isostaticity and marginal stability~\cite{Wyart_2005,Wyart2_2005,Wyart3_2005} and with mean-field theory~\cite{Wyart_2010,Degiuli_2014}.
Isostaticity and marginal stability have also been confirmed for the BP of Lennard--Jones~(LJ) glass, a representative model of van der Waals glasses~\cite{Xu_2007,Shimada_2018,Wang_2019}, and hard-sphere glass, a representative model of colloidal glasses~\cite{Brito_2006,Brito_2009}.
Across these simulations, quasi-localized vibrations~(QLVs) emerge alongside the BP as a consequence of marginal stability driven by frustration~\cite{Xu_2010,Mizuno_2017,Lerner_2017,Shimada2_2018}.
Thus, the BP is now well rationalized in packing-based glasses, including particle packings, van der Waals glasses, and colloidal glasses, in terms of isostaticity and marginal stability.

In contrast, the BP remains less well understood in network-forming glasses, that is, covalent network glasses such as oxide glasses, especially when compared with packing-based systems including van der Waals and colloidal glasses.
Extensive light, inelastic X-ray, and neutron scattering experiments have measured the dynamical structure factor of covalent network glasses and consistently observed the BP~\cite{Buchenau_1986,Arai_1999,Harris_2000,Monaco_2006,Ruffle_2008,Baldi_2008,Baldi_2010,Baldi_2011,Baldi_2016}.
The concept of isostaticity was originally proposed to interpret these experimental observations in covalent network glasses~\cite{Phillips_1979,Dohler_1980,Phillips2_1981,Thorpe_1983}.
However, a detailed validation of isostaticity, analogous to that achieved for packing, van der Waals, and colloidal glasses, has not yet been performed.
Moreover, marginal stability driven by frustration remains largely unexplored in this class of materials.
In addition, how isostaticity and marginal stability manifest in the experimentally measured dynamical structure factor has not been clarified.
Given the breadth of scattering data on covalent network glasses, establishing the BP mechanism in these materials and its expression in the dynamical structure factor is a crucial step toward bridging theoretical frameworks and experiments.

In this work, we focus on silica~($\text{SiO}_2$) glass, a representative oxide glass, to investigate its BP.
Previous numerical studies have examined the vDOS and QLVs in silica glass~\cite{Taraskin_1997,Richard_2020,Shcheblanov_2021,Guerra_2022}.
In contrast to these studies, our primary goal is to understand the BP in terms of isostaticity and marginal stability.
We further analyze the dynamical structure factor to clarify how isostaticity and marginal stability are expressed in this experimentally accessible observable.
We find that the BP in silica glass follows from near-isostatic constraints and marginal stability, as in HS and LJ glasses, providing strong evidence for the universality of these principles across diverse glassy materials.
We further show that these principles manifest as a wavenumber-independent broad band in the dynamical structure factor, and we demonstrate that this band is consistent with inelastic X-ray scattering data for silica glass.
These results advance understanding of the BP in network-forming glasses and offer practical criteria for interpreting scattering data.

%%%%%%%%%%%%%%%%%%%%%%%%%%%%%%%%%%%%%%%%%%%%%%%%%%%%%%%%%%%%%%%%%%%%%%%%%%%%%%%%%%%
\section{Methods}

%%%%%%%%%%%%%%%%%%%%%%%%%%%%%%%%%%%%%%%%%%%%%%%%%%%%%%%%%%%%%%%%%%%%%%%%%%%%%%%%%%%%%%%%%%%%%%%%%
\subsection{Silica glass}
We perform molecular dynamics~(MD) simulations of silica~($\text{SiO}_2$) glass.
Silica glass consists of $N_\text{Si}$ silicon~(Si) atoms and $N_\text{O}$ oxygen~(O) atoms in three-dimensional space, where $N_\text{O}=2N_\text{Si}$ and the total number of atoms is $N=N_\text{O}+N_\text{Si}=3N_\text{Si}$.
The masses of the Si atom and the O atom are denoted as $m_{\text{Si}}$ and $m_{\text{O}}$, respectively, with the ratio $m_{\text{Si}}/m_{\text{O}}=1.755$.
The interatomic forces are modeled using the BKS family of potentials, originally proposed in Ref.~\cite{Beest_1990} and subsequently modified and extended in several aspects~\cite{Wolf_1992,Wolf_1999,Carre_2007,Carre_2016,Sundararaman_2018}.
BKS-type models are pairwise-additive and comprise a short-range two-body term and a long-range Coulomb term, and therefore they do not include explicit many-body angular contributions that encode covalent directionality.
For a more explicit treatment of covalent bonding, many-body potentials with angular terms have been developed, notably the Vashishta potential~\cite{Vashishta_1990}, which adds three-body interactions associated with O--Si--O and Si--O--Si angles.
In this study we adopt the SHIK parameterization of the BKS family~\cite{Sundararaman_2018}, which has been rigorously tested to reproduce experimental data and first-principles simulations, including thermodynamic quantities, structural properties (radial distribution function, static structure factor, bond-angle distribution), and elastic moduli~\cite{Wolf_1992,Wolf_1999,Carre_2007,Carre_2016,Sundararaman_2018}.
Previous numerical studies have investigated mechanical responses and vibrational states of silica glasses using BKS-type models~\cite{Mantisi_2012,Shcheblanov_2015,Damart_2017}, and a recent simulation explored the effects of densification on the BP~\cite{Hamdaoui_2025}.

The potential comprises two contributions: the short-range term and the long-range Coulomb term.
The short-range term is defined as
\begin{equation}
v_S(r) = A_{\alpha \beta} \exp \left( - B_{\alpha \beta} r \right) - \frac{C_{\alpha \beta}}{r^6} + \frac{D_{\alpha \beta}}{r^{24}},
\end{equation}
where $r$ is the interparticle distance, and $\alpha$ and $\beta$ denote either O or Si.
The parameters $A_{\alpha \beta}$, $B_{\alpha \beta}$, $C_{\alpha \beta}$, and $D_{\alpha \beta}$ are specified in Table~\ref{table3}.
The long-range Coulomb term is given by
\begin{equation}
v_L(r) = \frac{q_\alpha q_\beta}{r},
\end{equation}
where the charge of silicon is set as $q_\text{Si} = 1.7755\,e$ (with $e$ the elementary charge) and, to ensure charge neutrality, the charge of oxygen is set as $q_\text{O} = - q_\text{Si}/2$.
The potentials $v_S(r)$ and $v_L(r)$ are truncated at $r=r_{cS}=8$~\AA\ and $r=r_{cL}=10$~\AA, respectively. 
For $v_L(r)$, this truncation is referred to as Wolf truncation~\cite{Wolf_1992,Wolf_1999}.
To prevent discontinuities in the potential and its force (the first derivative of the potential) at the cutoff distances, $v_S(r)$ and $v_L(r)$ are smoothed as follows:
\begin{equation}
\begin{aligned}
\phi_S(r) &= \left[ v_S(r) - v_S(r_{cS}) - (r - r_{cS}) v'_S(r_{cS}) \right] G_{cS}(r), \\
\phi_L(r) &= \left[ v_L(r) - v_L(r_{cL}) \right] G_{cL}(r),
\end{aligned}
\end{equation}
with the smoothing functions
\begin{equation}
\begin{aligned}
G_{cS}(r) &= \exp \left[ - \frac{\gamma_S^2}{(r - r_{cS})^2} \right], \\
G_{cL}(r) &= \exp \left[ - \frac{\gamma_L^2}{(r - r_{cL})^2} \right],
\end{aligned}
\end{equation}
where $\gamma_S=\gamma_L = 0.2$~\AA.

%%%%%%%%%%%%%%%%%%%%%%%%%%%%%%%%%%%%%%%%%%%%%%%%%%%%%%%
\begin{table}[t]
\caption{\label{table3}
{Parameters for the potential of silica glass.}
$q_\text{Si}=1.7755\,e$, $q_\text{O}=-q_\text{Si}/2$, $r_{cS}=8.0$~\AA, $r_{cL}=10.0$~\AA, and $\gamma_S=\gamma_L=0.2$~\AA.
}
\centering
\renewcommand{\arraystretch}{1.1}
\begin{tabular}{|c|c|c|c|c|}
\hline
 $\alpha$--$\beta$ & $A_{\alpha \beta}$~(eV) & $B_{\alpha \beta}$~(\AA$^{-1}$) & $C_{\alpha \beta}$~(eV\AA$^{6}$) & $D_{\alpha \beta}$~(eV\AA$^{24}$) \\
\hline
 Si--O  & $23107.8$ & $5.098$ & $139.7$ & $66.0$ \\
\hline
 O--O   & $1120.5$  & $2.893$ & $26.1$  & $16800.0$ \\
\hline
 Si--Si & $2797.9$  & $4.407$ & $0.0$   & $3423204.0$ \\
\hline
\end{tabular}
\end{table}
%%%%%%%%%%%%%%%%%%%%%%%%%%%%%%%%%%%%%%%%%%%%%%%%%%%%%%%

We performed three independent MD simulations using LAMMPS~\cite{Plimpton_1995}.
We fixed the mass density of silica glass at $\rho=2.20$~g/cm$^3$ and controlled the temperature with a Nose-Hoover thermostat~\cite{Nose_1984,Hoover_1985}.
Initial Si and O positions were randomized and the system was equilibrated at $T=3500$~K for $100$~ps to obtain a homogeneous liquid.
The system was then quenched to $T=300$~K at a rate of $1$~K/ps, followed by equilibration at $T=300$~K for $100$~ps.
Subsequently, all atomic velocities were set to zero and the configuration was relaxed (energy minimization) to obtain the inherent structure $\vec{r}=[\vec{r}_{1},\vec{r}_{2},\ldots,\vec{r}_{N}]$.

To access a broad range of wavenumbers $q$ and frequencies $\omega$, we used several system sizes ranging from $N=1.5\times10^4$ to $1.2\times10^5$ atoms.
All reported quantities are averages over the three independent configurations.

In addition to silica glass, we analyze HS, LJ, and soft-sphere~(SS) glasses in three dimensions.
The HS model, extensively studied in the context of the jamming transition~\cite{Durian_1995,Hern_2003,Hecke_2010}, is a packing glass known to exhibit isostaticity and marginal stability together with the BP and QLVs~\cite{Wyart_2005,Wyart2_2005,Wyart3_2005,Silbert_2005,Xu_2010,Charbonneau_2016,Mizuno_2017,Mizuno_2018}.
We consider two HS preparations, a high-connectivity sample~(HSH) generated at high pressure and a low-connectivity sample~(HSL) generated at low pressure.
Furthermore, LJ and SS glasses, representative van der Waals systems, have been investigated for their BP and QLVs in prior work~\cite{Xu_2007,Shimada_2018,Wang_2019,Monaco2_2009,Marruzzo_2013,Mizuno_2014,Mizuno2_2013}.

%%%%%%%%%%%%%%%%%%%%%%%%%%%%%%%%%%%%%%%%%%%%%%%%%%%%%%%%%%%%%%%%%%%%%%%%%%%%%%%%%%%%%%%%%%%%%%%%%
\subsection{HS glass}
HS glass is a one-component system in three dimensions that has been extensively studied in the context of the jamming transition~\cite{Durian_1995,Hern_2003,Hecke_2010,Silbert_2005,Xu_2010,Charbonneau_2016,Mizuno_2017,Mizuno_2018}.
Two particles $i$ and $j$ interact via the harmonic potential,
\begin{equation}\label{pot-simple}
\phi(r)=\frac{\epsilon}{2}\left(1-\frac{r}{\sigma}\right)^2 H(\sigma-r),
\end{equation}
where $r=r_{ij}$ is the interparticle distance, $\sigma$ is the particle diameter, and $H$ is the Heaviside step function.
All particles have the same mass $m$.
Physical quantities are measured in units of length $\sigma$, mass $m$, and energy $\epsilon$; accordingly, $q$ and $\omega$ are reported in units of $\sigma^{-1}$ and $\sqrt{\epsilon/(m\sigma^2)}$, respectively.
We prepared HS glasses at pressures $P=0.05$ and $0.005$, referred to as HSH and HSL in this paper.
At the higher pressure $P=0.05$ (HSH), particles are more densely packed and exhibit higher connectivity, whereas at the lower pressure $P=0.005$ (HSL), particles are less dense and have lower connectivity.
As with silica glass, we simulated multiple system sizes, ranging from $N=1.6\times10^4$ to $1.024\times10^6$ particles, to access broad ranges of $q$ and $\omega$.

%%%%%%%%%%%%%%%%%%%%%%%%%%%%%%%%%%%%%%%%%%%%%%%%%%%%%%%%%%%%%%%%%%%%%%%%%%%%%%%%%%%%%%%%%%%%%%%%%
\subsection{LJ glass}
LJ glass is a one-component system in three dimensions, studied in previous works~\cite{Monaco2_2009,Shimada_2018,Mizuno_2013}.
Two particles $i$ and $j$ interact via the Lennard--Jones (LJ) potential,
\begin{equation}\label{eq.potlj}
v(r)=4\epsilon\left[\left(\frac{\sigma}{r}\right)^{12}-\left(\frac{\sigma}{r}\right)^{6}\right],
\end{equation}
where $r=r_{ij}$ is the interparticle distance and $\sigma$ is the particle diameter.
The potential is truncated at $r=r_c=2.5\sigma$.
To avoid artificial anharmonicities caused by the discontinuity at $r=r_c$~\cite{Mizuno2_2016}, we employ the smoothed form
\begin{equation}\label{eq.potsmooth}
\phi(r)=v(r)-v(r_c)-(r-r_c)\,v'(r_c),
\end{equation}
which ensures that both the potential and the force (the derivative of the potential) vanish smoothly at $r=r_c$.
All particles have the same mass $m$.
Physical quantities are measured in units of length $\sigma$, mass $m$, and energy $\epsilon$.
The number density is set to $\hat{\rho}=N/V=1.015$, where $V$ is the system volume.
We simulated several system sizes, ranging from $N=1.6\times10^4$ to $1.024\times10^6$.

%%%%%%%%%%%%%%%%%%%%%%%%%%%%%%%%%%%%%%%%%%%%%%%%%%%%%%%%%%%%%%%%%%%%%%%%%%%%%%%%%%%%%%%%%%%%%%%%%
\subsection{SS glass}
SS glass is a binary mixture of large ($L$) and small ($S$) particles in three dimensions, which we have previously studied in Refs.~\cite{Mizuno2_2013,Mizuno_2014}.
Particles $i$ and $j$ of types $\alpha$ and $\beta$ ($\alpha,\beta\in\{L,S\}$) interact via a 12-inverse-power-law potential,
\begin{equation}\label{eq:sspotential}
v(r)=\epsilon\left(\frac{\sigma_{\alpha\beta}}{r}\right)^{12},
\end{equation}
where $r=r_{ij}$, $\sigma_{\alpha\beta}=(\sigma_\alpha+\sigma_\beta)/2$, and $\sigma_\alpha$ is the diameter of the large or small species ($\sigma_L$ or $\sigma_S$).
The size ratio is $\sigma_S/\sigma_L=0.7$, and the composition is equimolar with $x_{L,S}=N_{L,S}/N=1/2$ and $N=N_L+N_S$.
As in the LJ case, we employ the smoothed form $\phi(r)$ defined in Eq.~(\ref{eq.potsmooth}).
All particles have the same mass $m$.
Physical quantities are measured in units of length $\sigma = \left(\sum_{\alpha,\beta=L,S} x_\alpha x_\beta \sigma_{\alpha \beta}^3 \right)^{1/3}$, mass $m$, and energy $\epsilon$.
The number density is set to $\hat{\rho}=N/V=1.015$.
We simulated several system sizes, ranging from $N=1.6\times10^4$ to $1.024\times10^6$.

%%%%%%%%%%%%%%%%%%%%%%%%%%%%%%%%%%%%%%%%%%%%%%%%%%%%%%%
\begin{table*}[t]
\caption{\label{table1}
{Physical quantities including elastic moduli and Debye values.}
For silica glass, the quantities are measured as mass density $\rho$~(g/cm$^3$), elastic moduli $K,G$~(GPa), sound speeds $c_\alpha$~(m/s), wavenumber $q$~(\AA$^{-1}$), Debye level $A_D$~(THz$^{-3}$), and frequency $\omega$~(THz).
See also Table~S1 of SI for values of the overconstrained-network system and the isostatic-network system.
}
\centering
\renewcommand{\arraystretch}{2.0}
\setlength{\tabcolsep}{3pt}
\begin{tabular}{|c|c|c|c|c|c|c|c|c|c|c|c|c|c|c|c|c|c|c|}
\hline
 Glass & $\rho$ & $K$ & $K_A$ & $K_N$ & ${\displaystyle \frac{K_N}{K_A}}~(\%)$ & $G$ & $G_A$ & $G_N$ & ${\displaystyle \frac{G_N}{G_A}}~(\%)$ & $\nu$ & $c_L$ & $c_T$ & ${\displaystyle \frac{c_L}{c_T}}$ & $q_D$ & $A_D$ & $\omega_D$ & $\omega_\text{BP}$ & $\omega_0$ \\
\hline
Silica & $2.20$ & $40.9$ & $172$ & $131$ & $76.3$ & $30.5$ & $104$ & $73.6$ & $73.7$ & $0.202$ & $6090$ & $3720$ & $1.64$ & $1.58$ & $0.00257$ & $10.5$ & $1.21$ & $0.46$ \\
\hline
HSH & $1.40$ & $0.544$ & $0.674$ & $0.131$ & $19.4$ & $0.122$ & $0.344$ & $0.222$ & $64.4$ & $0.395$ & $0.712$ & $0.296$ & $2.40$ & $4.36$ & $0.965$ & $1.46$ & $0.0970$ & $0.033$ \\
\hline
HSL & $1.25$ & $0.332$ & $0.478$ & $0.145$ & $30.4$ & $0.0406$ & $0.281$ & $0.240$ & $85.5$ & $0.441$ & $0.556$ & $0.180$ & $3.09$ & $4.20$ & $4.70$ & $0.81$ & $0.0379$ & $-$ \\
\hline
LJ & $1.015$ & $61.2$ & $61.7$ & $0.530$ & $0.859$ & $13.6$ & $35.8$ & $22.2$ & $61.9$ & $0.396$ & $8.84$ & $3.67$ & $2.41$ & $3.92$ & $0.000699$ & $16.3$ & $1.05$ & $0.38$ \\
\hline
SS & $1.015$ & $40.8$ & $40.8$ & $0.00$ & $0.00$ & $6.21$ & $14.7$ & $8.50$ & $57.8$ & $0.428$ & $6.96$ & $2.47$ & $2.81$ & $3.92$ & $0.00225$ & $11.0$ & $0.798$ & $0.35$ \\
\hline
\end{tabular}
\end{table*}
%%%%%%%%%%%%%%%%%%%%%%%%%%%%%%%%%%%%%%%%%%%%%%%%%%%%%%%

%%%%%%%%%%%%%%%%%%%%%%%%%%%%%%%%%%%%%%%%%%%%%%%%%%%%%%%%%%%%%%%%%%%%%%%%%%%%%%%%%%%%%%%%%%%%%%%%%
\subsection{Vibrational modes}
Using the inherent structure $\vec{r}=[\vec{r}_{1},\vec{r}_{2},\cdots,\vec{r}_{N}]$, we perform a standard normal-mode analysis by solving the eigenvalue problem of the dynamical matrix (a $3N\times3N$ matrix) to obtain the eigenvalues $\lambda_k$ and eigenvectors $\vec{e}_k=[\vec{e}_{k,1},\vec{e}_{k,2},\cdots,\vec{e}_{k,N}]$ for modes $k=1,2,\cdots,3N$~\cite{MizunoIkeda2022}.
We remove the three zero-frequency translational modes from the analysis.
The eigenfrequencies are $\omega_k=\sqrt{\lambda_k}$, and the eigenvectors are orthonormalized such that $\vec{e}_k\cdot\vec{e}_\ell=\sum_{i=1}^{N}\vec{e}_{k,i}\cdot\vec{e}_{\ell,i}=\delta_{k\ell}$, where $\delta_{k,l}$ is the Kronecker delta.

In the present work, we analyze several system sizes, ranging from $N\sim10^4$ to $10^6$, as in Refs.~\cite{Mizuno_2017,Shimada_2018}.
We compute all vibrational modes for the smallest system and only the low-frequency modes for larger systems.
We then merge the mode datasets from different sizes as a function of $\omega_k$; results from different sizes connect smoothly, thereby extending access to the lower-frequency regime.
$g(\omega)$ in Eq.~(\ref{eq.vdos}) and $S_\alpha(q,\omega)$ in Eq.~(\ref{eq.sqomega}) are computed from these combined datasets, and results from different sizes are presented together in the figures.

%%%%%%%%%%%%%%%%%%%%%%%%%%%%%%%%%%%%%%%%%%%%%%%%%%%%%%%%%%%%%%%%%%%%%%%%%%%%%%%%%%%%%%%%%%%%%%%%%
\subsubsection{vDOS}
From the set of eigenfrequencies $\{\omega_k\}$ for vibrational modes $k=1,2,\ldots,3N$, we compute the vDOS $g(\omega)$ as
\begin{equation}\label{eq.vdos}
g(\omega)=\frac{1}{3N}\sum_{k=1}^{3N}\delta(\omega-\omega_k),
\end{equation}
where $\delta$ denotes the Dirac delta function.

%%%%%%%%%%%%%%%%%%%%%%%%%%%%%%%%%%%%%%%%%%%%%%%%%%%%%%%%%%%%%%%%%%%%%%%%%%%%%%%%%%%%%%%%%%%%%%%%%
\subsubsection{Debye vDOS}
In an isotropic elastic medium, vibrational modes are phonons~\cite{Ashcroft_1976}.
Continuum elasticity gives the linear dispersions $\omega=c_T q$ and $\omega=c_L q$, where $c_T$ and $c_L$ are the transverse and longitudinal sound speeds, respectively.
Counting phonon states yields the Debye vDOS,
\begin{equation}\label{equofdebye}
g_D(\omega)=A_D\,\omega^{2}=\frac{3}{\omega_D^3}\,\omega^{2}\propto\omega^{2},
\end{equation}
where $A_D=3/\omega_D^3$ is the Debye level, and the Debye frequency $\omega_D$ is
\begin{equation}
\omega_D=\left(\frac{c_L^{-3}+2c_T^{-3}}{3}\right)^{-1/3} q_D
=\left(\frac{18\pi^2\hat{\rho}}{c_L^{-3}+2c_T^{-3}}\right)^{1/3},
\end{equation}
with the Debye wavenumber $q_D=(6\pi^2\hat{\rho})^{1/3}$ and the number density $\hat{\rho}=N/V$.

%%%%%%%%%%%%%%%%%%%%%%%%%%%%%%%%%%%%%%%%%%%%%%%%%%%%%%%%%%%%%%%%%%%%%%%%%%%%%%%%%%%%%%%%%%%%%%%%%
\subsubsection{Stretching and compression of Si--O bonds in silica glass}
For silica glass, we quantify the stretching and compression of Si--O bonds in vibrational mode $k$ via the dimensionless measure $\delta e_{k,\text{Si--O}}^{2}$, defined as
\begin{equation}
\delta e_{k,\text{Si--O}}^{2}
= \frac{m_\text{O}}{N_{\text{Si--O}}}
\sum_{\langle ij\rangle \in \mathcal{B}_{\text{Si--O}}}
\left|
\left(
\frac{\vec{e}_{k,i}}{\sqrt{m_\text{Si}}}
-
\frac{\vec{e}_{k,j}}{\sqrt{m_\text{O}}}
\right)\!\cdot\!{\vec{n}}_{ij}
\right|^{2},
\end{equation}
where $\mathcal{B}_{\text{Si--O}}$ is the set of nearest-neighbor Si--O pairs, $N_{\text{Si--O}}=|\mathcal{B}_{\text{Si--O}}|=4N_\text{Si}$ is the total number of Si--O bonds, and ${\vec{n}}_{ij}=(\vec{r}_i-\vec{r}_j)/|\vec{r}_i-\vec{r}_j|$ is the unit vector along the Si--O bond, with $i$ the Si atom and $j$ the O atom.
This quantity captures the mean-squared relative displacement of a Si atom and its O neighbor projected along the bond direction; the mass factors render it dimensionless.

%%%%%%%%%%%%%%%%%%%%%%%%%%%%%%%%%%%%%%%%%%%%%%%%%%%%%%%%%%%%%%%%%%%%%%%%%%%%%%%%%%%%%%%%%%%%%%%%%
\subsubsection{Participation ratio}
We measure the fraction of particles that participate in vibrational mode $k$ using the participation ratio $\mathcal{P}_k$~\cite{MizunoIkeda2022}:
\begin{equation}
\mathcal{P}_k
= \frac{1}{N}\left[\sum_{i=1}^{N}\left(\vec{e}_{k,i}\cdot\vec{e}_{k,i}\right)^{2}\right]^{-1}.
\end{equation}
For reference, $\mathcal{P}_k=1$ corresponds to a perfectly extended mode with equal amplitude on all particles, whereas $\mathcal{P}_k=1/N$ corresponds to a mode localized on a single particle.

%%%%%%%%%%%%%%%%%%%%%%%%%%%%%%%%%%%%%%%%%%%%%%%%%%%%%%%%%%%%%%%%%%%%%%%%%%%%%%%%%%%%%%%%%%%%%%%%%
\subsection{Dynamical structure factors}
From the data of $\omega_k$ and $\vec{e}_k=[\vec{e}_{k,1},\vec{e}_{k,2},\cdots,\vec{e}_{k,N}]$, we compute the dynamical structure factors $S_\alpha(q,\omega)$, where $\alpha\in\{T,L\}$ denotes the transverse and longitudinal polarizations, respectively, following~\cite{Grest_1982,MizunoIkeda2022}:
\begin{equation}\label{eq.sqomega}
S_\alpha(q,\omega)=\frac{k_B T}{2N}\frac{q^2}{\omega^2}\sum_{k=1}^{3N}F_{k,\alpha}(q)\,\delta\!\left(\omega-\omega_k\right),
\end{equation}
with
\begin{equation}\label{eq.sqomega2}
\begin{aligned}
F_{k,T}(q)&=\left|\sum_{i=1}^{N}\left(\frac{\vec{e}_{k,i}}{\sqrt{m_i}}\times\hat{\vec{q}}\right)\exp\!\big(\mathrm{i}\vec{q}\cdot\vec{r}_i\big)\right|^{2},\\
F_{k,L}(q)&=\left|\sum_{i=1}^{N}\left(\frac{\vec{e}_{k,i}}{\sqrt{m_i}}\cdot\hat{\vec{q}}\right)\exp\!\big(\mathrm{i}\vec{q}\cdot\vec{r}_i\big)\right|^{2},
\end{aligned}
\end{equation}
where $k_B$ is the Boltzmann constant; $q=|\vec{q}|$ and $\omega$ are the wavenumber and frequency; and $\hat{\vec{q}}=\vec{q}/q$.
It is important to note that scattering experiments typically access the longitudinal component $S_L(q,\omega)$~\cite{Suck_1992}.

Using $S_T(q,\omega)$ and $S_L(q,\omega)$ in Eqs.~(\ref{eq.sqomega}) and~(\ref{eq.sqomega2}), we obtain
\begin{equation}\label{eq.vdossqo1}
\begin{aligned}
&\int_{0}^{q_D}\!\!\left\{\frac{S_T(q,\omega)}{k_B T}+\frac{S_L(q,\omega)}{k_B T}\right\}\,dq \\
&= \frac{1}{2N\omega^{2}}\sum_{k=1}^{3N}\delta\!\left(\omega-\omega_k\right)
\left(\sum_{i,j=1}^{N}\frac{\vec{e}_{k,i}\cdot\vec{e}_{k,j}}{\sqrt{m_i m_j}}
\int_{0}^{q_D}\!e^{\mathrm{i}\vec{q}\cdot(\vec{r}_i-\vec{r}_j)}q^{2}\,dq\right).
\end{aligned}
\end{equation}
Assuming an isotropic elastic medium, we have
\begin{equation}\label{eq.isoem}
\begin{aligned}
&\int_{0}^{q_D}\!e^{\mathrm{i}\vec{q}\cdot(\vec{r}_i-\vec{r}_j)}q^{2}\,dq
=\frac{1}{4\pi}\int_{0\le|\vec{q}|\le q_D}\!e^{\mathrm{i}\vec{q}\cdot(\vec{r}_i-\vec{r}_j)}\,d^{3}\vec{q}\\
&=\frac{1}{4\pi}\delta_{i,j}\!\left(\int_{0\le|\vec{q}|\le q_D}\!d^{3}\vec{q}\right)
=\frac{q_D^{3}}{3}\,\delta_{i,j}.
\end{aligned}
\end{equation}
Substituting Eq.~(\ref{eq.isoem}) into Eq.~(\ref{eq.vdossqo1}) and using $g(\omega)$ from Eq.~(\ref{eq.vdos}), we obtain
\begin{equation}\label{eq.vdossqo2}
\begin{aligned}
&\int_{0}^{q_D}\!\!\left\{\frac{S_T(q,\omega)}{k_B T}+\frac{S_L(q,\omega)}{k_B T}\right\}\,dq \\
&= \frac{q_D^{3}}{2\omega^{2}}\frac{1}{3N}\sum_{k=1}^{3N}\delta\!\left(\omega-\omega_k\right)
\left(\sum_{i=1}^{N}\frac{|\vec{e}_{k,i}|^{2}}{m_i}\right)
= \frac{q_D^{3}}{2M(\omega)}\,\frac{g(\omega)}{\omega^{2}},
\end{aligned}
\end{equation}
where we define the effective mass $M(\omega)$ via $M^{-1}=\sum_{i=1}^{N}|\vec{e}_{k,i}|^{2}/m_i$, evaluated at $\omega=\omega_k$.
Thus $M(\omega)$ depends on the mode $k$ and hence on the frequency.
Equation~(\ref{eq.vdossqo2}) corresponds to Eq.~(\ref{eq_vdostl}).

In theoretical analyses~\cite{Schirmacher_2006,Marruzzo_2013,Schirmacher_2015,Mizuno_2018,Wyart_2010,Degiuli_2014,Baldi_2011}, Eq.~(\ref{eq_vdostl}) is applied to compute the vDOS from the dynamical structure factor (or from Green's functions).
As shown above, Eq.~(\ref{eq_vdostl}) rests on the isotropic-medium assumption in Eq.~(\ref{eq.isoem}), which is generally a good approximation in the low-frequency regime.

%%%%%%%%%%%%%%%%%%%%%%%%%%%%%%%%%%%%%%%%%%%%%%%%%%%%%%%%%%%%%%%%%%%%%%%%%%%%%%%%%%%%%%%%%%%%%%%%%
\subsection{Elastic moduli and Debye values}
We calculate the elastic moduli, the bulk modulus $K$ and the shear modulus $G$, using the harmonic formulation based on the linear response theory~\cite{Lemaitre_2006,Mizuno3_2016,MizunoIkeda2022}.
From these moduli, we compute Debye quantities such as the Debye frequency $\omega_D$ and the Debye level $A_D$.

Table~\ref{table1} summarizes the elastic moduli and related quantities.
The elasticity of glasses involves not only affine but also nonaffine deformation~\cite{Lemaitre_2006,Zaccone_2011,Mizuno_2013,Mizuno3_2016}.
Consequently, an elastic modulus $M$ decomposes as $M = M_A - M_N$, where $M_A$ and $M_N$ are the affine and nonaffine contributions.
HS (HSH and HSL), LJ, and SS glasses exhibit strong nonaffine contributions to their shear moduli, whereas the nonaffine contribution to the bulk modulus is comparatively small; notably, in LJ and SS glasses the bulk modulus is almost entirely determined by the affine component.

By contrast, silica glass exhibits significant nonaffine contributions to both bulk and shear moduli: the nonaffine components exceed $70\%$ of the affine components.
Such strong nonaffinity in both moduli arises from the lack of centrosymmetry in the tetrahedral network structure~\cite{Milkus_2016,Krausser_2017}, which has also been observed in amorphous silicon~\cite{Minamitani_2025} and in noncentrosymmetric crystals such as $\alpha$-quartz~\cite{Cui_2019}.
As a result, the bulk and shear moduli become closer in magnitude.
The Poisson's ratio $\nu = (3K - 2G)/(6K + 2G)$ is approximately $0.2$, about half the value $\nu \approx 0.4$ observed in HS, LJ, and SS glasses.
Values near $\nu=0.2$ categorize silica glass as a strong glass, whereas $\nu \approx 0.4$ indicates that HS, LJ, and SS glasses are fragile glasses, as discussed in Refs.~\cite{Greaves_2011,Duval_2013}.

%%%%%%%%%%%%%%%%%%%%%%%%%%%%%%%%%%%%%%%%%%%%%%%%%%%%%%%
\begin{figure*}[t]
\centering
\includegraphics[width=1.0\textwidth]{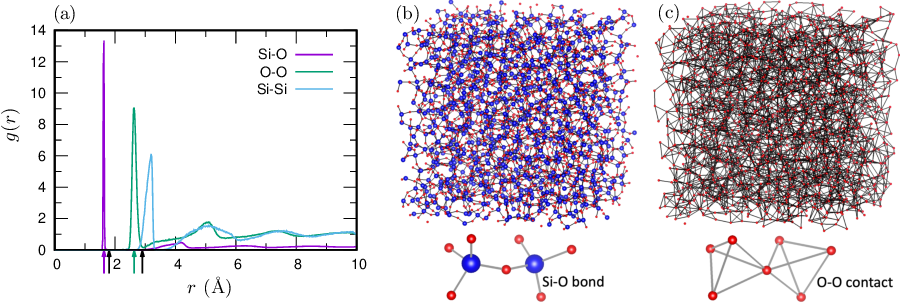}
\caption{\label{fig3_shik_gr}
{Atomic configuration in the inherent structure of silica glass.}
(a) The radial distribution function $g(r)$ is plotted as a function of distance $r$ for Si--O (purple), O--O (green), and Si--Si (cyan) pairs.
For clarity, the Si--O curve is scaled by a factor of $0.2$ (\textit{i.e.,} $0.2\,g(r)$).
The purple and green arrows mark the first peaks at $r\approx1.6$~\AA\ (Si--O) and $r\approx2.6$~\AA\ (O--O), which indicate the typical nearest-neighbor separations.
The black arrows indicate the distance cutoffs used to define connectivity: $r\le1.8$~\AA\ for Si--O {bonds} and $r\le2.9$~\AA\ for O--O {contacts}, chosen at the first minima following these peaks.
(b) Atomic configuration rendered with Si--O bonds.
(c) Atomic configuration rendered with O--O contacts; only O atoms are shown to highlight the O network that outlines the tetrahedral units (Si centers not shown).
The overall structure consists of corner-sharing $\text{SiO}_4$ tetrahedra.
}
\end{figure*}
%%%%%%%%%%%%%%%%%%%%%%%%%%%%%%%%%%%%%%%%%%%%%%%%%%%%%%%

%%%%%%%%%%%%%%%%%%%%%%%%%%%%%%%%%%%%%%%%%%%%%%%%%%%%%%%
\begin{figure}[t]
\centering
\includegraphics[width=0.425\textwidth]{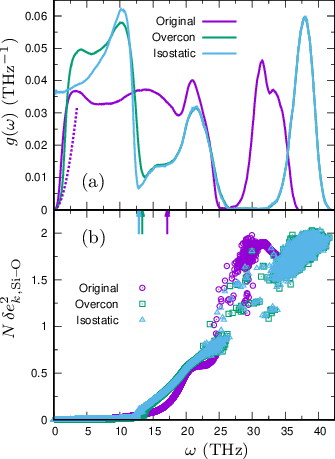}
\caption{\label{fig3_shik_dosbond}
{Vibrational states in silica glass.}
(a) $g(\omega)$ and (b) $N\,\delta e_{k,\text{Si--O}}^{2}$ are plotted as functions of frequency $\omega$ for the original system (purple), the overconstrained-network system (green), and the isostatic-network system (cyan).
Vertical arrows mark the frequencies above which the cumulative number of modes reaches $4N_{\text{Si}}$ out of the total $9N_{\text{Si}}=3N$ modes.
In (a), the purple dotted line indicates the Debye vDOS, $A_D\,\omega^{2}$, for the original system.
}
\end{figure}
%%%%%%%%%%%%%%%%%%%%%%%%%%%%%%%%%%%%%%%%%%%%%%%%%%%%%%%

%%%%%%%%%%%%%%%%%%%%%%%%%%%%%%%%%%%%%%%%%%%%%%%%%%%%%%%
\begin{figure*}[t]
\centering
\includegraphics[width=1.0\textwidth]{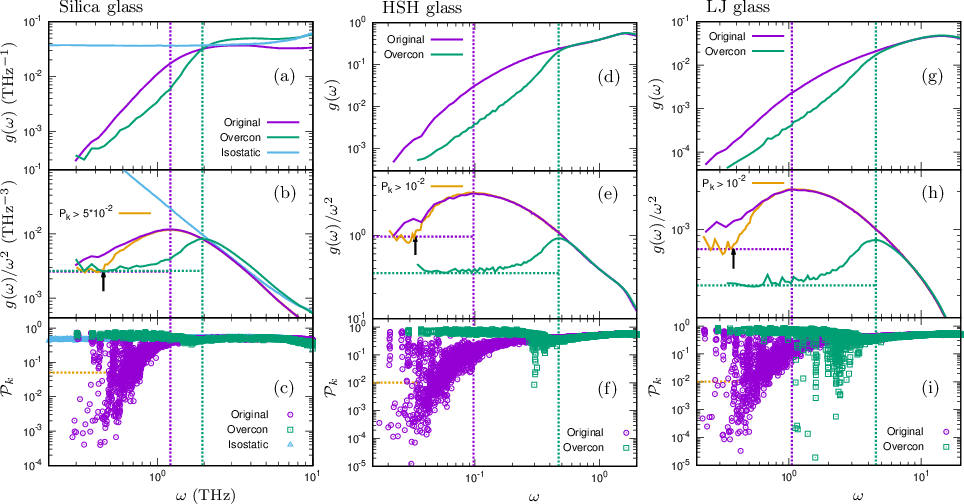}
\caption{\label{fig2_total}
{Vibrational states in the low-frequency regime.}
(a-c) silica glass, (d-f) HSH glass, and (g-i) LJ glass.
$g(\omega)$, $g(\omega)/\omega^{2}$, and $\mathcal{P}_k$ are plotted as functions of $\omega$ for the original system (purple) and the overconstrained-network system (green).
For silica glass (a-c), the isostatic-network system (cyan) is also shown.
Vertical lines indicate the BP frequency $\omega_\text{BP}$ for the original and overconstrained-network systems, and horizontal dotted lines in (b,e,h) indicate the Debye level $A_D$.
Note that, for the isostatic-network system of silica glass, $\omega_\text{BP}\to 0$ and $A_D\to \infty$.
Panels (b,e,h) also show, in orange, the vDOS of extended modes $g_{\text{EXT}}(\omega)$, defined by the participation-ratio threshold $\mathcal{P}_k>\mathcal{P}_\text{th}$ with $\mathcal{P}_\text{th}=5\times10^{-2}$ for silica glass and $\mathcal{P}_\text{th}=10^{-2}$ for HSH and LJ glasses; these thresholds are marked by the horizontal dotted lines in (c,f,i).
Arrows in (b,e,h) mark the continuum-limit frequency $\omega_0$ at which $g_{\text{EXT}}(\omega)/\omega^{2}$ converges to $A_D$.
The vDOS of QLVs, $g_{\text{QLV}}(\omega)$ for modes with $\mathcal{P}_k\le \mathcal{P}_\text{th}$, is shown in Fig.~\ref{fig3_qlv}.
See also Fig.~S1 in the SI for HSL and SS glasses.
}
\end{figure*}
%%%%%%%%%%%%%%%%%%%%%%%%%%%%%%%%%%%%%%%%%%%%%%%%%%%%%%%

%%%%%%%%%%%%%%%%%%%%%%%%%%%%%%%%%%%%%%%%%%%%%%%%%%%%%%%
\begin{figure}[t]
\centering
\includegraphics[width=0.4\textwidth]{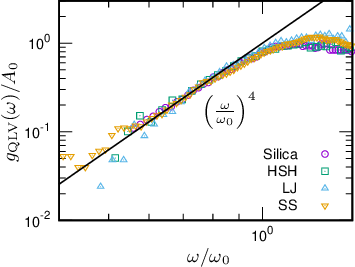}
\caption{\label{fig3_qlv}
{Vibrational density of states of quasi-localized vibrations~(QLVs).}
$g_{\text{QLV}}(\omega)/A_0$ is plotted as a function of $\omega/\omega_0$ for silica glass (circles), HSH glass (squares), LJ glass (upward triangles), and SS glass (downward triangles).
The solid line indicates $g_{\text{QLV}}(\omega)=A_0(\omega/\omega_0)^4 \propto \omega^4$ for $\omega<\omega_0$.
The prefactor is $A_0=4.62\times10^{-4}$, $1.38\times10^{-3}$, $6.90\times10^{-5}$, and $1.18\times10^{-4}$~THz$^{-1}$ for silica, HSH, LJ, and SS glasses, respectively.
}
\end{figure}
%%%%%%%%%%%%%%%%%%%%%%%%%%%%%%%%%%%%%%%%%%%%%%%%%%%%%%%

%%%%%%%%%%%%%%%%%%%%%%%%%%%%%%%%%%%%%%%%%%%%%%%%%%%%%%%
\begin{figure*}[t]
\centering
\includegraphics[width=0.96\textwidth]{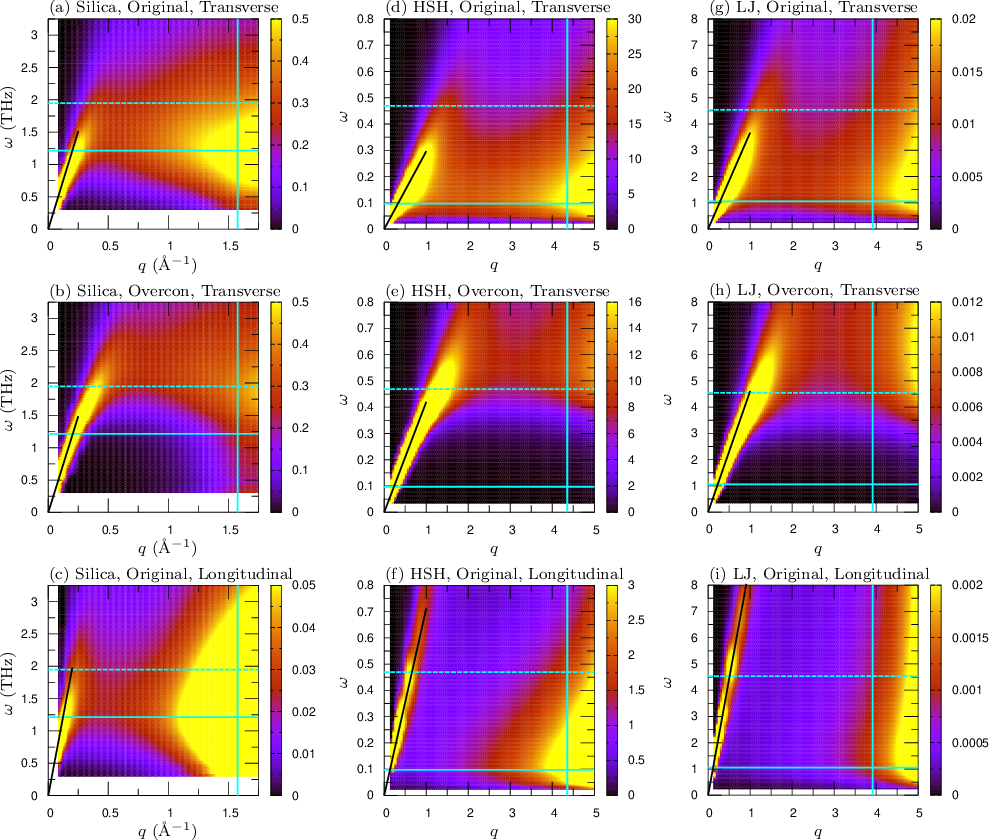}
\vspace*{-1.0mm}
\caption{\label{fig4_total_debyey}
{Dynamical structure factors.}
(a-c) silica glass, (d-f) HSH glass, and (g-i) LJ glass.
Transverse $S_T(q,\omega)/(k_B T)$ is shown as a function of $q$ and $\omega$ for the original systems in (a,d,g) and for the overconstrained-network systems in (b,e,h), whereas longitudinal $S_L(q,\omega)/(k_B T)$ is shown for the original systems in (c,f,i).
For silica glass, values are reported in units of eV$\,$THz$^{-1}$.
The vertical line marks the Debye wavenumber $q_D$.
Horizontal solid and dashed lines indicate the BP frequency $\omega_\text{BP}$ for the original and overconstrained-network systems, respectively.
The black solid curve shows the linear dispersion $\omega=c_\alpha q$ (with $\alpha=T$ in the $S_T$ panels and $\alpha=L$ in the $S_L$ panels), corresponding to phonon excitations.
Data for the isostatic-network system of silica glass are provided in Fig.~S2 of the SI.
Additional results for HSL and SS glasses are provided in Fig.~S3 of the SI.
}
\end{figure*}
%%%%%%%%%%%%%%%%%%%%%%%%%%%%%%%%%%%%%%%%%%%%%%%%%%%%%%%

%%%%%%%%%%%%%%%%%%%%%%%%%%%%%%%%%%%%%%%%%%%%%%%%%%%%%%%
\begin{figure}[t]
\centering
\includegraphics[width=0.475\textwidth]{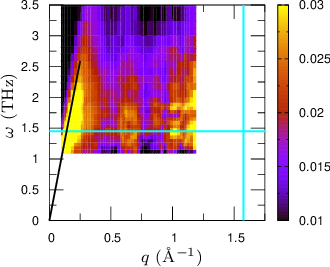}
\vspace*{-1.0mm}
\caption{\label{fig:ixs_silica}
{Inelastic X-ray scattering data of the inelastic part of the longitudinal dynamical structure factor.}
We plot $S_L(q,\omega)/\left[\hbar\,\omega\,(n(\omega,T)+1)\right]$ (which reduces to $S_L(q,\omega)/(k_B T)$ in the classical limit) in units of eV\,THz$^{-1}$.
Data correspond to ambient-pressure silica glass at density $\rho=2.20$~g/cm$^{3}$ and temperature $T=1620$~K.
The vertical line marks the Debye wavenumber $q_D =1.58$~\AA$^{-1}$.
The horizontal line marks the BP frequency $\omega_\text{BP} = 1.45$~THz.
The black curve shows the linear dispersion $\omega = c_L q$, using the longitudinal sound speed $c_L \approx 6500$~m/s measured by Brillouin light scattering~\cite{Baldi_2010}.
The spectrum is limited to frequencies $\omega \gtrsim 1$~THz.
This panel can be directly compared with the simulation in Fig.\ref{fig4_total_debyey}(c); although the temperatures differ (experiment $T=1620$~K and simulation $T=0$~K), the density matches and both display a phonon ridge following $\omega=c_L q$ together with a broad, non-phononic band around the BP.
}
\end{figure}
%%%%%%%%%%%%%%%%%%%%%%%%%%%%%%%%%%%%%%%%%%%%%%%%%%%%%%%

%%%%%%%%%%%%%%%%%%%%%%%%%%%%%%%%%%%%%%%%%%%%%%%%%%%%%%%%%%%%%%%%%%%%%%%%%%%%%%%%%%%%%%%%%%%%%%%%%
\section{Results}

Figure~\ref{fig3_shik_gr}(a) shows the radial distribution function $g(r)$ of the inherent structure.
The first peaks mark the {typical nearest-neighbor separations} for Si--O and O--O, at $r\approx1.6$~\AA\ and $r\approx2.6$~\AA, respectively (purple and green arrows).
Throughout, we refer to Si--O nearest-neighbor pairs as {Si--O bonds} because they represent covalent Si--O linkages, whereas we refer to O--O nearest-neighbor pairs as {O--O contacts} because there is no covalent O--O bond; these contacts arise from the geometry of O--Si--O linkages, with the directionality of Si--O covalency represented implicitly by the pairwise potential~\cite{Sundararaman_2018,Beest_1990}.
Specifically, we define Si--O bonds as pairs with $r\le 1.8$~\AA\ and O--O contacts as pairs with $r\le 2.9$~\AA\ (black arrows).
Panels~(b) and~(c) of Fig.~\ref{fig3_shik_gr} visualize the atomic network using these definitions.
These visualizations confirm a tetrahedral network in which each Si atom sits at the center and four O atoms occupy the vertices, yielding four Si--O bonds and six O--O contacts per tetrahedral unit.

In what follows, we present results for silica glass, HSH glass, and LJ glass in the main text.
Complementary data for HSL and SS glasses are provided in the Supporting Information~(SI) to corroborate and strengthen our conclusions.

%%%%%%%%%%%%%%%%%%%%%%%%%%%%%%%%%%%%%%%%%%%%%%%%%%%%%%%
\begin{table}[t]
\caption{\label{table2}
{Characteristics of Si--O bonds and O--O contacts in silica glass.}
``Separation" lists the typical nearest-neighbor separations.
``Spring constant" lists $\phi''_{S}$ and $\phi''_{L}$, the second derivatives of the short-range and long-range Coulomb pair potentials, respectively.
Neighbor~(pair) definitions use distance cutoffs taken at $r\le1.8$~\AA\ for Si--O {bonds} and $r\le2.9$~\AA\ for O--O {contacts}.
See also Fig.~\ref{fig3_shik_gr}.
}
\centering
\renewcommand{\arraystretch}{1.1}
\begin{tabular}{|c|c|c|c|}
\hline
 Pair & Separation~(\AA) & Spring constant~(eV\AA$^{-2}$) & Number of pairs \\
\hline
Si--O & $1.6$ & $\phi''_{S}=35.8$ & $4.00\,N_\text{Si}$ \\
      &       & $\phi''_{L}=-11.1$ & \\
\hline
O--O  & $2.6$ & $\phi''_{S}=4.55$ & $3.00\,N_\text{O}$ \\
      &       & $\phi''_{L}=1.29$  & \\
\hline
\end{tabular}
\end{table}
%%%%%%%%%%%%%%%%%%%%%%%%%%%%%%%%%%%%%%%%%%%%%%%%%%%%%%%

%%%%%%%%%%%%%%%%%%%%%%%%%%%%%%%%%%%%%%%%%%%%%%%%%%%%%%%%%%%%%%%%%%%%%%%%%%%%%%%%%%%%%%%%%%%%%%%%%
\subsection{Isostaticity}
Figure~\ref{fig3_shik_dosbond}(a) displays the vDOS $g(\omega)$ for silica glass (purple curve), which agrees with previously reported results~\cite{Carre_2007,Carre_2016,Sundararaman_2018,Mantisi_2012,Shcheblanov_2015,Damart_2017,Hamdaoui_2025}.
For reference, the Debye vDOS $A_D\,\omega^{2}$ is also shown; in the low-frequency regime $g(\omega)$ clearly exceeds this Debye vDOS, revealing excess modes.
We further resolve the low-frequency behavior in Fig.~\ref{fig2_total}(a,b).
In Fig.~\ref{fig2_total}(b), the reduced vDOS $g(\omega)/\omega^{2}$ lies above the Debye level $A_D$, thereby confirming the presence of the BP.
We define the BP frequency $\omega_{\text{BP}}$ as the frequency at which the reduced vDOS $g(\omega)/\omega^{2}$ attains its maximum.
For silica glass, this yields $\omega_{\text{BP}}\approx 1.21$~THz, consistent with scattering experiments~\cite{Buchenau_1986,Arai_1999,Harris_2000,Monaco_2006,Ruffle_2008,Baldi_2008,Baldi_2010,Baldi_2011,Baldi_2016}.

To understand the vDOS and the BP, we begin by examining isostaticity in silica glass.
As shown in Fig.~\ref{fig3_shik_gr}, the Si and O atoms form a tetrahedral network with typical nearest-neighbor separations of $1.6$~\AA\ for Si--O bonds and $2.6$~\AA\ for O--O contacts.
Accordingly, we construct a network system by connecting pairs with unstressed (neither pre-stretched nor pre-compressed) springs: Si--O pairs within the cutoff $1.8$~\AA\ are linked, and O--O pairs within $2.9$~\AA\ are likewise connected.
Table~\ref{table2} lists typical values of the spring constants $\phi''_{S}$ and $\phi''_{L}$, \textit{i.e.,} the second derivatives of the short-range and Coulomb potentials, respectively, evaluated at $r=1.6$~\AA\ for Si--O and $r=2.6$~\AA\ for O--O, together with the counts of springs corresponding to Si--O bonds and O--O contacts.
We first focus on the short-range interaction $\phi''_{S}$, neglecting the Coulomb contribution $\phi''_{L}$; that is, we analyze a network system in which these pairs are connected by springs with stiffness $\phi''_{S}$.

We can demonstrate that the network system constructed above is isostatic, meaning that the number of constraints $N_{\text{const}}$ equals the number of degrees of freedom $N_{\text{dof}}$.
First, the total number of degrees of freedom is $N_{\text{dof}}=3(N_{\text{Si}}+N_{\text{O}})=9N_{\text{Si}}$.
Next, following Ref.~\cite{Thorpe_1983}, we count constraints as follows.
To match the situation in Ref.~\cite{Thorpe_1983} within our spring representation, we regard the O--O springs along tetrahedral edges as {effective} angular constraints associated with the $\angle\text{O--Si--O}$ bending~\cite{Sundararaman_2018,Beest_1990}, and we treat only the Si--O springs as central-force bonds (this situation is visualized in Fig.~\ref{fig3_shik_gr}(b)).
The total number of constraints is then
\begin{equation}\label{eq.count}
N_{\text{const}}=\sum_{b} N_b\!\left[\frac{b}{2}+(2b-3)\right],
\end{equation}
where $b$ is the coordination number and $N_b$ is the number of atoms with coordination $b$; the first term counts bond-stretching (two-body) constraints, and the second term counts bond-bending (three-body) constraints centered on the atom.
Oxygen has $b=2$ and contributes only the bond-stretching part $b/2=1$ per O (we do not assign O-centered angular constraints), whereas silicon has $b=4$ and contributes $b/2=2$ bond-stretching constraints and $(2b-3)=5$ angular constraints associated with $\angle\text{O--Si--O}$.
Substituting into Eq.~(\ref{eq.count}) gives
\begin{equation}
N_{\text{const}}=N_{\text{O}}\cdot 1 + N_{\text{Si}}\cdot (2+5)=9N_{\text{Si}}=N_{\text{dof}},
\end{equation}
which confirms that the network representation is isostatic.

We present the vDOS for this isostatic-network system in Fig.~\ref{fig3_shik_dosbond}(a) and Fig.~\ref{fig2_total}(a) {(cyan curves)}.
The vDOS remains finite as $\omega\to 0$, indicating a gapless spectrum at isostaticity that comprises both strictly zero-frequency floppy modes and many additional soft, low-frequency modes of isostatic origin~\cite{Wyart_2005,Wyart2_2005,Wyart3_2005}.

A notable point is that Si--O bonds are significantly stiffer than O--O contacts: $\phi''_{S}=35.8$~eV\,\AA$^{-2}$ for Si--O versus $\phi''_{S}=4.55$~eV\,\AA$^{-2}$ for O--O, as listed in Table~\ref{table2}.
Consequently, vibrations that involve stretching and compression of Si--O bonds appear at high frequencies~\cite{Damart_2017}.
Since each tetrahedral unit contains four Si--O bonds, the total number of such modes can reach $4N_{\text{Si}}$.
To probe these modes, we quantify Si--O bond stretching and compression using the dimensionless measure $\delta e_{k,\text{Si--O}}^{2}$ (see \textit{Materials and Methods}).
Figure~\ref{fig3_shik_dosbond}(b) plots $N\,\delta e_{k,\text{Si--O}}^{2}$ versus $\omega$.
$N\,\delta e_{k,\text{Si--O}}^{2}$ is nearly zero in the low-$\omega$ regime, including the BP region, indicating minimal Si--O bond deformation there.
Above $\omega\approx12.5$~THz, however, $N\,\delta e_{k,\text{Si--O}}^{2}$ increases, signaling the onset of bond-stretching and bond-compression character.
Thus $\omega\approx12.5$~THz marks a boundary: modes above this frequency predominantly occupy the $4N_{\text{Si}}$ sector associated with Si--O bond stretching and compression, whereas modes below it comprise the remaining $5N_{\text{Si}}$ modes.

Because the short-range interaction $\phi''_{S}$ couples not only atoms within a tetrahedral unit but also atoms belonging to different units, it is natural to consider spring-network variants that deviate from isostaticity.
These interunit couplings correspond to van der Waals interactions and are often referred to as weak interactions~\cite{Phillips2_1981,Thorpe_1983,Cai_1989}.
Specifically, we examine a network in which all O--O contacts are connected by unstressed springs.
In this case the system is overconstrained, $N_{\text{const}}>N_{\text{dof}}$, and the vDOS tends to zero as $\omega\to 0$.
We refer to this overconstrained variant as the overconstrained-network system, in contrast to the isostatic-network system defined above.
As shown in Fig.~\ref{fig3_shik_dosbond}, the excess constraints shift the low-frequency floppy modes of the isostatic-network system to higher frequencies in the overconstrained-network system (compare the cyan and green curves).
By contrast, the high-frequency modes associated with stretching and compression of Si--O bonds above $\omega\approx12.5$~THz are minimally affected, and the $N\,\delta e_{k,\text{Si--O}}^{2}$ data show little difference between the two network systems.

Finally, to transition from the overconstrained-network system back to the original atomistic system, we must reintroduce effects neglected so far: the short-range Si--Si interaction, the long-range Coulomb interaction, and the frustration introduced when unstressed springs are replaced by stressed (stretched or compressed) springs~\cite{Wyart_2005,Wyart2_2005,Wyart3_2005,Xu_2007,Xu_2010,Mizuno_2017,Lerner_2017,Shimada2_2018}.
While their consequences for low-frequency modes are discussed in the next section, here we focus on their impact at high frequencies.
As seen in Fig.~\ref{fig3_shik_dosbond}, the overconstrained-network system exhibits a clear separation: the band of modes dominated by stretching and compression of Si--O bonds (up to $4N_{\text{Si}}$ modes) is well separated from the remaining $5N_{\text{Si}}$ modes.
In the original system, this separation is less pronounced.
A key reason is that the Coulomb contribution, characterized by $\phi''_{L}$, reduces the effective stiffness of Si--O bonds: $\phi''_{S}=35.8$~eV\,\AA$^{-2}$ is lowered to $\phi''_{S}+\phi''_{L}=35.8-11.1=24.7$~eV\,\AA$^{-2}$ (Table~\ref{table2}).
This weakening shifts the Si--O bond-stretching/compression band toward lower frequencies.
Nevertheless, the trend persists in the original system: the $N\,\delta e_{k,\text{Si--O}}^{2}$ data show that bond stretching and compression continue to dominate the high-frequency region, whereas modes without such bond deformation dominate at low frequencies.

%%%%%%%%%%%%%%%%%%%%%%%%%%%%%%%%%%%%%%%%%%%%%%%%%%%%%%%%%%%%%%%%%%%%%%%%%%%%%%%%%%%%%%%%%%%%%%%%%
\subsection{Marginal stability}
We now examine marginal stability in silica glass.
Since the concept of marginal stability is well established in packing-based glasses~\cite{Durian_1995,Hern_2003,Hecke_2010,Wyart_2005,Wyart2_2005,Wyart3_2005,Wyart_2010,Degiuli_2014}, we discuss it alongside data for HSH and LJ glasses.
Figure~\ref{fig2_total} presents $g(\omega)$, $g(\omega)/\omega^{2}$, and the participation ratio $\mathcal{P}_k$ (see \textit{Materials and Methods}).
The quantity $\mathcal{P}_k$ measures the fraction of particles participating in mode $k$~\cite{MizunoIkeda2022}.
As limiting cases, $\mathcal{P}_k=1$ corresponds to a fully extended mode in which all particles vibrate equally, whereas $\mathcal{P}_k=1/N$ corresponds to a localized mode involving a single particle.

Let us first examine HSH glass in panels (d-f) and LJ glass in panels (g-i).
For each system, we also analyze a overconstrained-network counterpart in which interacting pairs (contacts for HSH and neighbors within the cutoff for LJ) are connected by unstressed (neither pre-stretched nor pre-compressed) springs with stiffness set by the pairwise force constants $\phi''$.
This construction removes pre-stress and the associated frustration present in the original systems and corresponds to the ``unstressed" system in previous work~\cite{Wyart_2005,Wyart2_2005,Wyart3_2005,Xu_2007,Xu_2010,Mizuno_2017,Lerner_2017,Shimada2_2018}.
Examining panels (e,h), both overconstrained-network systems (green) exhibit a clear BP.
Below the BP frequency $\omega_\text{BP}$, $g(\omega)/\omega^2$ converges to the Debye level $A_D$, \textit{i.e.,} $g(\omega)\to A_D \omega^2$, and the vibrational states are extended phonons with large $\mathcal{P}_k$
\footnote{
Below $\omega_\text{BP}$, we also observe modes with small $\mathcal{P}_k$ indicative of localization, as seen in panels (f,i).
These modes plausibly originate in regions of low local connectivity (low coordination number, \textit{i.e.,} few contacts or interacting neighbors), although a definitive identification requires further study.
}.
Above $\omega_\text{BP}$, a band of soft modes of isostatic origin appears.
In the isostatic limit, the low-frequency sector comprises both strictly floppy modes at zero frequency and additional soft, nonzero-frequency modes that likewise originate from isostaticity (though they are not classified as floppy).
When excess constraints are introduced, both modes are lifted to finite frequencies and merge into a weakly dispersive, non-phononic band.
This band remains spatially extended with large $\mathcal{P}_k$, and its accumulation produces the excess over the Debye law, \textit{i.e.,} the BP.

Then, when frustration is introduced by replacing unstressed springs with stressed (stretched or compressed) springs, the band of soft modes~(of isostatic origin) shifts toward lower frequencies, and the BP accordingly moves downward with decreasing $\omega_{\text{BP}}$ (purple).
In addition, QLVs with low $\mathcal{P}_k$ emerge at the low-frequency edge below $\omega_{\text{BP}}$.
Following previous work~\cite{Mizuno_2017,Shimada_2018,Wang_2019}, we partition modes into extended modes with $\mathcal{P}_k>\mathcal{P}_{\text{th}}$ and QLVs with $\mathcal{P}_k<\mathcal{P}_{\text{th}}$, and we compute their vDOSs, $g_{\text{EXT}}(\omega)$ and $g_{\text{QLV}}(\omega)$, separately.
Here we take $\mathcal{P}_{\text{th}}=10^{-2}$ for HSH and LJ glasses.
We observe that $g_{\text{EXT}}(\omega)$ converges to the Debye law $A_D\,\omega^2$ at a characteristic frequency $\omega_0$ (orange curves in panels (e,h)).
The same behavior is found for SS glass in Fig.~S1 of the SI.
Furthermore, for HSH, LJ, and SS glasses, $g_{\text{QLV}}(\omega)$ follows $A_0(\omega/\omega_0)^4\propto\omega^4$, as shown in Fig.~\ref{fig3_qlv}, consistent with previous studies~\cite{Mizuno_2017,Shimada_2018,Wang_2019}.
Importantly, the QLVs exhibit a gapless vDOS with a power-law dependence on $\omega$, indicating that the introduction of frustration drives the systems toward marginal stability.

The discussion of HSH and LJ glasses above extends to silica glass.
In the isostatic-network system (cyan), the vDOS approaches a nonzero constant as $\omega\to 0$ (\textit{i.e.,} $g(\omega)\propto \omega^{0}$); accordingly, $g(\omega)/\omega^{2}\propto \omega^{-2}\to\infty$ and the BP collapses to zero frequency.
The participation-ratio data show that these low-frequency states are spatially extended.
Taken together, these results demonstrate the presence of extended floppy modes at zero frequency, together with additional soft, nonzero-frequency modes of isostatic origin~\cite{Wyart_2005,Wyart2_2005,Wyart3_2005}, as discussed above.
Turning to the overconstrained-network system (green), the excess constraints cause the vDOS to vanish as $\omega\to 0$.
As seen in panel (b), $g(\omega)/\omega^{2}$ displays a clear BP, analogous to HSH and LJ glasses.
Below $\omega_\text{BP}$, $g(\omega)/\omega^{2}\to A_D$ (hence $g(\omega)\to A_D\, \omega^{2}$), and the vibrational states are extended phonons with large $\mathcal{P}_k$.
Above $\omega_\text{BP}$, an isostaticity-derived, weakly dispersive, non-phononic band appears at finite frequencies; it is spatially extended with large $\mathcal{P}_k$ and produces the excess over the Debye law, \textit{i.e.,} the BP.

Finally, to move from the overconstrained-network system to the original system, we reinstate all effects neglected in the overconstrained-network construction, including the short-range Si--Si interaction, the long-range Coulomb interaction, and the pre-stress that generates frustration
\footnote{
While all of the short-range Si--Si interaction, the long-range Coulomb interaction, and the pre-stress are needed for a faithful description, the essential ingredient that controls the approach to marginal stability is the pre-stress and the associated frustration; the short-range Si--Si and long-range Coulomb terms primarily set the magnitude and sign of the pre-stress and thereby tune the BP position.
}.
With these ingredients restored, the BP shifts to lower frequencies with a reduced $\omega_\text{BP}$.
Notably, QLVs also emerge at the low-frequency edge below $\omega_\text{BP}$.
Following the procedure used above for HSH and LJ glasses~\cite{Mizuno_2017,Shimada_2018,Wang_2019}, we compute $g_{\text{EXT}}(\omega)$ and $g_{\text{QLV}}(\omega)$ by setting $\mathcal{P}_{\text{th}}=5\times10^{-2}$.
As shown in Fig.~\ref{fig2_total}(b) and Fig.~\ref{fig3_qlv}, for $\omega<\omega_0$ we find $g_{\text{EXT}}(\omega)\to A_D\,\omega^2$ \textit{i.e.,} $g(\omega)$ follows the Debye law, while $g_{\text{QLV}}(\omega)$ follows the gapless form $A_0(\omega/\omega_0)^4\propto\omega^4$.
Previous simulations~\cite{Richard_2020,Guerra_2022} have also reported $g_{\text{QLV}}\propto\omega^4$ using different statistical analyses of QLVs
\footnote{
Although a quartic law $g_{\text{QLV}}(\omega)\propto\omega^{4}$ is often reported, the precise exponent remains under discussion~\cite{Schirmacher_2024,Xu_2024}.
What matters here is that, irrespective of the exact exponent, $g_{\text{QLV}}(\omega)$ is gapless and follows a power law as $\omega\to 0$.
Such a gapless power-law spectrum is a hallmark of marginal stability.
}.
Taken together, these results show that, once the short-range Si--Si and long-range Coulomb interactions are restored and the associated pre-stress (frustration) is present, the system is driven toward marginal stability.

Silica glass, HS glass, LJ glass, and SS glass all share the characteristic that they become marginally stable states driven by frustration.
The key difference lies in the origin of that frustration.
In HS, LJ, and SS glasses, it arises from short-range interactions \textit{i.e.,} contact forces and van der Waals forces.
By contrast, silica glass involves not only short-range interactions but also long-range Coulomb interactions, and frustration from both contributions collectively drives the approach to marginal stability.
Because short-range frustration is mainly associated with repulsive forces, it tends to destabilize the system, whereas the attractive Coulomb interaction between Si and O atoms acts to stabilize and partly offset this effect.
Consequently, in silica glass the difference in BP frequency between the overconstrained-network system and the original system is modest, about a factor of $1.5$ (compare the green and purple vertical lines in Fig.~\ref{fig2_total}(b)).
In HS, LJ, and SS glasses, the corresponding shift is larger, by a factor of $5$ to $8$.

%%%%%%%%%%%%%%%%%%%%%%%%%%%%%%%%%%%%%%%%%%%%%%%%%%%%%%%%%%%%%%%%%%%%%%%%%%%%%%%%%%%%%%%%%%%%%%%%%
\subsection{Dynamical structure factor}
Finally, we examine how isostaticity and marginal stability are reflected in the dynamical structure factor $S_\alpha(q,\omega)$ (with $\alpha=T,L$), which is accessible via inelastic scattering experiments.
In the low-frequency regime, the vDOS can be estimated by integrating $S_\alpha(q,\omega)$ over wavenumber up to the Debye wavenumber $q_D$~\cite{Baldi_2011,Schirmacher_2006,Marruzzo_2013,Schirmacher_2015,Mizuno_2018,Wyart_2010,Degiuli_2014},
\begin{equation}\label{eq_vdostl}
\frac{g(\omega)}{\omega^2}
= \frac{2\,M(\omega)}{q_D^3}
\int_0^{q_D}\!\left\{\frac{S_T(q,\omega)}{k_B T}+\frac{S_L(q,\omega)}{k_B T}\right\}\,dq,
\end{equation}
where $k_B$ is the Boltzmann constant and $M(\omega)$ is the effective mass (see \textit{Materials and Methods}).
Setting $\omega=\omega_\text{BP}$ allows us to resolve how excitations at each $q$ in $S_T(q,\omega_\text{BP})$ and $S_L(q,\omega_\text{BP})$ contribute to $g(\omega_\text{BP})/\omega_\text{BP}^2$, \textit{i.e.,} the BP.
This viewpoint underlies several theoretical analyses~\cite{Schirmacher_2006,Marruzzo_2013,Schirmacher_2015,Wyart_2010,Degiuli_2014,Mizuno_2018}, which employ Green's functions related to $S_\alpha(q,\omega)$ via the fluctuation--dissipation theorem.

Figure~\ref{fig4_total_debyey} presents $S_T(q,\omega)$ and $S_L(q,\omega)$, focusing on the low-frequency and low-wavenumber ($q\lesssim q_D$) regime.
First, we examine the transverse $S_T(q,\omega)$, where data for the overconstrained-network system are shown alongside those for the original system.
We begin with HSH glass in panels (d,e) and LJ glass in panels (g,h).
For the overconstrained-network (unstressed) systems in panels (e,h), phonon excitations appear at low $q$ and low $\omega$ along the linear dispersion $\omega=c_T q$, where $c_T$ is the transverse sound speed.
In addition, a broad, approximately wavenumber-independent band emerges around and above the BP and extends up to $q_D$, indicating non-phononic excitations.
This band originates from isostaticity-derived modes that, in the presence of excess constraints, are lifted to finite frequencies (see Fig.~\ref{fig2_total}).
Therefore, the BP is built from two components: linearly dispersing phonons that follow $\omega=c_T q$, and a wavenumber-independent band of isostaticity-derived modes.

Turning to the original systems that include pre-stress (and hence frustration) in panels (d,g), the entire non-phononic band shifts to lower frequencies, with a corresponding decrease in $\omega_\text{BP}$.
At the low-frequency edge below $\omega_\text{BP}$, a band of QLV excitations appears~\cite{Mizuno_2017,Shimada_2018,Wang_2019}.
Phonon ridges along the linear dispersion $\omega=c_T q$ remain visible, but they are broader than in the overconstrained-network systems.
This broadening is attributed to isostaticity-derived modes shifting into the low-frequency range and hybridizing with the phonons, which increases the phonon linewidths.
Similar behavior is observed for HSL and SS glasses, as shown in Fig.~S3 of the SI.
In this way, marginal stability manifests as a wavenumber-independent band that is shifted downward by frustration, together with the emergence of QLV excitations at the low-frequency edge.
Notably, a broad, wavenumber-independent band in the dynamical structure factor has also been reported recently in simulations of LJ glasses~\cite{Hu_2022}, consistent with the present results.

The discussion above for HSH and LJ glasses extends to silica glass in panels (a,b) of Fig.~\ref{fig4_total_debyey}.
For silica glass, additional data for the isostatic-network system are provided in Fig.~S2 of the SI.
First, in the isostatic-network system in Fig.~S2, exact isostaticity generates a nearly wavenumber-independent band that reaches the zero-frequency limit; correspondingly, $S_T(q,\omega)$ accumulates substantial, nearly $q$-independent spectral weight as $\omega\to 0$, while $g(\omega)/\omega^{2} \to\infty$ as indicated by Eq.~(\ref{eq_vdostl}).
In the overconstrained-network system in Fig.~\ref{fig4_total_debyey}(b), excess constraints lift this isostaticity-derived band to finite frequencies around the BP, and a phonon branch appears along $\omega=c_T q$.
In the original system in Fig.~\ref{fig4_total_debyey}(a), restoring pre-stress from short-range and long-range Coulomb interactions shifts the entire non-phononic band to lower frequencies, and $\omega_\text{BP}$ decreases accordingly.
At the low-frequency edge below $\omega_\text{BP}$, QLVs form a broad band.
As discussed above, in silica glass the attractive Coulomb interaction between Si and O stabilizes the structure and partially offsets the destabilizing effect of short-range frustration.
Consequently, the downward shift of the band is smaller in silica glass than in HSH, LJ, and SS glasses.
Despite these quantitative differences, the underlying physics is the same across systems.

An intriguing aspect is the behavior of the longitudinal channel, $S_L(q,\omega)$.
Across silica, HSH, LJ, and SS glasses, $S_L(q,\omega)$ is markedly weaker than $S_T(q,\omega)$ in the low-frequency regime, including the BP region, indicating that the BP predominantly reflects transverse motion~\cite{Monaco2_2009,Mizuno2_2013,Mizuno_2014,Mizuno_2018}.
In silica glass, however, $S_L(q,\omega)$ shows a similar $\omega$--$q$ structure to $S_T(q,\omega)$: a broad, nearly wavenumber-independent band appears around the BP as shown in Fig.~\ref{fig4_total_debyey}(c), whereas the corresponding feature is much less apparent in HSH and LJ glasses (Figs.~\ref{fig4_total_debyey}(f) and (i)) and in SS glass (Fig.~S3).
This behavior likely reflects closer proximity to isostaticity in silica glass, which enhances the population of isostaticity-derived soft modes with both transverse and longitudinal character.
Moreover, silica glass exhibits strong nonaffine elasticity not only under shear but also under volumetric deformation (see \textit{Materials and Methods}), further reducing the contrast between longitudinal and transverse responses.

Since scattering experiments probe the longitudinal channel, we now compare our simulations with inelastic X-ray scattering data~\cite{Baldi_2008,Baldi_2010,Baldi_2011,Baldi_2016}.
Figure~\ref{fig:ixs_silica} shows the measured longitudinal dynamical structure factor for ambient-pressure silica glass at density $\rho=2.20$~g/cm$^{3}$ and temperature $T=1620$~K.
The plotted quantity is the inelastic part of the experimentally determined dynamical structure factor, $S_L(q,\omega)/\left[\hbar\,\omega\,(n(\omega,T)+1)\right]$ (which reduces to $S_L(q,\omega)/(k_B T)$ in the classical limit), in units of eV\,THz$^{-1}$, where $n(\omega,T)$ is the Bose--Einstein population factor and $\hbar = h/2\pi$ with $h$ the Planck constant.
Details on the procedure used to properly normalize the inelastic X-ray scattering data are provided in Ref.~\cite{Baldi_2016}.
The data are limited to frequencies $\omega \gtrsim 1$ THz because subtracting the elastic line becomes delicate at lower frequencies.
We also note that our simulation data in Fig.\ref{fig4_total_debyey}(c) correspond to $T=0$~K; accordingly, the BP frequency is $\omega_\text{BP} \approx 1.21$~THz in simulation but $\omega_\text{BP} \approx 1.45$ THz in the experiment, consistent with the expected upward shift at elevated temperature $T=1620$~K.
Despite this offset, the qualitative behavior is fully consistent between simulation and experiment:
(i) at low wavenumber and frequency, longitudinal phonons follow the linear dispersion $\omega=c_L q$; and
(ii) a broad, wavenumber-independent (dispersionless) non-phononic band is visible around the BP, extending from $q\approx 0.3$\AA$^{-1}$ up to $\approx 1.2$~\AA$^{-1}$.
By direct analogy with the simulations, we identify this dispersionless band in the inelastic X-ray scattering data as the band of isostaticity-derived modes.

%%%%%%%%%%%%%%%%%%%%%%%%%%%%%%%%%%%%%%%%%%%%%%%%%%%%%%%%%%%%%%%%%%%%%%%%%%%%%%%%%%%%%%%%%%%%%%%%%
\section{Conclusion}
We have explained the BP in silica glass in terms of isostaticity and marginal stability.
When the tetrahedral network structure is extracted, it forms an isostatic-network system where the number of constraints equals the number of degrees of freedom.
This isostatic network comprises both strictly zero-frequency floppy modes and many additional soft, low-frequency modes of isostatic origin, and the vDOS remains finite as $\omega\to 0$, \textit{i.e.,} $g(\omega)\propto\omega^{0}$.
In practice, interunit couplings of van der Waals type add constraints beyond the degrees of freedom.
The resulting spring-network is overconstrained and displays phonons at low frequency whose vDOS follows the Debye law, \textit{i.e.,} $g(\omega)\simeq A_D\omega^{2}$, while the isostaticity-derived soft modes are shifted to finite frequencies and form the non-phononic excess that constitutes the BP.
This behavior of the overconstrained-network system is the same mechanism operative in HS, LJ, and SS glasses.
Reinstating all effects neglected in the overconstrained-network construction, including short-range Si--Si interactions, long-range Coulomb interactions, and the associated pre-stress that generates frustration, shifts the non-phononic band downward, producing a low-frequency BP.
Concurrently, QLVs appear at the low-frequency edge below the BP and exhibit a gapless power-law vDOS $\propto\omega^{4}$ as $\omega\to 0$.
Together, these features indicate that frustration drives silica glass to a marginally stable state, as also observed in HS, LJ, and SS glasses.

Furthermore, we find that isostaticity and marginal stability are encoded in the dynamical structure factor as a broad, nearly wavenumber-independent band around the BP.
In this frequency range, isostaticity-derived soft modes produce a dispersionless band.
In silica glass, the band is sharply visible in both the transverse and longitudinal dynamical structure factor, whereas in HS, LJ, and SS glasses it is prominent in the transverse channel but much less apparent in the longitudinal one.
This contrast is especially useful because scattering experiments generally cannot access the transverse component at low frequencies and instead probe the longitudinal channel.
Indeed, a direct comparison between inelastic X-ray scattering measurements and our simulations of silica glass shows good agreement in the longitudinal dynamical structure factor, including both the low-wavenumber phonon branch and the broad dispersionless band around the BP.
On this basis, we identify the dispersionless band observed in the inelastic X-ray scattering data as the band of isostaticity-derived modes.
Thus, the BP can be analyzed directly via the longitudinal component of the dynamical structure factor, informing interpretations of existing measurements~\cite{Buchenau_1986,Yamamuro_1996,Arai_1999,Harris_2000,Monaco_2006,Ruffle_2008,Monaco_2009,Baldi_2008,Baldi_2010,Baldi_2011,Ruta_2012,Baldi_2016} and guiding future investigations.

In conclusion, isostaticity and marginal stability are broadly applicable to both packing-type and network-forming glasses and capture fundamental aspects of the physics of amorphous solids.
These principles manifest as a dispersionless, nearly wavenumber-independent band in the dynamical structure factor (including the longitudinal channel accessible to inelastic scattering), providing a direct, testable signature.
At the end of the present work, we carried out an effective-medium mean-field analysis based on random spring networks, following Refs.~\cite{Wyart_2010,Degiuli_2014}.
Within this framework we computed the vDOS and the dynamical structure factor; the results and explanations are reported in the SI.
For the overconstrained-network system (often termed the ``unstressed system" in the effective-medium literature), the analysis predicts an accumulation of isostaticity-derived soft modes at finite frequencies that produces the BP.
When pre-stress is included to emulate the frustration present in the original system, these soft modes shift to lower frequencies and the BP moves downward.
Crucially, the theory predicts that these soft modes imprint themselves on the dynamical structure factor as a broad, wavenumber-independent band around the BP.
These effective-medium predictions are in excellent agreement with our simulations and with the inelastic X-ray scattering data on silica glass, establishing a consistent picture across theory, simulation, and experiment.

%%%%%%%%%%%%%%%%%%%%%%%%%%%%%%%%%%%%%%%%%%%%%%%%%%%%%%%%%%%%%%%%%%%%%%%%%%%%%%%%%%%
\section*{Acknowledgments}
We thank Atsushi Ikeda, Kumpei Shiraishi, Osamu Yamamuro, and Maiko Kofu for useful discussions.
This work was supported by JSPS KAKENHI Grant Numbers 22K03543, 23H04495, 25H01519.

%%%%%%%%%%%%%%%%%%%%%%%%%%%%%%%%%%%%%%%%%%%%%%%%%%%%%%%%%%%%%%%%%%%%%%%%%%%%%%%%%%%
\section*{Author contributions statement}
H.M. designed the research with help from T.M. and E.M.; E.M. performed molecular dynamics simulations of silica glass; G.B. acquired and curated the experimental inelastic X-ray scattering data; H.M. conducted the theoretical analysis and wrote the paper; and all authors discussed the results and commented on the manuscript.

%%%%%%%%%%%%%%%%%%%%%%%%%%%%%%%%%%%%%%%%%%%%%%%%%%%%%%%%%%%%%%%%%%%%%%%%%%%%%%%%%%%
\section*{Author declarations}
The authors declare no conflicts of interest.

%%%%%%%%%%%%%%%%%%%%%%%%%%%%%%%%%%%%%%%%%%%%%%%%%%%%%%%%%%%%%%%%%%%%%%%%%%%%%%%%%%%%%%%%%%%%%%%%%%%%%%%%%%%%%%%%%%%%%%%%%%%
\bibliographystyle{apsrev4-2}
\bibliography{reference}

%apsrev4-2.bst 2019-01-14 (MD) hand-edited version of apsrev4-1.bst
%Control: key (0)
%Control: author (72) initials jnrlst
%Control: editor formatted (1) identically to author
%Control: production of article title (-1) disabled
%Control: page (0) single
%Control: year (1) truncated
%Control: production of eprint (0) enabled
\begin{thebibliography}{106}%
\makeatletter
\providecommand \@ifxundefined [1]{%
 \@ifx{#1\undefined}
}%
\providecommand \@ifnum [1]{%
 \ifnum #1\expandafter \@firstoftwo
 \else \expandafter \@secondoftwo
 \fi
}%
\providecommand \@ifx [1]{%
 \ifx #1\expandafter \@firstoftwo
 \else \expandafter \@secondoftwo
 \fi
}%
\providecommand \natexlab [1]{#1}%
\providecommand \enquote  [1]{``#1''}%
\providecommand \bibnamefont  [1]{#1}%
\providecommand \bibfnamefont [1]{#1}%
\providecommand \citenamefont [1]{#1}%
\providecommand \href@noop [0]{\@secondoftwo}%
\providecommand \href [0]{\begingroup \@sanitize@url \@href}%
\providecommand \@href[1]{\@@startlink{#1}\@@href}%
\providecommand \@@href[1]{\endgroup#1\@@endlink}%
\providecommand \@sanitize@url [0]{\catcode `\\12\catcode `\$12\catcode
  `\&12\catcode `\#12\catcode `\^12\catcode `\_12\catcode `\%12\relax}%
\providecommand \@@startlink[1]{}%
\providecommand \@@endlink[0]{}%
\providecommand \url  [0]{\begingroup\@sanitize@url \@url }%
\providecommand \@url [1]{\endgroup\@href {#1}{\urlprefix }}%
\providecommand \urlprefix  [0]{URL }%
\providecommand \Eprint [0]{\href }%
\providecommand \doibase [0]{https://doi.org/}%
\providecommand \selectlanguage [0]{\@gobble}%
\providecommand \bibinfo  [0]{\@secondoftwo}%
\providecommand \bibfield  [0]{\@secondoftwo}%
\providecommand \translation [1]{[#1]}%
\providecommand \BibitemOpen [0]{}%
\providecommand \bibitemStop [0]{}%
\providecommand \bibitemNoStop [0]{.\EOS\space}%
\providecommand \EOS [0]{\spacefactor3000\relax}%
\providecommand \BibitemShut  [1]{\csname bibitem#1\endcsname}%
\let\auto@bib@innerbib\@empty
%</preamble>
\bibitem [{\citenamefont {Ashcroft}\ and\ \citenamefont
  {Mermin}(1976)}]{Ashcroft_1976}%
  \BibitemOpen
  \bibfield  {author} {\bibinfo {author} {\bibfnamefont {N.~W.}\ \bibnamefont
  {Ashcroft}}\ and\ \bibinfo {author} {\bibfnamefont {N.~D.}\ \bibnamefont
  {Mermin}},\ }\href@noop {} {\emph {\bibinfo {title} {Solid State Physics}}}\
  (\bibinfo  {publisher} {Harcourt College Publishers, New York},\ \bibinfo
  {year} {1976})\BibitemShut {NoStop}%
\bibitem [{\citenamefont {Ramos}(2022)}]{Ramos_2022}%
  \BibitemOpen
  \bibfield  {author} {\bibinfo {author} {\bibfnamefont {M.~A.}\ \bibnamefont
  {Ramos}},\ }\href@noop {} {\emph {\bibinfo {title} {Low-Temperature Thermal
  and Vibrational Properties of Disordered Solids: A Half-Century of Universal
  “Anomalies” of Glasses}}}\ (\bibinfo  {publisher} {World Scientific Pub Co
  Inc},\ \bibinfo {year} {2022})\BibitemShut {NoStop}%
\bibitem [{\citenamefont {Nakayama}(2002)}]{Nakayama_2002}%
  \BibitemOpen
  \bibfield  {author} {\bibinfo {author} {\bibfnamefont {T.}~\bibnamefont
  {Nakayama}},\ }\href {https://doi.org/10.1088/0034-4885/65/8/203} {\bibfield
  {journal} {\bibinfo  {journal} {Reports on Progress in Physics}\ }\textbf
  {\bibinfo {volume} {65}},\ \bibinfo {pages} {1195} (\bibinfo {year}
  {2002})}\BibitemShut {NoStop}%
\bibitem [{\citenamefont {Buchenau}\ \emph {et~al.}(1986)\citenamefont
  {Buchenau}, \citenamefont {Prager}, \citenamefont {N\"ucker}, \citenamefont
  {Dianoux}, \citenamefont {Ahmad},\ and\ \citenamefont
  {Phillips}}]{Buchenau_1986}%
  \BibitemOpen
  \bibfield  {author} {\bibinfo {author} {\bibfnamefont {U.}~\bibnamefont
  {Buchenau}}, \bibinfo {author} {\bibfnamefont {M.}~\bibnamefont {Prager}},
  \bibinfo {author} {\bibfnamefont {N.}~\bibnamefont {N\"ucker}}, \bibinfo
  {author} {\bibfnamefont {A.~J.}\ \bibnamefont {Dianoux}}, \bibinfo {author}
  {\bibfnamefont {N.}~\bibnamefont {Ahmad}},\ and\ \bibinfo {author}
  {\bibfnamefont {W.~A.}\ \bibnamefont {Phillips}},\ }\href
  {https://doi.org/10.1103/PhysRevB.34.5665} {\bibfield  {journal} {\bibinfo
  {journal} {Phys. Rev. B}\ }\textbf {\bibinfo {volume} {34}},\ \bibinfo
  {pages} {5665} (\bibinfo {year} {1986})}\BibitemShut {NoStop}%
\bibitem [{\citenamefont {Yamamuro}\ \emph {et~al.}(1996)\citenamefont
  {Yamamuro}, \citenamefont {Matsuo}, \citenamefont {Takeda}, \citenamefont
  {Kanaya}, \citenamefont {Kawaguchi},\ and\ \citenamefont
  {Kaji}}]{Yamamuro_1996}%
  \BibitemOpen
  \bibfield  {author} {\bibinfo {author} {\bibfnamefont {O.}~\bibnamefont
  {Yamamuro}}, \bibinfo {author} {\bibfnamefont {T.}~\bibnamefont {Matsuo}},
  \bibinfo {author} {\bibfnamefont {K.}~\bibnamefont {Takeda}}, \bibinfo
  {author} {\bibfnamefont {T.}~\bibnamefont {Kanaya}}, \bibinfo {author}
  {\bibfnamefont {T.}~\bibnamefont {Kawaguchi}},\ and\ \bibinfo {author}
  {\bibfnamefont {K.}~\bibnamefont {Kaji}},\ }\href
  {https://doi.org/10.1063/1.471928} {\bibfield  {journal} {\bibinfo  {journal}
  {The Journal of Chemical Physics}\ }\textbf {\bibinfo {volume} {105}},\
  \bibinfo {pages} {732} (\bibinfo {year} {1996})}\BibitemShut {NoStop}%
\bibitem [{\citenamefont {Arai}\ \emph {et~al.}(1999)\citenamefont {Arai},
  \citenamefont {Inamura}, \citenamefont {Otomo}, \citenamefont {Kitamura},
  \citenamefont {Bennington},\ and\ \citenamefont {Hannon}}]{Arai_1999}%
  \BibitemOpen
  \bibfield  {author} {\bibinfo {author} {\bibfnamefont {M.}~\bibnamefont
  {Arai}}, \bibinfo {author} {\bibfnamefont {Y.}~\bibnamefont {Inamura}},
  \bibinfo {author} {\bibfnamefont {T.}~\bibnamefont {Otomo}}, \bibinfo
  {author} {\bibfnamefont {N.}~\bibnamefont {Kitamura}}, \bibinfo {author}
  {\bibfnamefont {S.~M.}\ \bibnamefont {Bennington}},\ and\ \bibinfo {author}
  {\bibfnamefont {A.~C.}\ \bibnamefont {Hannon}},\ }\href
  {https://doi.org/https://doi.org/10.1016/S0921-4526(98)01354-4} {\bibfield
  {journal} {\bibinfo  {journal} {Physica B: Condensed Matter}\ }\textbf
  {\bibinfo {volume} {263-264}},\ \bibinfo {pages} {268} (\bibinfo {year}
  {1999})}\BibitemShut {NoStop}%
\bibitem [{\citenamefont {Harris}\ \emph {et~al.}(2000)\citenamefont {Harris},
  \citenamefont {Dove},\ and\ \citenamefont {Parker}}]{Harris_2000}%
  \BibitemOpen
  \bibfield  {author} {\bibinfo {author} {\bibfnamefont {M.~J.}\ \bibnamefont
  {Harris}}, \bibinfo {author} {\bibfnamefont {M.~T.}\ \bibnamefont {Dove}},\
  and\ \bibinfo {author} {\bibfnamefont {J.~M.}\ \bibnamefont {Parker}},\
  }\href {https://doi.org/10.1180/002646100549490} {\bibfield  {journal}
  {\bibinfo  {journal} {Mineralogical Magazine}\ }\textbf {\bibinfo {volume}
  {64}},\ \bibinfo {pages} {435^^e2^^80^^93440} (\bibinfo {year}
  {2000})}\BibitemShut {NoStop}%
\bibitem [{\citenamefont {Monaco}\ \emph {et~al.}(2006)\citenamefont {Monaco},
  \citenamefont {Chumakov}, \citenamefont {Monaco}, \citenamefont {Crichton},
  \citenamefont {Meyer}, \citenamefont {Comez}, \citenamefont {Fioretto},
  \citenamefont {Korecki},\ and\ \citenamefont {R\"uffer}}]{Monaco_2006}%
  \BibitemOpen
  \bibfield  {author} {\bibinfo {author} {\bibfnamefont {A.}~\bibnamefont
  {Monaco}}, \bibinfo {author} {\bibfnamefont {A.~I.}\ \bibnamefont
  {Chumakov}}, \bibinfo {author} {\bibfnamefont {G.}~\bibnamefont {Monaco}},
  \bibinfo {author} {\bibfnamefont {W.~A.}\ \bibnamefont {Crichton}}, \bibinfo
  {author} {\bibfnamefont {A.}~\bibnamefont {Meyer}}, \bibinfo {author}
  {\bibfnamefont {L.}~\bibnamefont {Comez}}, \bibinfo {author} {\bibfnamefont
  {D.}~\bibnamefont {Fioretto}}, \bibinfo {author} {\bibfnamefont
  {J.}~\bibnamefont {Korecki}},\ and\ \bibinfo {author} {\bibfnamefont
  {R.}~\bibnamefont {R\"uffer}},\ }\href
  {https://doi.org/10.1103/PhysRevLett.97.135501} {\bibfield  {journal}
  {\bibinfo  {journal} {Phys. Rev. Lett.}\ }\textbf {\bibinfo {volume} {97}},\
  \bibinfo {pages} {135501} (\bibinfo {year} {2006})}\BibitemShut {NoStop}%
\bibitem [{\citenamefont {Niss}\ \emph {et~al.}(2007)\citenamefont {Niss},
  \citenamefont {Begen}, \citenamefont {Frick}, \citenamefont {Ollivier},
  \citenamefont {Beraud}, \citenamefont {Sokolov}, \citenamefont {Novikov},\
  and\ \citenamefont {Alba-Simionesco}}]{Niss_2007}%
  \BibitemOpen
  \bibfield  {author} {\bibinfo {author} {\bibfnamefont {K.}~\bibnamefont
  {Niss}}, \bibinfo {author} {\bibfnamefont {B.}~\bibnamefont {Begen}},
  \bibinfo {author} {\bibfnamefont {B.}~\bibnamefont {Frick}}, \bibinfo
  {author} {\bibfnamefont {J.}~\bibnamefont {Ollivier}}, \bibinfo {author}
  {\bibfnamefont {A.}~\bibnamefont {Beraud}}, \bibinfo {author} {\bibfnamefont
  {A.}~\bibnamefont {Sokolov}}, \bibinfo {author} {\bibfnamefont {V.~N.}\
  \bibnamefont {Novikov}},\ and\ \bibinfo {author} {\bibfnamefont
  {C.}~\bibnamefont {Alba-Simionesco}},\ }\href
  {https://doi.org/10.1103/PhysRevLett.99.055502} {\bibfield  {journal}
  {\bibinfo  {journal} {Phys. Rev. Lett.}\ }\textbf {\bibinfo {volume} {99}},\
  \bibinfo {pages} {055502} (\bibinfo {year} {2007})}\BibitemShut {NoStop}%
\bibitem [{\citenamefont {Ruffl\'e}\ \emph {et~al.}(2008)\citenamefont
  {Ruffl\'e}, \citenamefont {Parshin}, \citenamefont {Courtens},\ and\
  \citenamefont {Vacher}}]{Ruffle_2008}%
  \BibitemOpen
  \bibfield  {author} {\bibinfo {author} {\bibfnamefont {B.}~\bibnamefont
  {Ruffl\'e}}, \bibinfo {author} {\bibfnamefont {D.~A.}\ \bibnamefont
  {Parshin}}, \bibinfo {author} {\bibfnamefont {E.}~\bibnamefont {Courtens}},\
  and\ \bibinfo {author} {\bibfnamefont {R.}~\bibnamefont {Vacher}},\ }\href
  {https://doi.org/10.1103/PhysRevLett.100.015501} {\bibfield  {journal}
  {\bibinfo  {journal} {Phys. Rev. Lett.}\ }\textbf {\bibinfo {volume} {100}},\
  \bibinfo {pages} {015501} (\bibinfo {year} {2008})}\BibitemShut {NoStop}%
\bibitem [{\citenamefont {Monaco}\ and\ \citenamefont
  {Giordano}(2009)}]{Monaco_2009}%
  \BibitemOpen
  \bibfield  {author} {\bibinfo {author} {\bibfnamefont {G.}~\bibnamefont
  {Monaco}}\ and\ \bibinfo {author} {\bibfnamefont {V.~M.}\ \bibnamefont
  {Giordano}},\ }\href@noop {} {\bibfield  {journal} {\bibinfo  {journal}
  {Proc. Natl. Acad. Sci. USA}\ }\textbf {\bibinfo {volume} {106}},\ \bibinfo
  {pages} {3659} (\bibinfo {year} {2009})}\BibitemShut {NoStop}%
\bibitem [{\citenamefont {Baldi}\ \emph {et~al.}(2008)\citenamefont {Baldi},
  \citenamefont {Giordano}, \citenamefont {Monaco}, \citenamefont {Sette},
  \citenamefont {Fabiani}, \citenamefont {Fontana},\ and\ \citenamefont
  {Ruocco}}]{Baldi_2008}%
  \BibitemOpen
  \bibfield  {author} {\bibinfo {author} {\bibfnamefont {G.}~\bibnamefont
  {Baldi}}, \bibinfo {author} {\bibfnamefont {V.~M.}\ \bibnamefont {Giordano}},
  \bibinfo {author} {\bibfnamefont {G.}~\bibnamefont {Monaco}}, \bibinfo
  {author} {\bibfnamefont {F.}~\bibnamefont {Sette}}, \bibinfo {author}
  {\bibfnamefont {E.}~\bibnamefont {Fabiani}}, \bibinfo {author} {\bibfnamefont
  {A.}~\bibnamefont {Fontana}},\ and\ \bibinfo {author} {\bibfnamefont
  {G.}~\bibnamefont {Ruocco}},\ }\href
  {https://doi.org/10.1103/PhysRevB.77.214309} {\bibfield  {journal} {\bibinfo
  {journal} {Phys. Rev. B}\ }\textbf {\bibinfo {volume} {77}},\ \bibinfo
  {pages} {214309} (\bibinfo {year} {2008})}\BibitemShut {NoStop}%
\bibitem [{\citenamefont {Baldi}\ \emph {et~al.}(2010)\citenamefont {Baldi},
  \citenamefont {Giordano}, \citenamefont {Monaco},\ and\ \citenamefont
  {Ruta}}]{Baldi_2010}%
  \BibitemOpen
  \bibfield  {author} {\bibinfo {author} {\bibfnamefont {G.}~\bibnamefont
  {Baldi}}, \bibinfo {author} {\bibfnamefont {V.~M.}\ \bibnamefont {Giordano}},
  \bibinfo {author} {\bibfnamefont {G.}~\bibnamefont {Monaco}},\ and\ \bibinfo
  {author} {\bibfnamefont {B.}~\bibnamefont {Ruta}},\ }\href
  {https://doi.org/10.1103/PhysRevLett.104.195501} {\bibfield  {journal}
  {\bibinfo  {journal} {Phys. Rev. Lett.}\ }\textbf {\bibinfo {volume} {104}},\
  \bibinfo {pages} {195501} (\bibinfo {year} {2010})}\BibitemShut {NoStop}%
\bibitem [{\citenamefont {Baldi}\ \emph {et~al.}(2011)\citenamefont {Baldi},
  \citenamefont {Giordano},\ and\ \citenamefont {Monaco}}]{Baldi_2011}%
  \BibitemOpen
  \bibfield  {author} {\bibinfo {author} {\bibfnamefont {G.}~\bibnamefont
  {Baldi}}, \bibinfo {author} {\bibfnamefont {V.~M.}\ \bibnamefont
  {Giordano}},\ and\ \bibinfo {author} {\bibfnamefont {G.}~\bibnamefont
  {Monaco}},\ }\href {https://doi.org/10.1103/PhysRevB.83.174203} {\bibfield
  {journal} {\bibinfo  {journal} {Phys. Rev. B}\ }\textbf {\bibinfo {volume}
  {83}},\ \bibinfo {pages} {174203} (\bibinfo {year} {2011})}\BibitemShut
  {NoStop}%
\bibitem [{\citenamefont {Ruta}\ \emph {et~al.}(2012)\citenamefont {Ruta},
  \citenamefont {Baldi}, \citenamefont {Scarponi}, \citenamefont {Fioretto},
  \citenamefont {Giordano},\ and\ \citenamefont {Monaco}}]{Ruta_2012}%
  \BibitemOpen
  \bibfield  {author} {\bibinfo {author} {\bibfnamefont {B.}~\bibnamefont
  {Ruta}}, \bibinfo {author} {\bibfnamefont {G.}~\bibnamefont {Baldi}},
  \bibinfo {author} {\bibfnamefont {F.}~\bibnamefont {Scarponi}}, \bibinfo
  {author} {\bibfnamefont {D.}~\bibnamefont {Fioretto}}, \bibinfo {author}
  {\bibfnamefont {V.~M.}\ \bibnamefont {Giordano}},\ and\ \bibinfo {author}
  {\bibfnamefont {G.}~\bibnamefont {Monaco}},\ }\href
  {https://doi.org/10.1063/1.4768955} {\bibfield  {journal} {\bibinfo
  {journal} {The Journal of Chemical Physics}\ }\textbf {\bibinfo {volume}
  {137}},\ \bibinfo {pages} {214502} (\bibinfo {year} {2012})}\BibitemShut
  {NoStop}%
\bibitem [{\citenamefont {Baldi}\ \emph {et~al.}(2016)\citenamefont {Baldi},
  \citenamefont {Giordano}, \citenamefont {Ruta},\ and\ \citenamefont
  {Monaco}}]{Baldi_2016}%
  \BibitemOpen
  \bibfield  {author} {\bibinfo {author} {\bibfnamefont {G.}~\bibnamefont
  {Baldi}}, \bibinfo {author} {\bibfnamefont {V.~M.}\ \bibnamefont {Giordano}},
  \bibinfo {author} {\bibfnamefont {B.}~\bibnamefont {Ruta}},\ and\ \bibinfo
  {author} {\bibfnamefont {G.}~\bibnamefont {Monaco}},\ }\href
  {https://doi.org/10.1103/PhysRevB.93.144204} {\bibfield  {journal} {\bibinfo
  {journal} {Phys. Rev. B}\ }\textbf {\bibinfo {volume} {93}},\ \bibinfo
  {pages} {144204} (\bibinfo {year} {2016})}\BibitemShut {NoStop}%
\bibitem [{\citenamefont {Zeller}\ and\ \citenamefont
  {Pohl}(1971)}]{Zeller_1971}%
  \BibitemOpen
  \bibfield  {author} {\bibinfo {author} {\bibfnamefont {R.~C.}\ \bibnamefont
  {Zeller}}\ and\ \bibinfo {author} {\bibfnamefont {R.~O.}\ \bibnamefont
  {Pohl}},\ }\href {https://doi.org/10.1103/PhysRevB.4.2029} {\bibfield
  {journal} {\bibinfo  {journal} {Phys. Rev. B}\ }\textbf {\bibinfo {volume}
  {4}},\ \bibinfo {pages} {2029} (\bibinfo {year} {1971})}\BibitemShut
  {NoStop}%
\bibitem [{\citenamefont {Anderson}\ \emph {et~al.}(1972)\citenamefont
  {Anderson}, \citenamefont {Halperin},\ and\ \citenamefont
  {Varma}}]{Anderson_1972}%
  \BibitemOpen
  \bibfield  {author} {\bibinfo {author} {\bibfnamefont {P.~W.}\ \bibnamefont
  {Anderson}}, \bibinfo {author} {\bibfnamefont {B.~I.}\ \bibnamefont
  {Halperin}},\ and\ \bibinfo {author} {\bibfnamefont {C.~M.}\ \bibnamefont
  {Varma}},\ }\href@noop {} {\bibfield  {journal} {\bibinfo  {journal}
  {Philosophical Magazine}\ }\textbf {\bibinfo {volume} {25}},\ \bibinfo
  {pages} {1} (\bibinfo {year} {1972})}\BibitemShut {NoStop}%
\bibitem [{\citenamefont {Cahill}\ and\ \citenamefont
  {Pohl}(1988)}]{Cahill_1988}%
  \BibitemOpen
  \bibfield  {author} {\bibinfo {author} {\bibfnamefont {D.}~\bibnamefont
  {Cahill}}\ and\ \bibinfo {author} {\bibfnamefont {R.~O.}\ \bibnamefont
  {Pohl}},\ }\href {https://doi.org/10.1146/annurev.pc.39.100188.000521}
  {\bibfield  {journal} {\bibinfo  {journal} {Annual Review of Physical
  Chemistry}\ }\textbf {\bibinfo {volume} {39}},\ \bibinfo {pages} {93}
  (\bibinfo {year} {1988})}\BibitemShut {NoStop}%
\bibitem [{\citenamefont {Monaco}\ and\ \citenamefont
  {Mossa}(2009)}]{Monaco2_2009}%
  \BibitemOpen
  \bibfield  {author} {\bibinfo {author} {\bibfnamefont {G.}~\bibnamefont
  {Monaco}}\ and\ \bibinfo {author} {\bibfnamefont {S.}~\bibnamefont {Mossa}},\
  }\href@noop {} {\bibfield  {journal} {\bibinfo  {journal} {Proceedings of the
  National Academy of Sciences}\ }\textbf {\bibinfo {volume} {106}},\ \bibinfo
  {pages} {16907} (\bibinfo {year} {2009})}\BibitemShut {NoStop}%
\bibitem [{\citenamefont {Mizuno}\ \emph {et~al.}(2014)\citenamefont {Mizuno},
  \citenamefont {Mossa},\ and\ \citenamefont {Barrat}}]{Mizuno_2014}%
  \BibitemOpen
  \bibfield  {author} {\bibinfo {author} {\bibfnamefont {H.}~\bibnamefont
  {Mizuno}}, \bibinfo {author} {\bibfnamefont {S.}~\bibnamefont {Mossa}},\ and\
  \bibinfo {author} {\bibfnamefont {J.-L.}\ \bibnamefont {Barrat}},\ }\href
  {https://doi.org/10.1073/pnas.1409490111} {\bibfield  {journal} {\bibinfo
  {journal} {Proceedings of the National Academy of Sciences}\ }\textbf
  {\bibinfo {volume} {111}},\ \bibinfo {pages} {11949} (\bibinfo {year}
  {2014})}\BibitemShut {NoStop}%
\bibitem [{\citenamefont {Tanguy}(2023)}]{Tanguy_2023}%
  \BibitemOpen
  \bibfield  {author} {\bibinfo {author} {\bibfnamefont {A.}~\bibnamefont
  {Tanguy}},\ }\href {https://doi.org/10.5802/crphys.162} {\bibfield  {journal}
  {\bibinfo  {journal} {Comptes Rendus. Physique}\ }\textbf {\bibinfo {volume}
  {24}},\ \bibinfo {pages} {73} (\bibinfo {year} {2023})}\BibitemShut {NoStop}%
\bibitem [{\citenamefont {Leonforte}\ \emph {et~al.}(2005)\citenamefont
  {Leonforte}, \citenamefont {Boissi\`ere}, \citenamefont {Tanguy},
  \citenamefont {Wittmer},\ and\ \citenamefont {Barrat}}]{Leonforte_2005}%
  \BibitemOpen
  \bibfield  {author} {\bibinfo {author} {\bibfnamefont {F.}~\bibnamefont
  {Leonforte}}, \bibinfo {author} {\bibfnamefont {R.}~\bibnamefont
  {Boissi\`ere}}, \bibinfo {author} {\bibfnamefont {A.}~\bibnamefont {Tanguy}},
  \bibinfo {author} {\bibfnamefont {J.~P.}\ \bibnamefont {Wittmer}},\ and\
  \bibinfo {author} {\bibfnamefont {J.-L.}\ \bibnamefont {Barrat}},\ }\href
  {https://doi.org/10.1103/PhysRevB.72.224206} {\bibfield  {journal} {\bibinfo
  {journal} {Phys. Rev. B}\ }\textbf {\bibinfo {volume} {72}},\ \bibinfo
  {pages} {224206} (\bibinfo {year} {2005})}\BibitemShut {NoStop}%
\bibitem [{\citenamefont {L\'eonforte}\ \emph {et~al.}(2006)\citenamefont
  {L\'eonforte}, \citenamefont {Tanguy}, \citenamefont {Wittmer},\ and\
  \citenamefont {Barrat}}]{Leonforte_2006}%
  \BibitemOpen
  \bibfield  {author} {\bibinfo {author} {\bibfnamefont {F.}~\bibnamefont
  {L\'eonforte}}, \bibinfo {author} {\bibfnamefont {A.}~\bibnamefont {Tanguy}},
  \bibinfo {author} {\bibfnamefont {J.~P.}\ \bibnamefont {Wittmer}},\ and\
  \bibinfo {author} {\bibfnamefont {J.-L.}\ \bibnamefont {Barrat}},\ }\href
  {https://doi.org/10.1103/PhysRevLett.97.055501} {\bibfield  {journal}
  {\bibinfo  {journal} {Phys. Rev. Lett.}\ }\textbf {\bibinfo {volume} {97}},\
  \bibinfo {pages} {055501} (\bibinfo {year} {2006})}\BibitemShut {NoStop}%
\bibitem [{\citenamefont {Tanguy}\ \emph {et~al.}(2010)\citenamefont {Tanguy},
  \citenamefont {Mantisi},\ and\ \citenamefont {Tsamados}}]{Tanguy_2010}%
  \BibitemOpen
  \bibfield  {author} {\bibinfo {author} {\bibfnamefont {A.}~\bibnamefont
  {Tanguy}}, \bibinfo {author} {\bibfnamefont {B.}~\bibnamefont {Mantisi}},\
  and\ \bibinfo {author} {\bibfnamefont {M.}~\bibnamefont {Tsamados}},\
  }\href@noop {} {\bibfield  {journal} {\bibinfo  {journal} {Europhys. Lett.}\
  }\textbf {\bibinfo {volume} {90}},\ \bibinfo {pages} {16004} (\bibinfo {year}
  {2010})}\BibitemShut {NoStop}%
\bibitem [{\citenamefont {Manning}\ and\ \citenamefont
  {Liu}(2011)}]{Manning_2011}%
  \BibitemOpen
  \bibfield  {author} {\bibinfo {author} {\bibfnamefont {M.~L.}\ \bibnamefont
  {Manning}}\ and\ \bibinfo {author} {\bibfnamefont {A.~J.}\ \bibnamefont
  {Liu}},\ }\href {https://doi.org/10.1103/PhysRevLett.107.108302} {\bibfield
  {journal} {\bibinfo  {journal} {Phys. Rev. Lett.}\ }\textbf {\bibinfo
  {volume} {107}},\ \bibinfo {pages} {108302} (\bibinfo {year}
  {2011})}\BibitemShut {NoStop}%
\bibitem [{\citenamefont {Schirmacher}(2006)}]{Schirmacher_2006}%
  \BibitemOpen
  \bibfield  {author} {\bibinfo {author} {\bibfnamefont {W.}~\bibnamefont
  {Schirmacher}},\ }\href@noop {} {\bibfield  {journal} {\bibinfo  {journal}
  {Europhys. Lett.}\ }\textbf {\bibinfo {volume} {73}},\ \bibinfo {pages} {892}
  (\bibinfo {year} {2006})}\BibitemShut {NoStop}%
\bibitem [{\citenamefont {Marruzzo}\ \emph {et~al.}(2013)\citenamefont
  {Marruzzo}, \citenamefont {Schirmacher}, \citenamefont {Fratalocchi},\ and\
  \citenamefont {Ruocco}}]{Marruzzo_2013}%
  \BibitemOpen
  \bibfield  {author} {\bibinfo {author} {\bibfnamefont {A.}~\bibnamefont
  {Marruzzo}}, \bibinfo {author} {\bibfnamefont {W.}~\bibnamefont
  {Schirmacher}}, \bibinfo {author} {\bibfnamefont {A.}~\bibnamefont
  {Fratalocchi}},\ and\ \bibinfo {author} {\bibfnamefont {G.}~\bibnamefont
  {Ruocco}},\ }\href@noop {} {\bibfield  {journal} {\bibinfo  {journal}
  {Scientific Reports}\ }\textbf {\bibinfo {volume} {3}},\ \bibinfo {pages}
  {1407} (\bibinfo {year} {2013})}\BibitemShut {NoStop}%
\bibitem [{\citenamefont {Mizuno}\ \emph
  {et~al.}(2013{\natexlab{a}})\citenamefont {Mizuno}, \citenamefont {Mossa},\
  and\ \citenamefont {Barrat}}]{Mizuno2_2013}%
  \BibitemOpen
  \bibfield  {author} {\bibinfo {author} {\bibfnamefont {H.}~\bibnamefont
  {Mizuno}}, \bibinfo {author} {\bibfnamefont {S.}~\bibnamefont {Mossa}},\ and\
  \bibinfo {author} {\bibfnamefont {J.-L.}\ \bibnamefont {Barrat}},\ }\href
  {http://stacks.iop.org/0295-5075/104/i=5/a=56001} {\bibfield  {journal}
  {\bibinfo  {journal} {EPL (Europhysics Letters)}\ }\textbf {\bibinfo {volume}
  {104}},\ \bibinfo {pages} {56001} (\bibinfo {year}
  {2013}{\natexlab{a}})}\BibitemShut {NoStop}%
\bibitem [{\citenamefont {Schirmacher}\ \emph {et~al.}(2015)\citenamefont
  {Schirmacher}, \citenamefont {Scopigno},\ and\ \citenamefont
  {Ruocco}}]{Schirmacher_2015}%
  \BibitemOpen
  \bibfield  {author} {\bibinfo {author} {\bibfnamefont {W.}~\bibnamefont
  {Schirmacher}}, \bibinfo {author} {\bibfnamefont {T.}~\bibnamefont
  {Scopigno}},\ and\ \bibinfo {author} {\bibfnamefont {G.}~\bibnamefont
  {Ruocco}},\ }\href {https://doi.org/10.1016/j.jnoncrysol.2014.09.054}
  {\bibfield  {journal} {\bibinfo  {journal} {Journal of Non-Crystalline
  Solids}\ }\textbf {\bibinfo {volume} {407}},\ \bibinfo {pages} {133}
  (\bibinfo {year} {2015})}\BibitemShut {NoStop}%
\bibitem [{\citenamefont {Zhang}\ \emph {et~al.}(2017)\citenamefont {Zhang},
  \citenamefont {Zheng}, \citenamefont {Wang}, \citenamefont {Zhang},
  \citenamefont {Jin}, \citenamefont {Hong}, \citenamefont {Wang},\ and\
  \citenamefont {Zhang}}]{Zhang_2017}%
  \BibitemOpen
  \bibfield  {author} {\bibinfo {author} {\bibfnamefont {L.}~\bibnamefont
  {Zhang}}, \bibinfo {author} {\bibfnamefont {J.}~\bibnamefont {Zheng}},
  \bibinfo {author} {\bibfnamefont {Y.}~\bibnamefont {Wang}}, \bibinfo {author}
  {\bibfnamefont {L.}~\bibnamefont {Zhang}}, \bibinfo {author} {\bibfnamefont
  {Z.}~\bibnamefont {Jin}}, \bibinfo {author} {\bibfnamefont {L.}~\bibnamefont
  {Hong}}, \bibinfo {author} {\bibfnamefont {Y.}~\bibnamefont {Wang}},\ and\
  \bibinfo {author} {\bibfnamefont {J.}~\bibnamefont {Zhang}},\ }\href@noop {}
  {\bibfield  {journal} {\bibinfo  {journal} {Nature Commun.}\ }\textbf
  {\bibinfo {volume} {8}},\ \bibinfo {pages} {67} (\bibinfo {year}
  {2017})}\BibitemShut {NoStop}%
\bibitem [{\citenamefont {Karpov}\ \emph {et~al.}(1983)\citenamefont {Karpov},
  \citenamefont {Klinger},\ and\ \citenamefont {Ignat'ev}}]{Karpov_1983}%
  \BibitemOpen
  \bibfield  {author} {\bibinfo {author} {\bibfnamefont {V.~G.}\ \bibnamefont
  {Karpov}}, \bibinfo {author} {\bibfnamefont {M.~I.}\ \bibnamefont
  {Klinger}},\ and\ \bibinfo {author} {\bibfnamefont {F.~N.}\ \bibnamefont
  {Ignat'ev}},\ }\href@noop {} {\bibfield  {journal} {\bibinfo  {journal} {Sov.
  Phys. JETP}\ }\textbf {\bibinfo {volume} {57}},\ \bibinfo {pages} {439}
  (\bibinfo {year} {1983})}\BibitemShut {NoStop}%
\bibitem [{\citenamefont {Buchenau}\ \emph {et~al.}(1991)\citenamefont
  {Buchenau}, \citenamefont {Galperin}, \citenamefont {Gurevich},\ and\
  \citenamefont {Schober}}]{Buchenau_1991}%
  \BibitemOpen
  \bibfield  {author} {\bibinfo {author} {\bibfnamefont {U.}~\bibnamefont
  {Buchenau}}, \bibinfo {author} {\bibfnamefont {Y.~M.}\ \bibnamefont
  {Galperin}}, \bibinfo {author} {\bibfnamefont {V.~L.}\ \bibnamefont
  {Gurevich}},\ and\ \bibinfo {author} {\bibfnamefont {H.~R.}\ \bibnamefont
  {Schober}},\ }\href {https://doi.org/10.1103/PhysRevB.43.5039} {\bibfield
  {journal} {\bibinfo  {journal} {Phys. Rev. B}\ }\textbf {\bibinfo {volume}
  {43}},\ \bibinfo {pages} {5039} (\bibinfo {year} {1991})}\BibitemShut
  {NoStop}%
\bibitem [{\citenamefont {Gurevich}\ \emph {et~al.}(2003)\citenamefont
  {Gurevich}, \citenamefont {Parshin},\ and\ \citenamefont
  {Schober}}]{Gurevich_2003}%
  \BibitemOpen
  \bibfield  {author} {\bibinfo {author} {\bibfnamefont {V.~L.}\ \bibnamefont
  {Gurevich}}, \bibinfo {author} {\bibfnamefont {D.~A.}\ \bibnamefont
  {Parshin}},\ and\ \bibinfo {author} {\bibfnamefont {H.~R.}\ \bibnamefont
  {Schober}},\ }\href {https://doi.org/10.1103/PhysRevB.67.094203} {\bibfield
  {journal} {\bibinfo  {journal} {Phys. Rev. B}\ }\textbf {\bibinfo {volume}
  {67}},\ \bibinfo {pages} {094203} (\bibinfo {year} {2003})}\BibitemShut
  {NoStop}%
\bibitem [{\citenamefont {Bouchbinder}\ \emph {et~al.}(2021)\citenamefont
  {Bouchbinder}, \citenamefont {Lerner}, \citenamefont {Rainone}, \citenamefont
  {Urbani},\ and\ \citenamefont {Zamponi}}]{Bouchbinder_2021}%
  \BibitemOpen
  \bibfield  {author} {\bibinfo {author} {\bibfnamefont {E.}~\bibnamefont
  {Bouchbinder}}, \bibinfo {author} {\bibfnamefont {E.}~\bibnamefont {Lerner}},
  \bibinfo {author} {\bibfnamefont {C.}~\bibnamefont {Rainone}}, \bibinfo
  {author} {\bibfnamefont {P.}~\bibnamefont {Urbani}},\ and\ \bibinfo {author}
  {\bibfnamefont {F.}~\bibnamefont {Zamponi}},\ }\href
  {https://doi.org/10.1103/PhysRevB.103.174202} {\bibfield  {journal} {\bibinfo
   {journal} {Phys. Rev. B}\ }\textbf {\bibinfo {volume} {103}},\ \bibinfo
  {pages} {174202} (\bibinfo {year} {2021})}\BibitemShut {NoStop}%
\bibitem [{\citenamefont {Beltukov}\ \emph {et~al.}(2013)\citenamefont
  {Beltukov}, \citenamefont {Kozub},\ and\ \citenamefont
  {Parshin}}]{Beltukov_2013}%
  \BibitemOpen
  \bibfield  {author} {\bibinfo {author} {\bibfnamefont {Y.~M.}\ \bibnamefont
  {Beltukov}}, \bibinfo {author} {\bibfnamefont {V.~I.}\ \bibnamefont
  {Kozub}},\ and\ \bibinfo {author} {\bibfnamefont {D.~A.}\ \bibnamefont
  {Parshin}},\ }\href {https://doi.org/10.1103/PhysRevB.87.134203} {\bibfield
  {journal} {\bibinfo  {journal} {Phys. Rev. B}\ }\textbf {\bibinfo {volume}
  {87}},\ \bibinfo {pages} {134203} (\bibinfo {year} {2013})}\BibitemShut
  {NoStop}%
\bibitem [{\citenamefont {Vogel}\ and\ \citenamefont
  {Fuchs}(2023)}]{Vogel_2023}%
  \BibitemOpen
  \bibfield  {author} {\bibinfo {author} {\bibfnamefont {F.}~\bibnamefont
  {Vogel}}\ and\ \bibinfo {author} {\bibfnamefont {M.}~\bibnamefont {Fuchs}},\
  }\href {https://doi.org/10.1103/PhysRevLett.130.236101} {\bibfield  {journal}
  {\bibinfo  {journal} {Phys. Rev. Lett.}\ }\textbf {\bibinfo {volume} {130}},\
  \bibinfo {pages} {236101} (\bibinfo {year} {2023})}\BibitemShut {NoStop}%
\bibitem [{\citenamefont {Phillips}(1979)}]{Phillips_1979}%
  \BibitemOpen
  \bibfield  {author} {\bibinfo {author} {\bibfnamefont {J.~C.}\ \bibnamefont
  {Phillips}},\ }\href {https://doi.org/10.1016/0022-3093(79)90033-4}
  {\bibfield  {journal} {\bibinfo  {journal} {Journal of Non-Crystalline
  Solids}\ }\textbf {\bibinfo {volume} {34}},\ \bibinfo {pages} {153} (\bibinfo
  {year} {1979})}\BibitemShut {NoStop}%
\bibitem [{\citenamefont {D^^c3^^b6hler}\ \emph {et~al.}(1980)\citenamefont
  {D^^c3^^b6hler}, \citenamefont {Dandoloff},\ and\ \citenamefont
  {Bilz}}]{Dohler_1980}%
  \BibitemOpen
  \bibfield  {author} {\bibinfo {author} {\bibfnamefont {G.~H.}\ \bibnamefont
  {D^^c3^^b6hler}}, \bibinfo {author} {\bibfnamefont {R.}~\bibnamefont
  {Dandoloff}},\ and\ \bibinfo {author} {\bibfnamefont {H.}~\bibnamefont
  {Bilz}},\ }\href {https://doi.org/10.1016/0022-3093(80)90010-1} {\bibfield
  {journal} {\bibinfo  {journal} {Journal of Non-Crystalline Solids}\ }\textbf
  {\bibinfo {volume} {42}},\ \bibinfo {pages} {87} (\bibinfo {year}
  {1980})}\BibitemShut {NoStop}%
\bibitem [{\citenamefont {Phillips}(1981)}]{Phillips2_1981}%
  \BibitemOpen
  \bibfield  {author} {\bibinfo {author} {\bibfnamefont {J.~C.}\ \bibnamefont
  {Phillips}},\ }\href {https://doi.org/10.1016/0022-3093(81)90172-1}
  {\bibfield  {journal} {\bibinfo  {journal} {Journal of Non-Crystalline
  Solids}\ }\textbf {\bibinfo {volume} {43}},\ \bibinfo {pages} {37} (\bibinfo
  {year} {1981})}\BibitemShut {NoStop}%
\bibitem [{\citenamefont {Thorpe}(1983)}]{Thorpe_1983}%
  \BibitemOpen
  \bibfield  {author} {\bibinfo {author} {\bibfnamefont {M.~F.}\ \bibnamefont
  {Thorpe}},\ }\href
  {https://doi.org/https://doi.org/10.1016/0022-3093(83)90424-6} {\bibfield
  {journal} {\bibinfo  {journal} {Journal of Non-Crystalline Solids}\ }\textbf
  {\bibinfo {volume} {57}},\ \bibinfo {pages} {355} (\bibinfo {year}
  {1983})}\BibitemShut {NoStop}%
\bibitem [{\citenamefont {Maxwell}(1864)}]{Maxwell_1864}%
  \BibitemOpen
  \bibfield  {author} {\bibinfo {author} {\bibfnamefont {J.~C.}\ \bibnamefont
  {Maxwell}},\ }\href {https://doi.org/10.1080/14786446408643668} {\bibfield
  {journal} {\bibinfo  {journal} {The London, Edinburgh, and Dublin
  Philosophical Magazine and Journal of Science}\ }\textbf {\bibinfo {volume}
  {27}},\ \bibinfo {pages} {294} (\bibinfo {year} {1864})}\BibitemShut
  {NoStop}%
\bibitem [{\citenamefont {Feng}\ \emph {et~al.}(1985)\citenamefont {Feng},
  \citenamefont {Thorpe},\ and\ \citenamefont {Garboczi}}]{Feng_1985}%
  \BibitemOpen
  \bibfield  {author} {\bibinfo {author} {\bibfnamefont {S.}~\bibnamefont
  {Feng}}, \bibinfo {author} {\bibfnamefont {M.~F.}\ \bibnamefont {Thorpe}},\
  and\ \bibinfo {author} {\bibfnamefont {E.}~\bibnamefont {Garboczi}},\ }\href
  {https://doi.org/10.1103/PhysRevB.31.276} {\bibfield  {journal} {\bibinfo
  {journal} {Phys. Rev. B}\ }\textbf {\bibinfo {volume} {31}},\ \bibinfo
  {pages} {276} (\bibinfo {year} {1985})}\BibitemShut {NoStop}%
\bibitem [{\citenamefont {He}\ and\ \citenamefont {Thorpe}(1985)}]{He_1985}%
  \BibitemOpen
  \bibfield  {author} {\bibinfo {author} {\bibfnamefont {H.}~\bibnamefont
  {He}}\ and\ \bibinfo {author} {\bibfnamefont {M.~F.}\ \bibnamefont
  {Thorpe}},\ }\href {https://doi.org/10.1103/PhysRevLett.54.2107} {\bibfield
  {journal} {\bibinfo  {journal} {Phys. Rev. Lett.}\ }\textbf {\bibinfo
  {volume} {54}},\ \bibinfo {pages} {2107} (\bibinfo {year}
  {1985})}\BibitemShut {NoStop}%
\bibitem [{\citenamefont {Cai}\ and\ \citenamefont {Thorpe}(1989)}]{Cai_1989}%
  \BibitemOpen
  \bibfield  {author} {\bibinfo {author} {\bibfnamefont {Y.}~\bibnamefont
  {Cai}}\ and\ \bibinfo {author} {\bibfnamefont {M.~F.}\ \bibnamefont
  {Thorpe}},\ }\href {https://doi.org/10.1103/PhysRevB.40.10535} {\bibfield
  {journal} {\bibinfo  {journal} {Phys. Rev. B}\ }\textbf {\bibinfo {volume}
  {40}},\ \bibinfo {pages} {10535} (\bibinfo {year} {1989})}\BibitemShut
  {NoStop}%
\bibitem [{\citenamefont {Jacobs}\ and\ \citenamefont
  {Thorpe}(1995)}]{Jacobs_1995}%
  \BibitemOpen
  \bibfield  {author} {\bibinfo {author} {\bibfnamefont {D.~J.}\ \bibnamefont
  {Jacobs}}\ and\ \bibinfo {author} {\bibfnamefont {M.~F.}\ \bibnamefont
  {Thorpe}},\ }\href {https://doi.org/10.1103/PhysRevLett.75.4051} {\bibfield
  {journal} {\bibinfo  {journal} {Phys. Rev. Lett.}\ }\textbf {\bibinfo
  {volume} {75}},\ \bibinfo {pages} {4051} (\bibinfo {year}
  {1995})}\BibitemShut {NoStop}%
\bibitem [{\citenamefont {Trachenko}\ \emph {et~al.}(1998)\citenamefont
  {Trachenko}, \citenamefont {Dove}, \citenamefont {Hammonds}, \citenamefont
  {Harris},\ and\ \citenamefont {Heine}}]{Trachenko_1998}%
  \BibitemOpen
  \bibfield  {author} {\bibinfo {author} {\bibfnamefont {K.}~\bibnamefont
  {Trachenko}}, \bibinfo {author} {\bibfnamefont {M.~T.}\ \bibnamefont {Dove}},
  \bibinfo {author} {\bibfnamefont {K.~D.}\ \bibnamefont {Hammonds}}, \bibinfo
  {author} {\bibfnamefont {M.~J.}\ \bibnamefont {Harris}},\ and\ \bibinfo
  {author} {\bibfnamefont {V.}~\bibnamefont {Heine}},\ }\href
  {https://doi.org/10.1103/PhysRevLett.81.3431} {\bibfield  {journal} {\bibinfo
   {journal} {Phys. Rev. Lett.}\ }\textbf {\bibinfo {volume} {81}},\ \bibinfo
  {pages} {3431} (\bibinfo {year} {1998})}\BibitemShut {NoStop}%
\bibitem [{\citenamefont {Trachenko}\ \emph {et~al.}(2000)\citenamefont
  {Trachenko}, \citenamefont {Dove}, \citenamefont {Harris},\ and\
  \citenamefont {Heine}}]{Trachenko_2000}%
  \BibitemOpen
  \bibfield  {author} {\bibinfo {author} {\bibfnamefont {K.~O.}\ \bibnamefont
  {Trachenko}}, \bibinfo {author} {\bibfnamefont {M.~T.}\ \bibnamefont {Dove}},
  \bibinfo {author} {\bibfnamefont {M.~J.}\ \bibnamefont {Harris}},\ and\
  \bibinfo {author} {\bibfnamefont {V.}~\bibnamefont {Heine}},\ }\href
  {https://doi.org/10.1088/0953-8984/12/37/304} {\bibfield  {journal} {\bibinfo
   {journal} {Journal of Physics: Condensed Matter}\ }\textbf {\bibinfo
  {volume} {12}},\ \bibinfo {pages} {8041} (\bibinfo {year}
  {2000})}\BibitemShut {NoStop}%
\bibitem [{\citenamefont {Durian}(1995)}]{Durian_1995}%
  \BibitemOpen
  \bibfield  {author} {\bibinfo {author} {\bibfnamefont {D.~J.}\ \bibnamefont
  {Durian}},\ }\href {https://doi.org/10.1103/PhysRevLett.75.4780} {\bibfield
  {journal} {\bibinfo  {journal} {Phys. Rev. Lett.}\ }\textbf {\bibinfo
  {volume} {75}},\ \bibinfo {pages} {4780} (\bibinfo {year}
  {1995})}\BibitemShut {NoStop}%
\bibitem [{\citenamefont {O'Hern}\ \emph {et~al.}(2003)\citenamefont {O'Hern},
  \citenamefont {Silbert}, \citenamefont {Liu},\ and\ \citenamefont
  {Nagel}}]{Hern_2003}%
  \BibitemOpen
  \bibfield  {author} {\bibinfo {author} {\bibfnamefont {C.~S.}\ \bibnamefont
  {O'Hern}}, \bibinfo {author} {\bibfnamefont {L.~E.}\ \bibnamefont {Silbert}},
  \bibinfo {author} {\bibfnamefont {A.~J.}\ \bibnamefont {Liu}},\ and\ \bibinfo
  {author} {\bibfnamefont {S.~R.}\ \bibnamefont {Nagel}},\ }\href
  {https://doi.org/10.1103/PhysRevE.68.011306} {\bibfield  {journal} {\bibinfo
  {journal} {Phys. Rev. E}\ }\textbf {\bibinfo {volume} {68}},\ \bibinfo
  {pages} {011306} (\bibinfo {year} {2003})}\BibitemShut {NoStop}%
\bibitem [{\citenamefont {van Hecke}(2009)}]{Hecke_2010}%
  \BibitemOpen
  \bibfield  {author} {\bibinfo {author} {\bibfnamefont {M.}~\bibnamefont {van
  Hecke}},\ }\href {https://doi.org/10.1088/0953-8984/22/3/033101} {\bibfield
  {journal} {\bibinfo  {journal} {Journal of Physics: Condensed Matter}\
  }\textbf {\bibinfo {volume} {22}},\ \bibinfo {pages} {033101} (\bibinfo
  {year} {2009})}\BibitemShut {NoStop}%
\bibitem [{\citenamefont {Wyart}(2005)}]{Wyart_2005}%
  \BibitemOpen
  \bibfield  {author} {\bibinfo {author} {\bibfnamefont {M.}~\bibnamefont
  {Wyart}},\ }\href {https://api.semanticscholar.org/CorpusID:119417013}
  {\bibfield  {journal} {\bibinfo  {journal} {Annales De Physique}\ }\textbf
  {\bibinfo {volume} {30}},\ \bibinfo {pages} {1} (\bibinfo {year}
  {2005})}\BibitemShut {NoStop}%
\bibitem [{\citenamefont {Wyart}\ \emph
  {et~al.}(2005{\natexlab{a}})\citenamefont {Wyart}, \citenamefont {Nagel},\
  and\ \citenamefont {Witten}}]{Wyart2_2005}%
  \BibitemOpen
  \bibfield  {author} {\bibinfo {author} {\bibfnamefont {M.}~\bibnamefont
  {Wyart}}, \bibinfo {author} {\bibfnamefont {S.~R.}\ \bibnamefont {Nagel}},\
  and\ \bibinfo {author} {\bibfnamefont {T.~A.}\ \bibnamefont {Witten}},\
  }\href {http://stacks.iop.org/0295-5075/72/i=3/a=486} {\bibfield  {journal}
  {\bibinfo  {journal} {EPL (Europhysics Letters)}\ }\textbf {\bibinfo {volume}
  {72}},\ \bibinfo {pages} {486} (\bibinfo {year}
  {2005}{\natexlab{a}})}\BibitemShut {NoStop}%
\bibitem [{\citenamefont {Wyart}\ \emph
  {et~al.}(2005{\natexlab{b}})\citenamefont {Wyart}, \citenamefont {Silbert},
  \citenamefont {Nagel},\ and\ \citenamefont {Witten}}]{Wyart3_2005}%
  \BibitemOpen
  \bibfield  {author} {\bibinfo {author} {\bibfnamefont {M.}~\bibnamefont
  {Wyart}}, \bibinfo {author} {\bibfnamefont {L.~E.}\ \bibnamefont {Silbert}},
  \bibinfo {author} {\bibfnamefont {S.~R.}\ \bibnamefont {Nagel}},\ and\
  \bibinfo {author} {\bibfnamefont {T.~A.}\ \bibnamefont {Witten}},\ }\href
  {https://doi.org/10.1103/PhysRevE.72.051306} {\bibfield  {journal} {\bibinfo
  {journal} {Phys. Rev. E}\ }\textbf {\bibinfo {volume} {72}},\ \bibinfo
  {pages} {051306} (\bibinfo {year} {2005}{\natexlab{b}})}\BibitemShut
  {NoStop}%
\bibitem [{\citenamefont {Wyart}(2010)}]{Wyart_2010}%
  \BibitemOpen
  \bibfield  {author} {\bibinfo {author} {\bibfnamefont {M.}~\bibnamefont
  {Wyart}},\ }\href {http://stacks.iop.org/0295-5075/89/i=6/a=64001} {\bibfield
   {journal} {\bibinfo  {journal} {EPL (Europhysics Letters)}\ }\textbf
  {\bibinfo {volume} {89}},\ \bibinfo {pages} {64001} (\bibinfo {year}
  {2010})}\BibitemShut {NoStop}%
\bibitem [{\citenamefont {DeGiuli}\ \emph {et~al.}(2014)\citenamefont
  {DeGiuli}, \citenamefont {Laversanne-Finot}, \citenamefont {D^^c3^^bcring},
  \citenamefont {Lerner},\ and\ \citenamefont {Wyart}}]{Degiuli_2014}%
  \BibitemOpen
  \bibfield  {author} {\bibinfo {author} {\bibfnamefont {E.}~\bibnamefont
  {DeGiuli}}, \bibinfo {author} {\bibfnamefont {A.}~\bibnamefont
  {Laversanne-Finot}}, \bibinfo {author} {\bibfnamefont {G.}~\bibnamefont
  {D^^c3^^bcring}}, \bibinfo {author} {\bibfnamefont {E.}~\bibnamefont
  {Lerner}},\ and\ \bibinfo {author} {\bibfnamefont {M.}~\bibnamefont
  {Wyart}},\ }\href {https://doi.org/10.1039/C4SM00561A} {\bibfield  {journal}
  {\bibinfo  {journal} {Soft Matter}\ }\textbf {\bibinfo {volume} {10}},\
  \bibinfo {pages} {5628} (\bibinfo {year} {2014})}\BibitemShut {NoStop}%
\bibitem [{\citenamefont {Silbert}\ \emph {et~al.}(2005)\citenamefont
  {Silbert}, \citenamefont {Liu},\ and\ \citenamefont {Nagel}}]{Silbert_2005}%
  \BibitemOpen
  \bibfield  {author} {\bibinfo {author} {\bibfnamefont {L.~E.}\ \bibnamefont
  {Silbert}}, \bibinfo {author} {\bibfnamefont {A.~J.}\ \bibnamefont {Liu}},\
  and\ \bibinfo {author} {\bibfnamefont {S.~R.}\ \bibnamefont {Nagel}},\ }\href
  {https://doi.org/10.1103/PhysRevLett.95.098301} {\bibfield  {journal}
  {\bibinfo  {journal} {Phys. Rev. Lett.}\ }\textbf {\bibinfo {volume} {95}},\
  \bibinfo {pages} {098301} (\bibinfo {year} {2005})}\BibitemShut {NoStop}%
\bibitem [{\citenamefont {Xu}\ \emph {et~al.}(2010)\citenamefont {Xu},
  \citenamefont {Vitelli}, \citenamefont {Liu},\ and\ \citenamefont
  {Nagel}}]{Xu_2010}%
  \BibitemOpen
  \bibfield  {author} {\bibinfo {author} {\bibfnamefont {N.}~\bibnamefont
  {Xu}}, \bibinfo {author} {\bibfnamefont {V.}~\bibnamefont {Vitelli}},
  \bibinfo {author} {\bibfnamefont {A.~J.}\ \bibnamefont {Liu}},\ and\ \bibinfo
  {author} {\bibfnamefont {S.~R.}\ \bibnamefont {Nagel}},\ }\href
  {https://doi.org/10.1209/0295-5075/90/56001} {\bibfield  {journal} {\bibinfo
  {journal} {Europhysics Letters}\ }\textbf {\bibinfo {volume} {90}},\ \bibinfo
  {pages} {56001} (\bibinfo {year} {2010})}\BibitemShut {NoStop}%
\bibitem [{\citenamefont {Charbonneau}\ \emph {et~al.}(2016)\citenamefont
  {Charbonneau}, \citenamefont {Corwin}, \citenamefont {Parisi}, \citenamefont
  {Poncet},\ and\ \citenamefont {Zamponi}}]{Charbonneau_2016}%
  \BibitemOpen
  \bibfield  {author} {\bibinfo {author} {\bibfnamefont {P.}~\bibnamefont
  {Charbonneau}}, \bibinfo {author} {\bibfnamefont {E.~I.}\ \bibnamefont
  {Corwin}}, \bibinfo {author} {\bibfnamefont {G.}~\bibnamefont {Parisi}},
  \bibinfo {author} {\bibfnamefont {A.}~\bibnamefont {Poncet}},\ and\ \bibinfo
  {author} {\bibfnamefont {F.}~\bibnamefont {Zamponi}},\ }\href
  {https://doi.org/10.1103/PhysRevLett.117.045503} {\bibfield  {journal}
  {\bibinfo  {journal} {Phys. Rev. Lett.}\ }\textbf {\bibinfo {volume} {117}},\
  \bibinfo {pages} {045503} (\bibinfo {year} {2016})}\BibitemShut {NoStop}%
\bibitem [{\citenamefont {Mizuno}\ \emph {et~al.}(2017)\citenamefont {Mizuno},
  \citenamefont {Shiba},\ and\ \citenamefont {Ikeda}}]{Mizuno_2017}%
  \BibitemOpen
  \bibfield  {author} {\bibinfo {author} {\bibfnamefont {H.}~\bibnamefont
  {Mizuno}}, \bibinfo {author} {\bibfnamefont {H.}~\bibnamefont {Shiba}},\ and\
  \bibinfo {author} {\bibfnamefont {A.}~\bibnamefont {Ikeda}},\ }\href
  {https://doi.org/10.1073/pnas.1709015114} {\bibfield  {journal} {\bibinfo
  {journal} {Proceedings of the National Academy of Sciences}\ }\textbf
  {\bibinfo {volume} {114}},\ \bibinfo {pages} {E9767} (\bibinfo {year}
  {2017})}\BibitemShut {NoStop}%
\bibitem [{\citenamefont {Mizuno}\ and\ \citenamefont
  {Ikeda}(2018)}]{Mizuno_2018}%
  \BibitemOpen
  \bibfield  {author} {\bibinfo {author} {\bibfnamefont {H.}~\bibnamefont
  {Mizuno}}\ and\ \bibinfo {author} {\bibfnamefont {A.}~\bibnamefont {Ikeda}},\
  }\href {https://doi.org/10.1103/PhysRevE.98.062612} {\bibfield  {journal}
  {\bibinfo  {journal} {Phys. Rev. E}\ }\textbf {\bibinfo {volume} {98}},\
  \bibinfo {pages} {062612} (\bibinfo {year} {2018})}\BibitemShut {NoStop}%
\bibitem [{\citenamefont {Xu}\ \emph {et~al.}(2007)\citenamefont {Xu},
  \citenamefont {Wyart}, \citenamefont {Liu},\ and\ \citenamefont
  {Nagel}}]{Xu_2007}%
  \BibitemOpen
  \bibfield  {author} {\bibinfo {author} {\bibfnamefont {N.}~\bibnamefont
  {Xu}}, \bibinfo {author} {\bibfnamefont {M.}~\bibnamefont {Wyart}}, \bibinfo
  {author} {\bibfnamefont {A.~J.}\ \bibnamefont {Liu}},\ and\ \bibinfo {author}
  {\bibfnamefont {S.~R.}\ \bibnamefont {Nagel}},\ }\href
  {https://doi.org/10.1103/PhysRevLett.98.175502} {\bibfield  {journal}
  {\bibinfo  {journal} {Phys. Rev. Lett.}\ }\textbf {\bibinfo {volume} {98}},\
  \bibinfo {pages} {175502} (\bibinfo {year} {2007})}\BibitemShut {NoStop}%
\bibitem [{\citenamefont {Shimada}\ \emph
  {et~al.}(2018{\natexlab{a}})\citenamefont {Shimada}, \citenamefont {Mizuno},\
  and\ \citenamefont {Ikeda}}]{Shimada_2018}%
  \BibitemOpen
  \bibfield  {author} {\bibinfo {author} {\bibfnamefont {M.}~\bibnamefont
  {Shimada}}, \bibinfo {author} {\bibfnamefont {H.}~\bibnamefont {Mizuno}},\
  and\ \bibinfo {author} {\bibfnamefont {A.}~\bibnamefont {Ikeda}},\ }\href
  {https://doi.org/10.1103/PhysRevE.97.022609} {\bibfield  {journal} {\bibinfo
  {journal} {Phys. Rev. E}\ }\textbf {\bibinfo {volume} {97}},\ \bibinfo
  {pages} {022609} (\bibinfo {year} {2018}{\natexlab{a}})}\BibitemShut
  {NoStop}%
\bibitem [{\citenamefont {Wang}\ \emph {et~al.}(2019)\citenamefont {Wang},
  \citenamefont {Ninarello}, \citenamefont {Guan}, \citenamefont {Berthier},
  \citenamefont {Szamel},\ and\ \citenamefont {Flenner}}]{Wang_2019}%
  \BibitemOpen
  \bibfield  {author} {\bibinfo {author} {\bibfnamefont {L.}~\bibnamefont
  {Wang}}, \bibinfo {author} {\bibfnamefont {A.}~\bibnamefont {Ninarello}},
  \bibinfo {author} {\bibfnamefont {P.}~\bibnamefont {Guan}}, \bibinfo {author}
  {\bibfnamefont {L.}~\bibnamefont {Berthier}}, \bibinfo {author}
  {\bibfnamefont {G.}~\bibnamefont {Szamel}},\ and\ \bibinfo {author}
  {\bibfnamefont {E.}~\bibnamefont {Flenner}},\ }\href@noop {} {\bibfield
  {journal} {\bibinfo  {journal} {Nature Communications}\ }\textbf {\bibinfo
  {volume} {10}},\ \bibinfo {pages} {26} (\bibinfo {year} {2019})}\BibitemShut
  {NoStop}%
\bibitem [{\citenamefont {Brito}\ and\ \citenamefont
  {Wyart}(2006)}]{Brito_2006}%
  \BibitemOpen
  \bibfield  {author} {\bibinfo {author} {\bibfnamefont {C.}~\bibnamefont
  {Brito}}\ and\ \bibinfo {author} {\bibfnamefont {M.}~\bibnamefont {Wyart}},\
  }\href {https://doi.org/10.1209/epl/i2006-10238-x} {\bibfield  {journal}
  {\bibinfo  {journal} {Europhysics Letters}\ }\textbf {\bibinfo {volume}
  {76}},\ \bibinfo {pages} {149} (\bibinfo {year} {2006})}\BibitemShut
  {NoStop}%
\bibitem [{\citenamefont {Brito}\ and\ \citenamefont
  {Wyart}(2009)}]{Brito_2009}%
  \BibitemOpen
  \bibfield  {author} {\bibinfo {author} {\bibfnamefont {C.}~\bibnamefont
  {Brito}}\ and\ \bibinfo {author} {\bibfnamefont {M.}~\bibnamefont {Wyart}},\
  }\href {https://doi.org/10.1063/1.3157261} {\bibfield  {journal} {\bibinfo
  {journal} {The Journal of Chemical Physics}\ }\textbf {\bibinfo {volume}
  {131}},\ \bibinfo {pages} {024504} (\bibinfo {year} {2009})}\BibitemShut
  {NoStop}%
\bibitem [{\citenamefont {Lerner}\ and\ \citenamefont
  {Bouchbinder}(2018)}]{Lerner_2017}%
  \BibitemOpen
  \bibfield  {author} {\bibinfo {author} {\bibfnamefont {E.}~\bibnamefont
  {Lerner}}\ and\ \bibinfo {author} {\bibfnamefont {E.}~\bibnamefont
  {Bouchbinder}},\ }\href {https://doi.org/10.1103/PhysRevE.97.032140}
  {\bibfield  {journal} {\bibinfo  {journal} {Phys. Rev. E}\ }\textbf {\bibinfo
  {volume} {97}},\ \bibinfo {pages} {032140} (\bibinfo {year}
  {2018})}\BibitemShut {NoStop}%
\bibitem [{\citenamefont {Shimada}\ \emph
  {et~al.}(2018{\natexlab{b}})\citenamefont {Shimada}, \citenamefont {Mizuno},
  \citenamefont {Wyart},\ and\ \citenamefont {Ikeda}}]{Shimada2_2018}%
  \BibitemOpen
  \bibfield  {author} {\bibinfo {author} {\bibfnamefont {M.}~\bibnamefont
  {Shimada}}, \bibinfo {author} {\bibfnamefont {H.}~\bibnamefont {Mizuno}},
  \bibinfo {author} {\bibfnamefont {M.}~\bibnamefont {Wyart}},\ and\ \bibinfo
  {author} {\bibfnamefont {A.}~\bibnamefont {Ikeda}},\ }\href
  {https://doi.org/10.1103/PhysRevE.98.060901} {\bibfield  {journal} {\bibinfo
  {journal} {Phys. Rev. E}\ }\textbf {\bibinfo {volume} {98}},\ \bibinfo
  {pages} {060901} (\bibinfo {year} {2018}{\natexlab{b}})}\BibitemShut
  {NoStop}%
\bibitem [{\citenamefont {Taraskin}\ and\ \citenamefont
  {Elliott}(1997)}]{Taraskin_1997}%
  \BibitemOpen
  \bibfield  {author} {\bibinfo {author} {\bibfnamefont {S.~N.}\ \bibnamefont
  {Taraskin}}\ and\ \bibinfo {author} {\bibfnamefont {S.~R.}\ \bibnamefont
  {Elliott}},\ }\href {https://doi.org/10.1103/PhysRevB.56.8605} {\bibfield
  {journal} {\bibinfo  {journal} {Phys. Rev. B}\ }\textbf {\bibinfo {volume}
  {56}},\ \bibinfo {pages} {8605} (\bibinfo {year} {1997})}\BibitemShut
  {NoStop}%
\bibitem [{\citenamefont {Richard}\ \emph {et~al.}(2020)\citenamefont
  {Richard}, \citenamefont {Gonz\'alez-L\'opez}, \citenamefont {Kapteijns},
  \citenamefont {Pater}, \citenamefont {Vaknin}, \citenamefont {Bouchbinder},\
  and\ \citenamefont {Lerner}}]{Richard_2020}%
  \BibitemOpen
  \bibfield  {author} {\bibinfo {author} {\bibfnamefont {D.}~\bibnamefont
  {Richard}}, \bibinfo {author} {\bibfnamefont {K.}~\bibnamefont
  {Gonz\'alez-L\'opez}}, \bibinfo {author} {\bibfnamefont {G.}~\bibnamefont
  {Kapteijns}}, \bibinfo {author} {\bibfnamefont {R.}~\bibnamefont {Pater}},
  \bibinfo {author} {\bibfnamefont {T.}~\bibnamefont {Vaknin}}, \bibinfo
  {author} {\bibfnamefont {E.}~\bibnamefont {Bouchbinder}},\ and\ \bibinfo
  {author} {\bibfnamefont {E.}~\bibnamefont {Lerner}},\ }\href
  {https://doi.org/10.1103/PhysRevLett.125.085502} {\bibfield  {journal}
  {\bibinfo  {journal} {Phys. Rev. Lett.}\ }\textbf {\bibinfo {volume} {125}},\
  \bibinfo {pages} {085502} (\bibinfo {year} {2020})}\BibitemShut {NoStop}%
\bibitem [{\citenamefont {Shcheblanov}\ \emph {et~al.}(2021)\citenamefont
  {Shcheblanov}, \citenamefont {Povarnitsyn}, \citenamefont {Wiles},
  \citenamefont {Elliott},\ and\ \citenamefont {Taraskin}}]{Shcheblanov_2021}%
  \BibitemOpen
  \bibfield  {author} {\bibinfo {author} {\bibfnamefont {N.~S.}\ \bibnamefont
  {Shcheblanov}}, \bibinfo {author} {\bibfnamefont {M.~E.}\ \bibnamefont
  {Povarnitsyn}}, \bibinfo {author} {\bibfnamefont {J.~D.}\ \bibnamefont
  {Wiles}}, \bibinfo {author} {\bibfnamefont {S.~R.}\ \bibnamefont {Elliott}},\
  and\ \bibinfo {author} {\bibfnamefont {S.~N.}\ \bibnamefont {Taraskin}},\
  }\href {https://doi.org/https://doi.org/10.1002/pssb.202000422} {\bibfield
  {journal} {\bibinfo  {journal} {physica status solidi (b)}\ }\textbf
  {\bibinfo {volume} {258}},\ \bibinfo {pages} {2000422} (\bibinfo {year}
  {2021})}\BibitemShut {NoStop}%
\bibitem [{\citenamefont {Guerra}\ \emph {et~al.}(2022)\citenamefont {Guerra},
  \citenamefont {Bonfanti}, \citenamefont {Procaccia},\ and\ \citenamefont
  {Zapperi}}]{Guerra_2022}%
  \BibitemOpen
  \bibfield  {author} {\bibinfo {author} {\bibfnamefont {R.}~\bibnamefont
  {Guerra}}, \bibinfo {author} {\bibfnamefont {S.}~\bibnamefont {Bonfanti}},
  \bibinfo {author} {\bibfnamefont {I.}~\bibnamefont {Procaccia}},\ and\
  \bibinfo {author} {\bibfnamefont {S.}~\bibnamefont {Zapperi}},\ }\href
  {https://doi.org/10.1103/PhysRevE.105.054104} {\bibfield  {journal} {\bibinfo
   {journal} {Phys. Rev. E}\ }\textbf {\bibinfo {volume} {105}},\ \bibinfo
  {pages} {054104} (\bibinfo {year} {2022})}\BibitemShut {NoStop}%
\bibitem [{\citenamefont {van Beest}\ \emph {et~al.}(1990)\citenamefont {van
  Beest}, \citenamefont {Kramer},\ and\ \citenamefont {van
  Santen}}]{Beest_1990}%
  \BibitemOpen
  \bibfield  {author} {\bibinfo {author} {\bibfnamefont {B.~W.~H.}\
  \bibnamefont {van Beest}}, \bibinfo {author} {\bibfnamefont {G.~J.}\
  \bibnamefont {Kramer}},\ and\ \bibinfo {author} {\bibfnamefont {R.~A.}\
  \bibnamefont {van Santen}},\ }\href
  {https://doi.org/10.1103/PhysRevLett.64.1955} {\bibfield  {journal} {\bibinfo
   {journal} {Phys. Rev. Lett.}\ }\textbf {\bibinfo {volume} {64}},\ \bibinfo
  {pages} {1955} (\bibinfo {year} {1990})}\BibitemShut {NoStop}%
\bibitem [{\citenamefont {Wolf}(1992)}]{Wolf_1992}%
  \BibitemOpen
  \bibfield  {author} {\bibinfo {author} {\bibfnamefont {D.}~\bibnamefont
  {Wolf}},\ }\href {https://doi.org/10.1103/PhysRevLett.68.3315} {\bibfield
  {journal} {\bibinfo  {journal} {Phys. Rev. Lett.}\ }\textbf {\bibinfo
  {volume} {68}},\ \bibinfo {pages} {3315} (\bibinfo {year}
  {1992})}\BibitemShut {NoStop}%
\bibitem [{\citenamefont {Wolf}\ \emph {et~al.}(1999)\citenamefont {Wolf},
  \citenamefont {Keblinski}, \citenamefont {Phillpot},\ and\ \citenamefont
  {Eggebrecht}}]{Wolf_1999}%
  \BibitemOpen
  \bibfield  {author} {\bibinfo {author} {\bibfnamefont {D.}~\bibnamefont
  {Wolf}}, \bibinfo {author} {\bibfnamefont {P.}~\bibnamefont {Keblinski}},
  \bibinfo {author} {\bibfnamefont {S.~R.}\ \bibnamefont {Phillpot}},\ and\
  \bibinfo {author} {\bibfnamefont {J.}~\bibnamefont {Eggebrecht}},\ }\href
  {https://doi.org/10.1063/1.478738} {\bibfield  {journal} {\bibinfo  {journal}
  {The Journal of Chemical Physics}\ }\textbf {\bibinfo {volume} {110}},\
  \bibinfo {pages} {8254} (\bibinfo {year} {1999})}\BibitemShut {NoStop}%
\bibitem [{\citenamefont {Carr^^c3^^a9}\ \emph {et~al.}(2007)\citenamefont
  {Carr^^c3^^a9}, \citenamefont {Berthier}, \citenamefont {Horbach},
  \citenamefont {Ispas},\ and\ \citenamefont {Kob}}]{Carre_2007}%
  \BibitemOpen
  \bibfield  {author} {\bibinfo {author} {\bibfnamefont {A.}~\bibnamefont
  {Carr^^c3^^a9}}, \bibinfo {author} {\bibfnamefont {L.}~\bibnamefont
  {Berthier}}, \bibinfo {author} {\bibfnamefont {J.}~\bibnamefont {Horbach}},
  \bibinfo {author} {\bibfnamefont {S.}~\bibnamefont {Ispas}},\ and\ \bibinfo
  {author} {\bibfnamefont {W.}~\bibnamefont {Kob}},\ }\href
  {https://doi.org/10.1063/1.2777136} {\bibfield  {journal} {\bibinfo
  {journal} {The Journal of Chemical Physics}\ }\textbf {\bibinfo {volume}
  {127}},\ \bibinfo {pages} {114512} (\bibinfo {year} {2007})}\BibitemShut
  {NoStop}%
\bibitem [{\citenamefont {Carr^^c3^^a9}\ \emph {et~al.}(2016)\citenamefont
  {Carr^^c3^^a9}, \citenamefont {Ispas}, \citenamefont {Horbach},\ and\
  \citenamefont {Kob}}]{Carre_2016}%
  \BibitemOpen
  \bibfield  {author} {\bibinfo {author} {\bibfnamefont {A.}~\bibnamefont
  {Carr^^c3^^a9}}, \bibinfo {author} {\bibfnamefont {S.}~\bibnamefont {Ispas}},
  \bibinfo {author} {\bibfnamefont {J.}~\bibnamefont {Horbach}},\ and\ \bibinfo
  {author} {\bibfnamefont {W.}~\bibnamefont {Kob}},\ }\href
  {https://doi.org/https://doi.org/10.1016/j.commatsci.2016.07.041} {\bibfield
  {journal} {\bibinfo  {journal} {Computational Materials Science}\ }\textbf
  {\bibinfo {volume} {124}},\ \bibinfo {pages} {323} (\bibinfo {year}
  {2016})}\BibitemShut {NoStop}%
\bibitem [{\citenamefont {Sundararaman}\ \emph {et~al.}(2018)\citenamefont
  {Sundararaman}, \citenamefont {Huang}, \citenamefont {Ispas},\ and\
  \citenamefont {Kob}}]{Sundararaman_2018}%
  \BibitemOpen
  \bibfield  {author} {\bibinfo {author} {\bibfnamefont {S.}~\bibnamefont
  {Sundararaman}}, \bibinfo {author} {\bibfnamefont {L.}~\bibnamefont {Huang}},
  \bibinfo {author} {\bibfnamefont {S.}~\bibnamefont {Ispas}},\ and\ \bibinfo
  {author} {\bibfnamefont {W.}~\bibnamefont {Kob}},\ }\href
  {https://doi.org/10.1063/1.5023707} {\bibfield  {journal} {\bibinfo
  {journal} {The Journal of Chemical Physics}\ }\textbf {\bibinfo {volume}
  {148}},\ \bibinfo {pages} {194504} (\bibinfo {year} {2018})}\BibitemShut
  {NoStop}%
\bibitem [{\citenamefont {Vashishta}\ \emph {et~al.}(1990)\citenamefont
  {Vashishta}, \citenamefont {Kalia}, \citenamefont {Rino},\ and\ \citenamefont
  {Ebbsj{\"o}}}]{Vashishta_1990}%
  \BibitemOpen
  \bibfield  {author} {\bibinfo {author} {\bibfnamefont {P.}~\bibnamefont
  {Vashishta}}, \bibinfo {author} {\bibfnamefont {R.~K.}\ \bibnamefont
  {Kalia}}, \bibinfo {author} {\bibfnamefont {J.~P.}\ \bibnamefont {Rino}},\
  and\ \bibinfo {author} {\bibfnamefont {I.}~\bibnamefont {Ebbsj{\"o}}},\
  }\href {https://doi.org/10.1103/PhysRevB.41.12197} {\bibfield  {journal}
  {\bibinfo  {journal} {Physical Review B}\ }\textbf {\bibinfo {volume} {41}},\
  \bibinfo {pages} {12197} (\bibinfo {year} {1990})}\BibitemShut {NoStop}%
\bibitem [{\citenamefont {Mantisi}\ \emph {et~al.}(2012)\citenamefont
  {Mantisi}, \citenamefont {Tanguy}, \citenamefont {Kermouche},\ and\
  \citenamefont {Barthel}}]{Mantisi_2012}%
  \BibitemOpen
  \bibfield  {author} {\bibinfo {author} {\bibfnamefont {B.}~\bibnamefont
  {Mantisi}}, \bibinfo {author} {\bibfnamefont {A.}~\bibnamefont {Tanguy}},
  \bibinfo {author} {\bibfnamefont {G.}~\bibnamefont {Kermouche}},\ and\
  \bibinfo {author} {\bibfnamefont {E.}~\bibnamefont {Barthel}},\ }\href
  {https://doi.org/10.1140/epjb/e2012-30317-6} {\bibfield  {journal} {\bibinfo
  {journal} {The European Physical Journal B}\ }\textbf {\bibinfo {volume}
  {85}},\ \bibinfo {pages} {304} (\bibinfo {year} {2012})}\BibitemShut
  {NoStop}%
\bibitem [{\citenamefont {Shcheblanov}\ \emph {et~al.}(2015)\citenamefont
  {Shcheblanov}, \citenamefont {Mantisi}, \citenamefont {Umari},\ and\
  \citenamefont {Tanguy}}]{Shcheblanov_2015}%
  \BibitemOpen
  \bibfield  {author} {\bibinfo {author} {\bibfnamefont {N.~S.}\ \bibnamefont
  {Shcheblanov}}, \bibinfo {author} {\bibfnamefont {B.}~\bibnamefont
  {Mantisi}}, \bibinfo {author} {\bibfnamefont {P.}~\bibnamefont {Umari}},\
  and\ \bibinfo {author} {\bibfnamefont {A.}~\bibnamefont {Tanguy}},\ }\href
  {https://doi.org/10.1016/j.jnoncrysol.2015.07.035} {\bibfield  {journal}
  {\bibinfo  {journal} {Journal of Non-Crystalline Solids}\ }\textbf {\bibinfo
  {volume} {428}},\ \bibinfo {pages} {6} (\bibinfo {year} {2015})}\BibitemShut
  {NoStop}%
\bibitem [{\citenamefont {Damart}\ \emph {et~al.}(2017)\citenamefont {Damart},
  \citenamefont {Tanguy},\ and\ \citenamefont {Rodney}}]{Damart_2017}%
  \BibitemOpen
  \bibfield  {author} {\bibinfo {author} {\bibfnamefont {T.}~\bibnamefont
  {Damart}}, \bibinfo {author} {\bibfnamefont {A.}~\bibnamefont {Tanguy}},\
  and\ \bibinfo {author} {\bibfnamefont {D.}~\bibnamefont {Rodney}},\ }\href
  {https://doi.org/10.1103/PhysRevB.95.054203} {\bibfield  {journal} {\bibinfo
  {journal} {Phys. Rev. B}\ }\textbf {\bibinfo {volume} {95}},\ \bibinfo
  {pages} {054203} (\bibinfo {year} {2017})}\BibitemShut {NoStop}%
\bibitem [{\citenamefont {El~Hamdaoui}\ \emph {et~al.}(2025)\citenamefont
  {El~Hamdaoui}, \citenamefont {Ghardi}, \citenamefont {Rushton}, \citenamefont
  {Hasnaoui},\ and\ \citenamefont {Ouaskit}}]{Hamdaoui_2025}%
  \BibitemOpen
  \bibfield  {author} {\bibinfo {author} {\bibfnamefont {A.}~\bibnamefont
  {El~Hamdaoui}}, \bibinfo {author} {\bibfnamefont {E.~M.}\ \bibnamefont
  {Ghardi}}, \bibinfo {author} {\bibfnamefont {M.~J.~D.}\ \bibnamefont
  {Rushton}}, \bibinfo {author} {\bibfnamefont {A.}~\bibnamefont {Hasnaoui}},\
  and\ \bibinfo {author} {\bibfnamefont {S.}~\bibnamefont {Ouaskit}},\ }\href
  {https://doi.org/https://doi.org/10.1111/jace.20631} {\bibfield  {journal}
  {\bibinfo  {journal} {Journal of the American Ceramic Society}\ }\textbf
  {\bibinfo {volume} {108}},\ \bibinfo {pages} {e20631} (\bibinfo {year}
  {2025})}\BibitemShut {NoStop}%
\bibitem [{\citenamefont {Plimpton}(1995)}]{Plimpton_1995}%
  \BibitemOpen
  \bibfield  {author} {\bibinfo {author} {\bibfnamefont {S.}~\bibnamefont
  {Plimpton}},\ }\href {https://doi.org/10.1006/jcph.1995.1039} {\bibfield
  {journal} {\bibinfo  {journal} {Journal of Computational Physics}\ }\textbf
  {\bibinfo {volume} {117}},\ \bibinfo {pages} {1} (\bibinfo {year}
  {1995})}\BibitemShut {NoStop}%
\bibitem [{\citenamefont {Nos{\'e}}(1984)}]{Nose_1984}%
  \BibitemOpen
  \bibfield  {author} {\bibinfo {author} {\bibfnamefont {S.}~\bibnamefont
  {Nos{\'e}}},\ }\href {https://doi.org/10.1063/1.447334} {\bibfield  {journal}
  {\bibinfo  {journal} {The Journal of Chemical Physics}\ }\textbf {\bibinfo
  {volume} {81}},\ \bibinfo {pages} {511} (\bibinfo {year} {1984})}\BibitemShut
  {NoStop}%
\bibitem [{\citenamefont {Hoover}(1985)}]{Hoover_1985}%
  \BibitemOpen
  \bibfield  {author} {\bibinfo {author} {\bibfnamefont {W.~G.}\ \bibnamefont
  {Hoover}},\ }\href {https://doi.org/10.1103/PhysRevA.31.1695} {\bibfield
  {journal} {\bibinfo  {journal} {Physical Review A}\ }\textbf {\bibinfo
  {volume} {31}},\ \bibinfo {pages} {1695} (\bibinfo {year}
  {1985})}\BibitemShut {NoStop}%
\bibitem [{\citenamefont {Mizuno}\ \emph
  {et~al.}(2013{\natexlab{b}})\citenamefont {Mizuno}, \citenamefont {Mossa},\
  and\ \citenamefont {Barrat}}]{Mizuno_2013}%
  \BibitemOpen
  \bibfield  {author} {\bibinfo {author} {\bibfnamefont {H.}~\bibnamefont
  {Mizuno}}, \bibinfo {author} {\bibfnamefont {S.}~\bibnamefont {Mossa}},\ and\
  \bibinfo {author} {\bibfnamefont {J.-L.}\ \bibnamefont {Barrat}},\ }\href
  {https://doi.org/10.1103/PhysRevE.87.042306} {\bibfield  {journal} {\bibinfo
  {journal} {Phys. Rev. E}\ }\textbf {\bibinfo {volume} {87}},\ \bibinfo
  {pages} {042306} (\bibinfo {year} {2013}{\natexlab{b}})}\BibitemShut
  {NoStop}%
\bibitem [{\citenamefont {Mizuno}\ \emph
  {et~al.}(2016{\natexlab{a}})\citenamefont {Mizuno}, \citenamefont {Silbert},
  \citenamefont {Sperl}, \citenamefont {Mossa},\ and\ \citenamefont
  {Barrat}}]{Mizuno2_2016}%
  \BibitemOpen
  \bibfield  {author} {\bibinfo {author} {\bibfnamefont {H.}~\bibnamefont
  {Mizuno}}, \bibinfo {author} {\bibfnamefont {L.~E.}\ \bibnamefont {Silbert}},
  \bibinfo {author} {\bibfnamefont {M.}~\bibnamefont {Sperl}}, \bibinfo
  {author} {\bibfnamefont {S.}~\bibnamefont {Mossa}},\ and\ \bibinfo {author}
  {\bibfnamefont {J.-L.}\ \bibnamefont {Barrat}},\ }\href
  {https://doi.org/10.1103/PhysRevE.93.043314} {\bibfield  {journal} {\bibinfo
  {journal} {Phys. Rev. E}\ }\textbf {\bibinfo {volume} {93}},\ \bibinfo
  {pages} {043314} (\bibinfo {year} {2016}{\natexlab{a}})}\BibitemShut
  {NoStop}%
\bibitem [{\citenamefont {Mizuno}\ and\ \citenamefont
  {Ikeda}(2022)}]{MizunoIkeda2022}%
  \BibitemOpen
  \bibfield  {author} {\bibinfo {author} {\bibfnamefont {H.}~\bibnamefont
  {Mizuno}}\ and\ \bibinfo {author} {\bibfnamefont {A.}~\bibnamefont {Ikeda}},\
  }\bibinfo {title} {{Computational Simulations of the Vibrational Properties
  of Glasses}},\ in\ \href {https://doi.org/10.1142/9781800612587_0010} {\emph
  {\bibinfo {booktitle} {{Low-Temperature Thermal and Vibrational Properties of
  Disordered Solids}}}},\ \bibinfo {editor} {edited by\ \bibinfo {editor}
  {\bibfnamefont {M.~A.}\ \bibnamefont {Ramos}}}\ (\bibinfo  {publisher}
  {{WORLD SCIENTIFIC (EUROPE)}},\ \bibinfo {year} {2022})\ Chap.~\bibinfo
  {chapter} {10}, pp.\ \bibinfo {pages} {375--433}\BibitemShut {NoStop}%
\bibitem [{\citenamefont {Grest}\ \emph {et~al.}(1982)\citenamefont {Grest},
  \citenamefont {Nagel},\ and\ \citenamefont {Rahman}}]{Grest_1982}%
  \BibitemOpen
  \bibfield  {author} {\bibinfo {author} {\bibfnamefont {G.~S.}\ \bibnamefont
  {Grest}}, \bibinfo {author} {\bibfnamefont {S.~R.}\ \bibnamefont {Nagel}},\
  and\ \bibinfo {author} {\bibfnamefont {A.}~\bibnamefont {Rahman}},\ }\href
  {https://doi.org/10.1103/PhysRevLett.49.1271} {\bibfield  {journal} {\bibinfo
   {journal} {Phys. Rev. Lett.}\ }\textbf {\bibinfo {volume} {49}},\ \bibinfo
  {pages} {1271} (\bibinfo {year} {1982})}\BibitemShut {NoStop}%
\bibitem [{\citenamefont {Suck}\ \emph {et~al.}(1992)\citenamefont {Suck},
  \citenamefont {Egelstaff}, \citenamefont {Robinson}, \citenamefont {Sivia},\
  and\ \citenamefont {Taylor}}]{Suck_1992}%
  \BibitemOpen
  \bibfield  {author} {\bibinfo {author} {\bibfnamefont {J.-B.}\ \bibnamefont
  {Suck}}, \bibinfo {author} {\bibfnamefont {P.~A.}\ \bibnamefont {Egelstaff}},
  \bibinfo {author} {\bibfnamefont {R.~A.}\ \bibnamefont {Robinson}}, \bibinfo
  {author} {\bibfnamefont {D.~S.}\ \bibnamefont {Sivia}},\ and\ \bibinfo
  {author} {\bibfnamefont {A.~D.}\ \bibnamefont {Taylor}},\ }\href
  {https://doi.org/10.1209/0295-5075/19/3/010} {\bibfield  {journal} {\bibinfo
  {journal} {Europhysics Letters}\ }\textbf {\bibinfo {volume} {19}},\ \bibinfo
  {pages} {207} (\bibinfo {year} {1992})}\BibitemShut {NoStop}%
\bibitem [{\citenamefont {Lemaitre}\ and\ \citenamefont
  {Maloney}(2006)}]{Lemaitre_2006}%
  \BibitemOpen
  \bibfield  {author} {\bibinfo {author} {\bibfnamefont {A.}~\bibnamefont
  {Lemaitre}}\ and\ \bibinfo {author} {\bibfnamefont {C.}~\bibnamefont
  {Maloney}},\ }\href {https://doi.org/10.1007/s10955-005-9015-5} {\bibfield
  {journal} {\bibinfo  {journal} {Journal of Statistical Physics}\ }\textbf
  {\bibinfo {volume} {123}},\ \bibinfo {pages} {415} (\bibinfo {year}
  {2006})}\BibitemShut {NoStop}%
\bibitem [{\citenamefont {Mizuno}\ \emph
  {et~al.}(2016{\natexlab{b}})\citenamefont {Mizuno}, \citenamefont {Saitoh},\
  and\ \citenamefont {Silbert}}]{Mizuno3_2016}%
  \BibitemOpen
  \bibfield  {author} {\bibinfo {author} {\bibfnamefont {H.}~\bibnamefont
  {Mizuno}}, \bibinfo {author} {\bibfnamefont {K.}~\bibnamefont {Saitoh}},\
  and\ \bibinfo {author} {\bibfnamefont {L.~E.}\ \bibnamefont {Silbert}},\
  }\href {https://doi.org/10.1103/PhysRevE.93.062905} {\bibfield  {journal}
  {\bibinfo  {journal} {Phys. Rev. E}\ }\textbf {\bibinfo {volume} {93}},\
  \bibinfo {pages} {062905} (\bibinfo {year} {2016}{\natexlab{b}})}\BibitemShut
  {NoStop}%
\bibitem [{\citenamefont {Zaccone}\ and\ \citenamefont
  {Scossa-Romano}(2011)}]{Zaccone_2011}%
  \BibitemOpen
  \bibfield  {author} {\bibinfo {author} {\bibfnamefont {A.}~\bibnamefont
  {Zaccone}}\ and\ \bibinfo {author} {\bibfnamefont {E.}~\bibnamefont
  {Scossa-Romano}},\ }\href {https://doi.org/10.1103/PhysRevB.83.184205}
  {\bibfield  {journal} {\bibinfo  {journal} {Phys. Rev. B}\ }\textbf {\bibinfo
  {volume} {83}},\ \bibinfo {pages} {184205} (\bibinfo {year}
  {2011})}\BibitemShut {NoStop}%
\bibitem [{\citenamefont {Milkus}\ and\ \citenamefont
  {Zaccone}(2016)}]{Milkus_2016}%
  \BibitemOpen
  \bibfield  {author} {\bibinfo {author} {\bibfnamefont {R.}~\bibnamefont
  {Milkus}}\ and\ \bibinfo {author} {\bibfnamefont {A.}~\bibnamefont
  {Zaccone}},\ }\href {https://doi.org/10.1103/PhysRevB.93.094204} {\bibfield
  {journal} {\bibinfo  {journal} {Phys. Rev. B}\ }\textbf {\bibinfo {volume}
  {93}},\ \bibinfo {pages} {094204} (\bibinfo {year} {2016})}\BibitemShut
  {NoStop}%
\bibitem [{\citenamefont {Krausser}\ \emph {et~al.}(2017)\citenamefont
  {Krausser}, \citenamefont {Milkus},\ and\ \citenamefont
  {Zaccone}}]{Krausser_2017}%
  \BibitemOpen
  \bibfield  {author} {\bibinfo {author} {\bibfnamefont {J.}~\bibnamefont
  {Krausser}}, \bibinfo {author} {\bibfnamefont {R.}~\bibnamefont {Milkus}},\
  and\ \bibinfo {author} {\bibfnamefont {A.}~\bibnamefont {Zaccone}},\ }\href
  {https://doi.org/10.1039/C7SM00843K} {\bibfield  {journal} {\bibinfo
  {journal} {Soft Matter}\ }\textbf {\bibinfo {volume} {13}},\ \bibinfo {pages}
  {6079} (\bibinfo {year} {2017})}\BibitemShut {NoStop}%
\bibitem [{\citenamefont {Minamitani}\ \emph {et~al.}(2025)\citenamefont
  {Minamitani}, \citenamefont {Nakamura}, \citenamefont {Obayashi},\ and\
  \citenamefont {Mizuno}}]{Minamitani_2025}%
  \BibitemOpen
  \bibfield  {author} {\bibinfo {author} {\bibfnamefont {E.}~\bibnamefont
  {Minamitani}}, \bibinfo {author} {\bibfnamefont {T.}~\bibnamefont
  {Nakamura}}, \bibinfo {author} {\bibfnamefont {I.}~\bibnamefont {Obayashi}},\
  and\ \bibinfo {author} {\bibfnamefont {H.}~\bibnamefont {Mizuno}},\
  }\href@noop {} {\bibfield  {journal} {\bibinfo  {journal} {arXiv:2407.17707}\
  } (\bibinfo {year} {2025})}\BibitemShut {NoStop}%
\bibitem [{\citenamefont {Cui}\ \emph {et~al.}(2019)\citenamefont {Cui},
  \citenamefont {Zaccone},\ and\ \citenamefont {Rodney}}]{Cui_2019}%
  \BibitemOpen
  \bibfield  {author} {\bibinfo {author} {\bibfnamefont {B.}~\bibnamefont
  {Cui}}, \bibinfo {author} {\bibfnamefont {A.}~\bibnamefont {Zaccone}},\ and\
  \bibinfo {author} {\bibfnamefont {D.}~\bibnamefont {Rodney}},\ }\href
  {https://doi.org/10.1063/1.5129025} {\bibfield  {journal} {\bibinfo
  {journal} {The Journal of Chemical Physics}\ }\textbf {\bibinfo {volume}
  {151}},\ \bibinfo {pages} {224509} (\bibinfo {year} {2019})}\BibitemShut
  {NoStop}%
\bibitem [{\citenamefont {Greaves}\ \emph {et~al.}(2011)\citenamefont
  {Greaves}, \citenamefont {Greer}, \citenamefont {Lakes},\ and\ \citenamefont
  {Rouxel}}]{Greaves_2011}%
  \BibitemOpen
  \bibfield  {author} {\bibinfo {author} {\bibfnamefont {G.~N.}\ \bibnamefont
  {Greaves}}, \bibinfo {author} {\bibfnamefont {A.~L.}\ \bibnamefont {Greer}},
  \bibinfo {author} {\bibfnamefont {R.~S.}\ \bibnamefont {Lakes}},\ and\
  \bibinfo {author} {\bibfnamefont {T.}~\bibnamefont {Rouxel}},\ }\href@noop {}
  {\bibfield  {journal} {\bibinfo  {journal} {Nature Mater.}\ }\textbf
  {\bibinfo {volume} {10}},\ \bibinfo {pages} {823} (\bibinfo {year}
  {2011})}\BibitemShut {NoStop}%
\bibitem [{\citenamefont {Duval}\ \emph {et~al.}(2013)\citenamefont {Duval},
  \citenamefont {Deschamps},\ and\ \citenamefont {Saviot}}]{Duval_2013}%
  \BibitemOpen
  \bibfield  {author} {\bibinfo {author} {\bibfnamefont {E.}~\bibnamefont
  {Duval}}, \bibinfo {author} {\bibfnamefont {T.}~\bibnamefont {Deschamps}},\
  and\ \bibinfo {author} {\bibfnamefont {L.}~\bibnamefont {Saviot}},\ }\href
  {https://doi.org/10.1063/1.4817778} {\bibfield  {journal} {\bibinfo
  {journal} {The Journal of Chemical Physics}\ }\textbf {\bibinfo {volume}
  {139}},\ \bibinfo {pages} {064506} (\bibinfo {year} {2013})}\BibitemShut
  {NoStop}%
\bibitem [{Note1()}]{Note1}%
  \BibitemOpen
  \bibinfo {note} {Below $\omega _\protect \text {BP}$, we also observe modes
  with small $\protect \mathcal {P}_k$ indicative of localization, as seen in
  panels (f,i). These modes plausibly originate in regions of low local
  connectivity (low coordination number, \protect \textit {i.e.,} few contacts
  or interacting neighbors), although a definitive identification requires
  further study.}\BibitemShut {Stop}%
\bibitem [{Note2()}]{Note2}%
  \BibitemOpen
  \bibinfo {note} {While all of the short-range Si--Si interaction, the
  long-range Coulomb interaction, and the pre-stress are needed for a faithful
  description, the essential ingredient that controls the approach to marginal
  stability is the pre-stress and the associated frustration; the short-range
  Si--Si and long-range Coulomb terms primarily set the magnitude and sign of
  the pre-stress and thereby tune the BP position.}\BibitemShut {Stop}%
\bibitem [{Note3()}]{Note3}%
  \BibitemOpen
  \bibinfo {note} {Although a quartic law $g_{\protect \text {QLV}}(\omega
  )\propto \omega ^{4}$ is often reported, the precise exponent remains under
  discussion~\cite {Schirmacher_2024,Xu_2024}. What matters here is that,
  irrespective of the exact exponent, $g_{\protect \text {QLV}}(\omega )$ is
  gapless and follows a power law as $\omega \to 0$. Such a gapless power-law
  spectrum is a hallmark of marginal stability.}\BibitemShut {Stop}%
\bibitem [{\citenamefont {Hu}\ and\ \citenamefont {Tanaka}(2022)}]{Hu_2022}%
  \BibitemOpen
  \bibfield  {author} {\bibinfo {author} {\bibfnamefont {Y.-C.}\ \bibnamefont
  {Hu}}\ and\ \bibinfo {author} {\bibfnamefont {H.}~\bibnamefont {Tanaka}},\
  }\href {https://doi.org/10.1038/s41567-022-01628-6} {\bibfield  {journal}
  {\bibinfo  {journal} {Nature Physics}\ }\textbf {\bibinfo {volume} {18}},\
  \bibinfo {pages} {669} (\bibinfo {year} {2022})}\BibitemShut {NoStop}%
\bibitem [{\citenamefont {Schirmacher}\ \emph {et~al.}(2024)\citenamefont
  {Schirmacher}, \citenamefont {Paoluzzi}, \citenamefont {Mocanu},
  \citenamefont {Khomenko}, \citenamefont {Szamel}, \citenamefont {Zamponi},\
  and\ \citenamefont {Ruocco}}]{Schirmacher_2024}%
  \BibitemOpen
  \bibfield  {author} {\bibinfo {author} {\bibfnamefont {W.}~\bibnamefont
  {Schirmacher}}, \bibinfo {author} {\bibfnamefont {M.}~\bibnamefont
  {Paoluzzi}}, \bibinfo {author} {\bibfnamefont {F.~C.}\ \bibnamefont
  {Mocanu}}, \bibinfo {author} {\bibfnamefont {D.}~\bibnamefont {Khomenko}},
  \bibinfo {author} {\bibfnamefont {G.}~\bibnamefont {Szamel}}, \bibinfo
  {author} {\bibfnamefont {F.}~\bibnamefont {Zamponi}},\ and\ \bibinfo {author}
  {\bibfnamefont {G.}~\bibnamefont {Ruocco}},\ }\href
  {https://doi.org/10.1038/s41467-024-46981-7} {\bibfield  {journal} {\bibinfo
  {journal} {Nature Communications}\ }\textbf {\bibinfo {volume} {15}},\
  \bibinfo {pages} {3107} (\bibinfo {year} {2024})}\BibitemShut {NoStop}%
\bibitem [{\citenamefont {Xu}\ \emph {et~al.}(2024)\citenamefont {Xu},
  \citenamefont {Zhang}, \citenamefont {Tong}, \citenamefont {Wang},\ and\
  \citenamefont {Xu}}]{Xu_2024}%
  \BibitemOpen
  \bibfield  {author} {\bibinfo {author} {\bibfnamefont {D.}~\bibnamefont
  {Xu}}, \bibinfo {author} {\bibfnamefont {S.}~\bibnamefont {Zhang}}, \bibinfo
  {author} {\bibfnamefont {H.}~\bibnamefont {Tong}}, \bibinfo {author}
  {\bibfnamefont {L.}~\bibnamefont {Wang}},\ and\ \bibinfo {author}
  {\bibfnamefont {N.}~\bibnamefont {Xu}},\ }\href
  {https://doi.org/10.1038/s41467-024-45671-8} {\bibfield  {journal} {\bibinfo
  {journal} {Nature Communications}\ }\textbf {\bibinfo {volume} {15}},\
  \bibinfo {pages} {1424} (\bibinfo {year} {2024})}\BibitemShut {NoStop}%
\end{thebibliography}%

%%%%%%%%%%%%%%%%%%%%%%%%%%%%%%%%%%%%%%%%%%%%%%%%%%%%%%%%%%%%%%%%%%%%%%%%%%%%%%%%%%%%%%%%%%%%%%%%%%%%%%%%%%%%%%%%%%%%%%%%%%%
%%%%%%%%%%%%%%%%%%%%%%%%%%%%%%%%%%%%%%%%%%%%%%%%%%%%%%%%%%%%%%%%%%%%%%%%%%%%%%%%%%%%%%%%%%%%%%%%%%%%%%%%%%%%%%%%%%%%%%%%%%%
%%%%%%%%%%%%%%%%%%%%%%%%%%%%%%%%%%%%%%%%%%%%%%%%%%%%%%%%%%%%%%%%%%%%%%%%%%%%%%%%%%%%%%%%%%%%%%%%%%%%%%%%%%%%%%%%%%%%%%%%%%%
\clearpage
\onecolumngrid
\renewcommand{\thetable}{S\arabic{table}}
\renewcommand{\thefigure}{S\arabic{figure}}
\renewcommand{\theequation}{S\arabic{equation}}
\setcounter{equation}{0}
\setcounter{figure}{0}
\setcounter{table}{0}
%
%%%%%%%%%%%%%%%%%%%%%%%%%%%%%%%%%%%%%%%%%%%%%%%%%%%%%%%%%%%%%%%%%%%%%%%%%%%%%%%%%%%%%%%%%%%%%%%%%%%%%%%%%%%%%%%%%%%%%%%%%
\noindent
{\Large \bf Supplementary Information for:}

\vspace*{3mm}
\noindent
{\large \bf Boson peak in covalent network glasses: Isostaticity and marginal stability}

\vspace*{3mm}
\noindent
{by \large Hideyuki Mizuno, Tatsuya Mori, Giacomo Baldi, and Emi Minamitani}

\vspace*{3mm}
\noindent
Email: hideyuki.mizuno@phys.c.u-tokyo.ac.jp

%%%%%%%%%%%%%%%%%%%%%%%%%%%%%%%%%%%%%%%%%%%%%%%%%%%%%%%%%%%%%%%%%%%%%%%%%%%%%%%%%%%%%%%%%%%%%%%%%%%%%%%%%%%%%%%%%%
\vspace*{1cm}
\section*{Supplementary data}
In the following, we report supplementary data, including vibrational states in HSL and SS glasses~(Fig.~S1), dynamical structure factors in the isostatic-network system of silica glass~(Fig.~S2), dynamical structure factors in HSL and SS glasses~(Fig.~S3), and physical quantities including elastic moduli and Debye values~(Table~S1).

%%%%%%%%%%%%%%%%%%%%%%%%%%%%%%%%%%%%%%%%%%%%%%%%%%%%%%%%%%%%%%%%%%%%%%%%%%%%%%%%%%%%%%%%%%%%%%%%%%%%%%%%%%%%%%%%%%
\newpage

%%%%%%%%%%%%%%%%%%%%%%%%%%%%%%%%%%%%%%%%%%%%%%%%%%%%%%%
\begin{figure*}[t]
\centering
\includegraphics[width=0.8\textwidth]{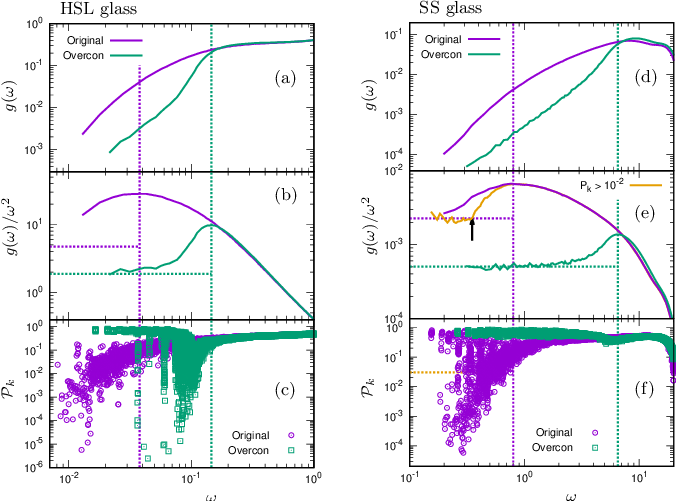}
\vspace*{0mm}
\caption{\label{fig2_total2}
{Vibrational states in the low-frequency regime in HSL and SS glasses.}
(a-c) HSL glass and (d-f) SS glass.
$g(\omega)$, $g(\omega)/\omega^2$, and $\mathcal{P}_k$ are plotted as functions of frequency $\omega$ for the original system (purple) and the overconstrained-network system (green).
The vertical lines mark the BP frequency $\omega_\text{BP}$ for the original and overconstrained-network systems, while the horizontal dotted lines in (b,e) indicate the Debye level $A_D$.
In addition, panel (e) for SS glass shows, in orange, the vDOS $g_\text{EXT}(\omega)$ of extended modes with $\mathcal{P}_k>\mathcal{P}_\text{th}=3\times10^{-2}$.
This threshold $\mathcal{P}_\text{th}$ is indicated by the horizontal dotted line in (f).
The arrow in (e) marks the continuum-limit frequency $\omega_0$ at which $g_\text{EXT}(\omega)/\omega^2$ converges to $A_D$.
}
\end{figure*}
%%%%%%%%%%%%%%%%%%%%%%%%%%%%%%%%%%%%%%%%%%%%%%%%%%%%%%%

%%%%%%%%%%%%%%%%%%%%%%%%%%%%%%%%%%%%%%%%%%%%%%%%%%%%%%%
\begin{figure*}[t]
\centering
\includegraphics[width=0.8\textwidth]{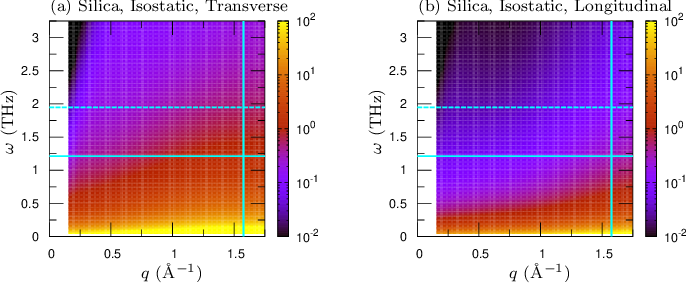}
\vspace*{0mm}
\caption{\label{fig4_shiku2_debyey}
{Dynamical structure factors in the isostatic-network system of silica glass.}
Panels (a) and (b) show the transverse $S_T(q,\omega)/(k_B T)$ and longitudinal $S_L(q,\omega)/(k_B T)$, respectively, as functions of $q$ and $\omega$.
Values are plotted in units of eV THz$^{-1}$.
The vertical line marks the Debye wavenumber $q_D$.
Horizontal solid and dotted lines indicate, for reference, the BP frequency $\omega_\text{BP}$ of the original system and the overconstrained-network system, respectively.
Note that $\omega_\text{BP}\to 0$ in the isostatic-network system.
}
\end{figure*}
%%%%%%%%%%%%%%%%%%%%%%%%%%%%%%%%%%%%%%%%%%%%%%%%%%%%%%%

%%%%%%%%%%%%%%%%%%%%%%%%%%%%%%%%%%%%%%%%%%%%%%%%%%%%%%%
\begin{figure*}[t]
\centering
\includegraphics[width=0.8\textwidth]{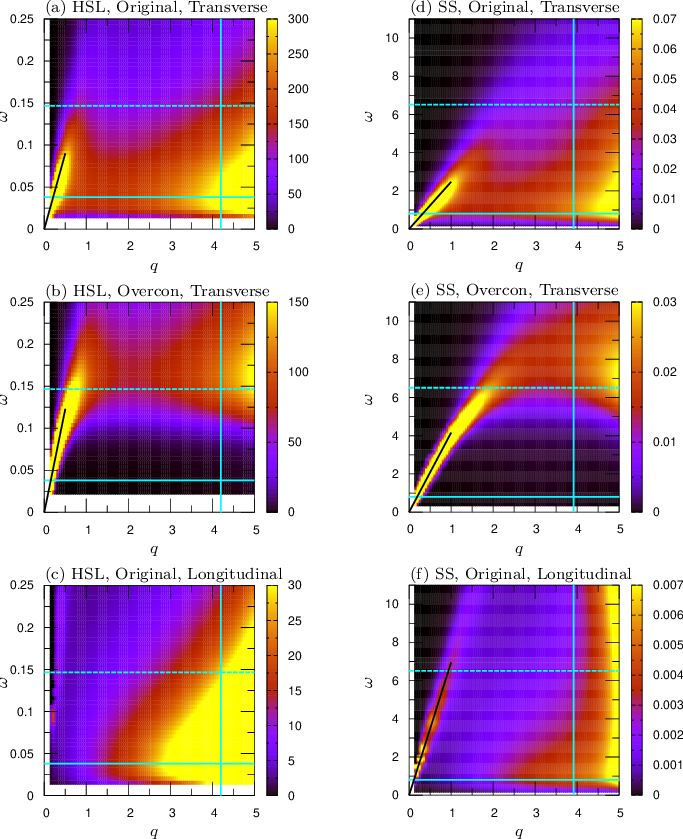}
\caption{\label{fig4_total2_debyey}
{Dynamical structure factors in HSL and SS glasses.}
(a-c) HSL glass and (d-f) SS glass.
Transverse $S_T(q,\omega)/(k_B T)$ is shown as a function of $q$ and $\omega$ for the original systems in (a,d) and the overconstrained-network systems in (b,e), whereas longitudinal $S_L(q,\omega)/(k_B T)$ is shown for the original systems in (c,f).
The vertical line marks the Debye wavenumber $q_D$.
Horizontal solid and dotted lines indicate the BP frequency $\omega_\text{BP}$ for the original and overconstrained-network systems, respectively.
The black solid curve shows the linear dispersion $\omega=c_\alpha q$ for phonon excitations, with $\alpha=T$ or $L$.
}
\end{figure*}
%%%%%%%%%%%%%%%%%%%%%%%%%%%%%%%%%%%%%%%%%%%%%%%%%%%%%%%

%%%%%%%%%%%%%%%%%%%%%%%%%%%%%%%%%%%%%%%%%%%%%%%%%%%%%%%
\begin{table*}[t]
\caption{\label{tables1}
{Physical quantities including elastic moduli and Debye values.}
For silica glass, the quantities are measured as mass density $\rho$ (g/cm$^3$), elastic moduli $K,G$ (GPa), sound speeds $c_\alpha$ (m/s), wavenumber $q$ (\AA$^{-1}$), Debye level $A_D$ (THz$^{-3}$), and frequency $\omega$ (THz).
In addition to values of the original system (reported in Table~3 of the main text), we list results for the overconstrained-network system and the isostatic-network system.
}
\centering
\renewcommand{\arraystretch}{2.0}
\setlength{\tabcolsep}{1.8pt}
\begin{tabular}{|c|c|c|c|c|c|c|c|c|c|c|c|c|c|c|c|c|c|c|}
\hline
 & & $\rho$ & $K$ & $K_A$ & $K_N$ & ${\displaystyle \frac{K_N}{K_A}}~(\%)$ & $G$ & $G_A$ & $G_N$ & ${\displaystyle \frac{G_N}{G_A}}~(\%)$ & $\nu$ & $c_L$ & $c_T$ & ${\displaystyle \frac{c_L}{c_T}}$ & $q_D$ & $A_D$ & $\omega_D$ & $\omega_\text{BP}$  \\
\hline
Silica & Original  & $2.20$ & $40.9$ & $172$ & $131$ & $76.3$ & $30.5$ & $104$ & $73.6$ & $73.7$ & $0.202$ & $6090$ & $3720$ & $1.64$ & $1.58$ & $0.00257$ & $10.5$ & $1.21$  \\
\cline{2-19}
       & Overcon & $2.20$ & $53.6$ & $203$ & $149$ & $73.6$ & $29.4$ & $121$ & $92.4$ & $75.8$ & $0.268$ & $6500$ & $3660$ & $1.78$ & $1.58$ & $0.00265$ & $10.4$ & $1.93$  \\
\cline{2-19}
       & Isostatic & $2.20$ & $\approx 0$ & $196$ & $196$ & $100$ & $\approx 0$ & $118$ & $118$ & $100$ & $-$ & $\approx 0$ & $\approx 0$ & $-$ & $1.58$ & $+\infty$ & $\approx 0$ & $\approx 0$  \\
\hline
HSH & Original & $1.40$ & $0.544$ & $0.674$ & $0.131$ & $19.4$ & $0.122$ & $0.344$ & $0.222$ & $64.4$ & $0.395$ & $0.712$ & $0.296$ & $2.40$ & $4.36$ & $0.965$ & $1.46$ & $0.0970$ \\
\cline{2-19}
    & Overcon & $1.40$ & $0.548$ & $0.641$ & $0.0932$ & $14.4$ & $0.246$ & $0.385$ & $0.138$ & $36.0$ & $0.305$ & $0.793$ & $0.420$ & $1.89$ & $4.36$ & $0.351$ & $2.04$ & $0.469$ \\
\hline
HSL & Original & $1.25$ & $0.332$ & $0.478$ & $0.145$ & $30.4$ & $0.0406$ & $0.281$ & $0.240$ & $85.5$ & $0.441$ & $0.556$ & $0.180$ & $3.09$ & $4.20$ & $4.70$ & $0.81$ & $0.0379$ \\
\cline{2-19}
    & Overcon & $1.25$ & $0.347$ & $0.475$ & $0.128$ & $26.9$ & $0.0757$ & $0.285$ & $0.209$ & $73.4$ & $0.398$ & $0.598$ & $0.246$ & $2.43$ & $4.20$ & $1.88$ & $1.17$ & $0.146$ \\
\hline
LJ & Original & $1.015$ & $61.2$ & $61.7$ & $0.530$ & $0.859$ & $13.6$ & $35.8$ & $22.2$ & $61.9$ & $0.396$ & $8.84$ & $3.67$ & $2.41$ & $3.92$ & $0.000699$ & $16.3$ & $1.05$ \\
\cline{2-19}
   & Overcon & $1.015$ & $60.6$ & $61.0$ & $0.415$ & $0.680$ & $21.6$ & $36.6$ & $15.0$ & $41.1$ & $0.341$ & $9.39$ & $4.61$ & $2.04$ & $3.92$ & $0.000360$ & $20.3$ & $4.54$ \\
\hline
SS & Original & $1.015$ & $40.8$ & $40.8$ & $0.00$ & $0.00$ & $6.21$ & $14.7$ & $8.50$ & $57.8$ & $0.428$ & $6.96$ & $2.47$ & $2.81$ & $3.92$ & $0.00225$ & $11.0$ & $0.798$ \\
\cline{2-19}
   & Overcon & $1.015$ & $35.4$ & $35.4$ & $0.00$ & $0.00$ & $17.6$ & $21.2$ & $3.66$ & $17.2$ & $0.287$ & $7.61$ & $4.16$ & $1.83$ & $3.92$ & $0.000499$ & $18.2$ & $6.51$ \\
\hline
\end{tabular}
\end{table*}
%%%%%%%%%%%%%%%%%%%%%%%%%%%%%%%%%%%%%%%%%%%%%%%%%%%%%%%

%%%%%%%%%%%%%%%%%%%%%%%%%%%%%%%%%%%%%%%%%%%%%%%%%%%%%%%%%%%%%%%%%%%%%%%%%%%%%%%%%%%%%%%%%%%%%%%%%%%%%%%%%%%%%%%%%%
\clearpage
\section*{Effective-medium mean-field analysis}
Here, we carry out an effective-medium theory (EMT) analysis based on random spring networks, following Refs.\cite{Wyart_2010,Degiuli_2014}.
In this framework, particles (nodes) are connected by linear springs to form a random network.
Two parameters control stability: the connectivity (contact number) $z$ and the level of pre-stress $e~(>0)$, which quantifies the internal forces carried by the springs.
Mechanical stability requires $z$ to exceed the isostatic threshold $z_c=2d$ in $d$ dimensions, \textit{i.e.,} the excess contact number $\delta z=z-z_c>0$.
In addition, if $e$ exceeds a critical value $e_c~(>0)$, the network becomes unstable; hence stability requires $e\le e_c$.
At $e=e_c$ the system lies exactly at the stability boundary, \textit{i.e.,} it is marginally stable.
However, this ``marginal stability" is in the sense of mean-field limit, and it does not match with the marginal stability of the quenched glasses studied in our simulations and experiments.
Under an effective-medium (mean-field) approximation, one can compute the complex, frequency-dependent effective spring constant $k_{\rm eff}(\omega)$ that characterizes the elastic response of the network.
For detailed derivations and the full formulation, see Refs.~\cite{Wyart_2010,Degiuli_2014}.

Given $k_{\rm eff}(\omega)$, the vDOS $g(\omega)$ and the dynamical structure factor $S(q,\omega)$ are computed via
\begin{align}
g(\omega) &= \frac{2m\omega}{\pi}\,\mathrm{Im}\!\left[\frac{3}{q_D^3}\int_0^{q_D}\frac{q^2\,dq}{k_{\rm eff}(\omega)\,q^2 - m\omega^2}\right], \\
S(q,\omega) &= \frac{k_B T}{\pi}\,\frac{q^2}{\omega}\,\mathrm{Im}\!\left[\frac{1}{k_{\rm eff}(\omega)\,q^2 - m\omega^2}\right],
\end{align}
where $q_D$ denotes the Debye wavenumber, and $\mathrm{Im}$ denotes the imaginary part.
In this theory longitudinal and transverse polarizations are not distinguished, so $S(q,\omega)=S_L(q,\omega)=S_T(q,\omega)$.
The two expressions obey
\begin{equation}
\frac{g(\omega)}{\omega^{2}} = \frac{2m}{q_D^{3}} \int_{0}^{q_D} 3\,\frac{S(q,\omega)}{k_B T}\,dq,
\end{equation}
which corresponds to Eq.~(3) in the main text upon identifying $3S\!\to\! S_T+S_L$ (sum over two transverse and one longitudinal polarizations) and, in the EMT setting with identical nodes, $m$ playing the role of the effective mass.
In what follows, we set the Debye wavenumber to $q_D = 4$ (dimensionless units of the theory) and $m=1$.

\begin{figure*}[t]
\centering
\includegraphics[width=0.9\textwidth]{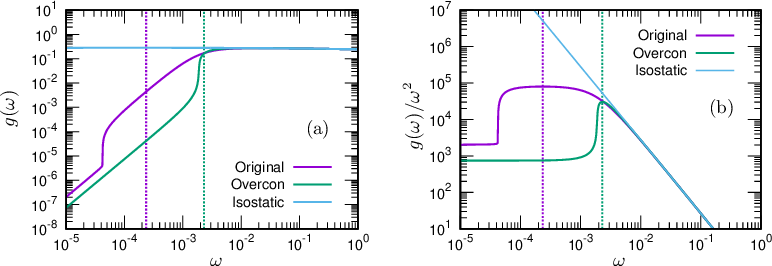}
\vspace*{0mm}
\caption{\label{figt1_emt}
{Vibrational density of states predicted by the effective-medium theory.}
Panels (a) and (b) show $g(\omega)$ and $g(\omega)/\omega^{2}$, respectively, for the original system (purple), the overconstrained-network system (green), and the isostatic-network system (cyan).
Parameter choices are as follows: the original system has $\delta z=10^{-2}$ and $e=e_c(1-5\times10^{-4})$; the overconstrained-network system has $\delta z=10^{-2}$ and $e=0$; and the isostatic-network system has $\delta z=0$ and $e=0$.
The vertical dotted lines mark the BP frequency $\omega_\text{BP}$ of the original system (purple) and the overconstrained-network system (green).
For the isostatic-network system, $\omega_\text{BP} \to 0$.
}
\end{figure*}

First, to eliminate pre-stress we set $e=0$ and tune the connectivity to $z=z_c$.
This corresponds to the isostatic-network system.
Figure~\ref{figt1_emt} shows $g(\omega)$ in (a) and $g(\omega)/\omega^2$ in (b) (cyan curves).
As $\omega\to 0$, $g(\omega)$ remains finite and $g(\omega)/\omega^2\to\infty$.
Thus, the low-frequency spectrum at isostaticity comprises strictly zero-frequency floppy modes together with many additional soft, low-frequency modes of isostatic origin.

Next, we introduce excess constraints by setting $\delta z=z-z_c=10^{-2}$ while keeping $e=0$.
This corresponds to the overconstrained-network system (often termed the ``unstressed system'' in the effective-medium literature).
In this case $g(\omega)\to 0$ as $\omega\to 0$, and $g(\omega)/\omega^2$ converges to the Debye level $A_D$, \textit{i.e.,} $g(\omega)\simeq A_D\,\omega^2$ (green curves).
Introducing excess constraints lifts both the strictly floppy modes and the additional isostaticity-derived soft modes to finite frequencies, where they merge into a weakly dispersive, non-phononic band that forms the BP.

Finally, we include pre-stress to mimic frustration by setting $e=e_c(1-5\times10^{-4})$ with $\delta z=10^{-2}$ held fixed.
This choice indicates $1-e/e_c=5\times10^{-4}$, \textit{i.e.}, the network is tuned arbitrarily close to the marginally stable point.
As noted above, the EMT treats vibrational modes as spatially extended and therefore does not capture localization or the characteristic $\omega^{4}$ vDOS of quasi-localized vibrations; consequently, the EMT notion of ``marginal stability" differs from that realized in quenched glasses in simulations and experiments.
In particular, at exactly $e=e_c$ the mean-field BP collapses to $\omega_{\text{BP}}\to 0$, which disagrees with simulations and experiments.
To enable a meaningful comparison with the original system in the main text (the quenched glasses studied in our simulations and experiments), we therefore adopt the slightly subcritical choice $e=e_c(1-5\times10^{-4})$.
Under this pre-stress, the isostaticity-derived soft-mode band shifts toward lower frequencies, and the BP correspondingly moves downward as $\omega_{\text{BP}}$ decreases (purple curves), in agreement with our simulations.

\begin{figure*}[t]
\centering
\includegraphics[width=0.9\textwidth]{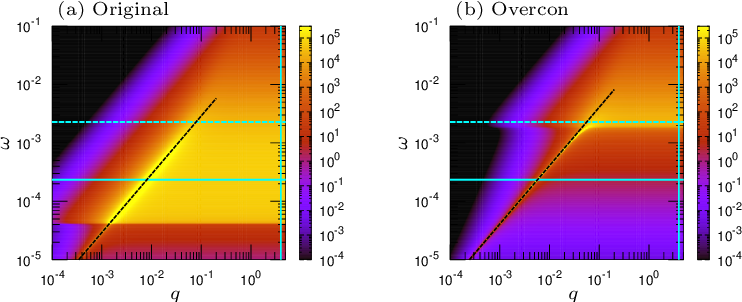}
\vspace*{0mm}
\caption{\label{figt2_emt}
{Dynamical structure factor predicted by the effective-medium theory.}
Panels (a) and (b) show $S(q,\omega)/(k_BT)$ for the original system and the overconstrained-network system, respectively.
Parameter choices: the original system has $\delta z=10^{-2}$ and $e=e_c(1-5\times10^{-4})$, and the overconstrained-network system has $\delta z=10^{-2}$ and $e=0$.
For reference, the vertical line marks the Debye wavenumber $q_D$, and the horizontal solid and dashed lines indicate the BP frequency $\omega_\text{BP}$ for the original system and the overconstrained-network system, respectively.
The black dashed curve shows the linear dispersion $\omega = c\, q$, corresponding to phonon excitations.
}
\end{figure*}

The same trends appear in the dynamical structure factor.
Figure~\ref{figt2_emt} shows $S(q,\omega)/k_BT$ for the original system ($\delta z=10^{-2}$, $e=e_c(1-5\times10^{-4})$) in (a) and for the overconstrained-network system ($\delta z=10^{-2}$, $e=0$) in (b).
At low wavenumbers and frequencies, phonons follow the linear dispersion $\omega=c\,q$, where $c$ is the sound speed.
In addition, a broad, wavenumber-independent (dispersionless) non-phononic band appears around the BP in both cases.
Compared with the overconstrained-network case, pre-stress in the original system shifts this band to lower frequencies; this downward shift enhances hybridization between isostaticity-derived soft modes and phonons, thereby broadening the phonon ridge along $\omega=c\,q$, consistent with our simulation results.

Taken together, these EMT predictions account for, and are consistent with, our simulations and the experimental observations, establishing a coherent picture across theory, simulation, and experiment.

%%%%%%%%%%%%%%%%%%%%%%%%%%%%%%%%%%%%%%%%%%%%%%%%%%%%%%%
%%%%%%%%%%%%%%%%%%%%%%%%%%%%%%%%%%%%%%%%%%%%%%%%%%%%%%%

%%%%%%%%%%%%%%%%%%%%%%%%%%%%%%%%%%%%%%%%%%%%%%%%%%%%%%%
%%%%%%%%%%%%%%%%%%%%%%%%%%%%%%%%%%%%%%%%%%%%%%%%%%%%%%%

\end{document}